%% file: main.tex
\documentclass[11pt,a4paper]{article}

\usepackage[utf8]{inputenc}

\input{0-title.tex}

\begin{document}
\maketitle
\input{0-abstract.tex}
\tableofcontents
\input{1-introduction.tex}
\input{2-inverseproblem.tex}
\input{3-illposedness.tex}
\input{4-regularization.tex}

\input{5-toymodel.tex}

\input{6-summary.tex}
\input{Appendix}

\input{Reference.tex}
\end{document}

%% file: 0-title.tex
\def\papertitle{\bf Inverse Problem Approach for Non-Perturbative QCD: Theoretical Foundation} 
\usepackage{authblk}
\title{\papertitle}

\author[1]{Ao-Sheng Xiong\footnote{Email: xiongash21@lzu.edu.cn}}
\author[1]{Fu-Sheng Yu\footnote{Email: yufsh@lzu.edu.cn, corresponding author}}
\author[1]{Yong Zheng\footnote{Email: yzheng2018@lzu.edu.cn}}
\affil[1]{Frontiers Science Center for Rare Isotopes, and School of Nuclear Science and Technology, Lanzhou University, Lanzhou 730000, China}
\author[2]{Ting Wei\footnote{Email: tingwei@lzu.edu.cn, corresponding author}}
\affil[2]{School of Mathematics and Statistics, Lanzhou University, Lanzhou 730000, China}
\usepackage[top=1in, bottom=1.25in, left=1in, right=1in]{geometry}
\usepackage{microtype}
\usepackage{lineno}
\usepackage{xspace}
\usepackage{booktabs}
\usepackage{comment}

\usepackage{caption} %these three command get the figure and table captions automatically small

\usepackage{graphicx}  % to include figures (can also use other packages)
\usepackage{color}
\usepackage{colortbl}
\DeclareGraphicsExtensions{.pdf,.PDF,png,.PNG}

\usepackage{amsmath} % Adds a large collection of math symbols
\usepackage{amsfonts,amssymb,amsthm,bm,braket,yhmath}
\usepackage{upgreek} % Adds in support for greek letters in roman typeset

\usepackage{hyperref}
\usepackage{hyperxmp}
\usepackage{hypcap} % Internal hyperlinks to floats.
\hypersetup{hidelinks} 
\usepackage{txfonts}
\usepackage{amssymb}
\usepackage{cases}
\usepackage{algorithm}
\usepackage{algpseudocode}
\usepackage{mathrsfs}
\usepackage{color}
\usepackage{amscd,amssymb,amsthm,amsmath,bm,graphicx,psfrag}
\usepackage{epsfig,mathrsfs}
\usepackage[bf,SL,BF]{subfigure}
\usepackage{amsmath}
\usepackage{pict2e,keyval,fp}
\usepackage[bf,SL,BF]{subfigure}
\newtheorem{thm}{Theorem}[section]

\theoremstyle{definition}
\newtheorem{defn}[thm]{Definition}

\numberwithin{equation}{section}
\def\kaon   {{\ensuremath{K}}\xspace}
\def\Kbar    {{\kern 0.2em\overline{\kern -0.2em \kaon}{}}\xspace}

\def\KorKbar    {\kern 0.18em\optbar{\kern -0.18em K}{}\xspace}

\def\Dbar    {{\kern 0.2em\overline{\kern -0.2em D}{}}\xspace}

\def\DorDbar    {\kern 0.18em\optbar{\kern -0.18em D}{}\xspace}

\def\Lbar        {{\ensuremath{\kern 0.1em\overline{\kern -0.1em\Lambda}}}\xspace}
\def\LorLbar    {\kern 0.18em\optbar{\kern -0.18em \PLambda}{}\xspace}

\newcommand{\tev}{\ensuremath{\mathrm{\,Te\kern -0.1em V}}\xspace}
\newcommand{\gev}{\ensuremath{\mathrm{\,Ge\kern -0.1em V}}\xspace}
\newcommand{\mev}{\ensuremath{\mathrm{\,Me\kern -0.1em V}}\xspace}
\newcommand{\kev}{\ensuremath{\mathrm{\,ke\kern -0.1em V}}\xspace}
\newcommand{\ev}{\ensuremath{\mathrm{\,e\kern -0.1em V}}\xspace}
\newcommand{\gevc}{\ensuremath{{\mathrm{\,Ge\kern -0.1em V\!/}c}}\xspace}
\newcommand{\mevc}{\ensuremath{{\mathrm{\,Me\kern -0.1em V\!/}c}}\xspace}
\newcommand{\gevcc}{\ensuremath{{\mathrm{\,Ge\kern -0.1em V\!/}c^2}}\xspace}
\newcommand{\gevgevcccc}{\ensuremath{{\mathrm{\,Ge\kern -0.1em V^2\!/}c^4}}\xspace}
\newcommand{\mevcc}{\ensuremath{{\mathrm{\,Me\kern -0.1em V\!/}c^2}}\xspace}

\usepackage{cite} % Allows for ranges in citations
\usepackage{mciteplus}

%% file: 0-abstract.tex
\begin{abstract}
A novel theoretical framework, the inverse problem approach, is proposed to calculate non-perturbative quantities in quantum chromodynamics (QCD). Based on the dispersion relation of quantum field theory, this approach determines unknown low-energy non-perturbative quantities from known high-energy perturbative inputs via solving an inverse problem. The resulting inverse problem is rigorously proven to be ill-posed, with the solutions being unique but unstable. To address this instability, the well-established Tikhonov regularization is employed, yielding stable approximate solutions that converge to the true values as input errors vanish. The key features of this approach are illustrated through three toy models, demonstrating that solution precision can be systematically improved through reduced input errors and optimized regularization strategies. 
\end{abstract}

%% file: 1-introduction.tex
\section{Introduction}\label{sec:Introduction}
The strong interaction, one of the four fundamental forces in nature, governs the majority of subatomic phenomena and is described by quantum chromodynamics (QCD).
At high-energy scales, the QCD coupling constant is sufficiently small to allow physical quantities to be calculated perturbatively, resulting in highly successful theoretical predictions. 
However, at low-energy scales, the coupling constant grows significantly, rendering perturbative methods inapplicable. This presents one of the most formidable challenges in theoretical physics, known as the non-perturbative problem of strong interaction.

The non-perturbative challenge of QCD is one of the most fundamental barriers to advancing particle physics and nuclear physics. 
It remains a central unsolved puzzle to understand the dynamical mechanism of color confinement and hadronization, which governs how quarks bind into observable hadrons such as protons and neutrons.
The spatial or momentum distributions of quarks and gluons within hadrons are crucial for interpreting high-energy scatterings, yet their underlying features are poorly understood, such as parton distribution functions/amplitudes and fragmentation functions.
The internal structures of exotic hadronic states observed in experiments remain under debate, with interpretations including compact multiquarks, molecular states, hybrid states, and nonresonant kinematic effects. 
The properties of QCD phase transition and the precise location of the critical point in the nuclear matter phase diagram remain major challenges for both theory and experiment.
Moreover, indirect probes of new physics beyond the Standard Model are intrinsically tied to non-perturbative QCD effects. For instance, the dominant theoretical uncertainties in the muon anomalous magnetic moment and the rare decays of $B\to K^{(*)}\ell^+\ell^-$ arise from the hadronic vacuum polarization and the hadronic transition form factors, respectively.

Over the past half century, several non-perturbative methods have been well developed and extensively utilized. Each of them has its advantages and disadvantages. 
Lattice QCD \cite{Wilson:1974sk}, based on the discretized Euclidean spacetime, is regarded as the most reliable and rigorous first-principles approach for non-perturbative calculations. However, it requires substantial computational resources and faces challenges in addressing real-time dynamics or properties of excited states. 
QCD sum rules \cite{Shifman:1978bx,Colangelo:2000dp} and the Dyson-Schwinger equations \cite{Roberts:1994dr} are broadly employed non-perturbative methods in the continuum Minkowski spacetime, but suffer some inevitable systematic uncertainties from the quark-hadron duality assumption and the truncation schemes, respectively. 
In addition, phenomenological models, including various quark models \cite{Chen:2016qju}, are inherently model-dependent, significantly constraining their predictive power.
As a result, the development of novel non-perturbative computational methods remains an active and important direction in the field.

In recent years, a novel non-perturbative QCD method was developed, the inverse problem approach. 
It was first proposed in 2020 during studies of $D^0-\overline D^0$ mixing \cite{Li:2020xrz}. 
In this system, the charm-quark expansion does not converge well, and perturbative estimates of the mass and width differences of neutral charmed mesons are several orders of magnitude smaller than the experimental measurements. 
Taking the charm-quark scale as non-perturbative, the relevant quantities can be obtained by solving the inverse problem of an integral equation from dispersion relation, with the input of high-energy perturbative calculations by the effective theory of heavy-quark expansion above the bottom-quark scale.
The theoretical results of the mixing parameters of $D^0-\overline D^0$ are then consistent with the experimental data \cite{Li:2020xrz}. 
Subsequently, this approach has been extended to investigate a variety of physical problems \cite{HnLi-spectrum}, demonstrating numerous advantages in its applications.
Firstly, charm physics is usually difficult in theory, since the charm-quark scale is neither high enough for the heavy-quark expansion nor low enough for the chiral perturbation theory.  
It can now benefit from the inverse problem approach as seen in $D^0-\overline D^0$ mixing.
Secondly, the whole non-perturbative region is solved simultaneously in this approach, so that the properties of excited states can be determined together with the ground state \cite{HnLi-spectrum}.  
The study of excited states is challenging for many QCD-inspired approaches, whereas the inverse problem framework may offer certain advantages in this respect.
%Thirdly, the predictive power of QCD sum rules is significantly limited by large uncertainties, primarily stemming from the assumption of global quark-hadron duality. This assumption neglects the excited states and continuum spectrum, thereby introducing a free parameter that cannot be determined from the first principles of QCD. 
%The inverse problem approach offers a significant improvement to QCD sum rules by directly solving the excited states and continuum spectrum \cite{HnLi-spectrum}, thus avoiding the model-dependent free parameter. This can increase the predictive capability and reliability of the method.

%
\begin{figure}
    \centering
    \includegraphics[width=1\linewidth]{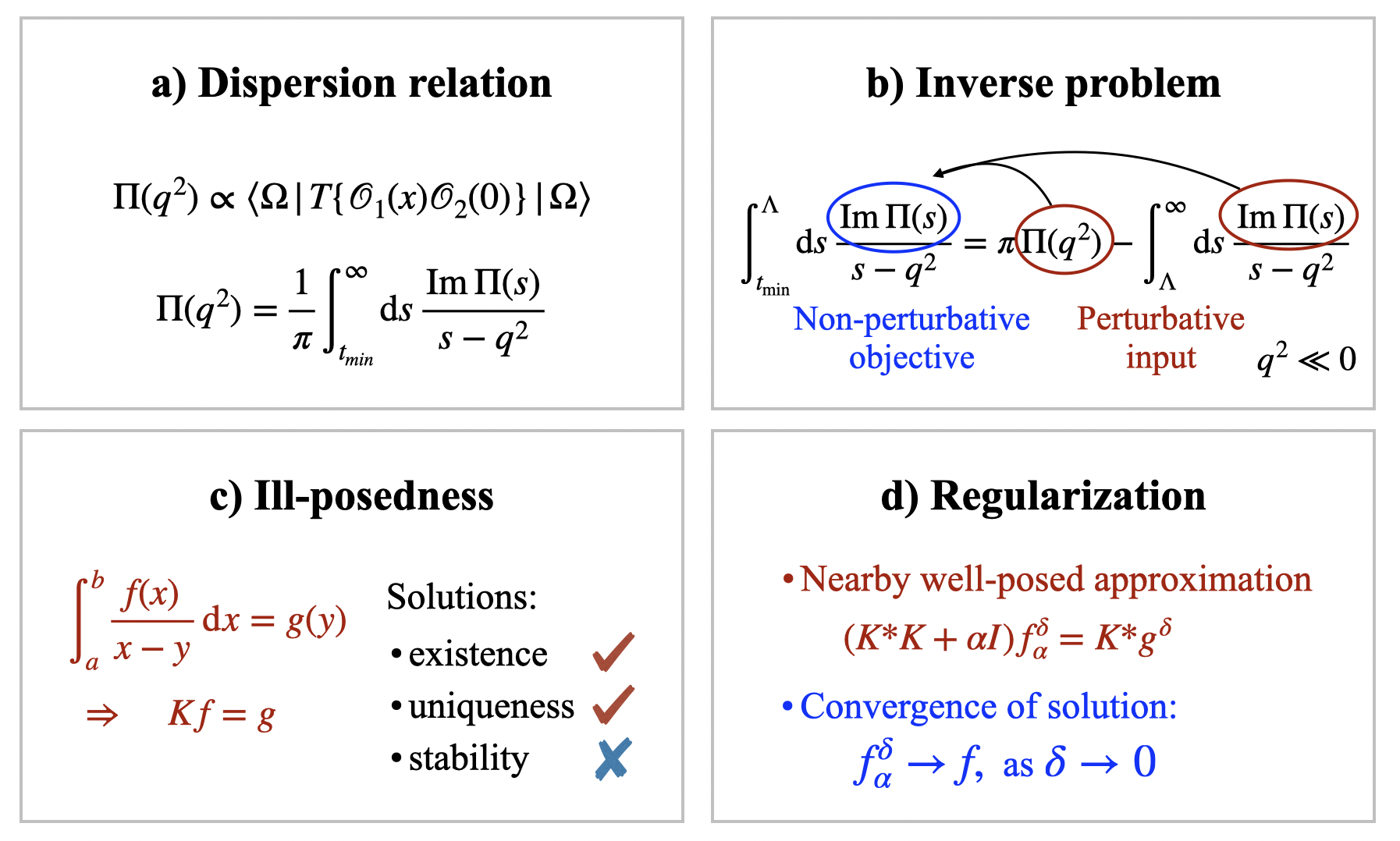}
    \caption{Illustration of the inverse problem approach of dispersion relation.}
    \label{fig:Inverse_Problem}
\end{figure}

However, the mathematical structure of the inverse problem approach in previous works has remained unclear, which is one of the main purposes of this work. 
Here, we systematically develop the theoretical framework of the inverse problem approach, aiming at building a solid mathematical foundation. 
The main ideas of the inverse problem approach are summarized below and illustrated schematically in Fig.\,\ref{fig:Inverse_Problem}.
It is based on an integral equation derived from the dispersion relation in quantum field theory, where the high- and low-energy scales are separated into distinct components.
Then the unknown non-perturbative quantities, represented as the integrand, are determined by solving an inverse problem, with the known perturbative calculations as input.
The inverse problem is a well-established field in mathematics, whose practical application generally requires a thorough investigation of its mathematical properties, particularly its inherent ill-posedness.
We prove that the dispersion-relation integral equation is ill-posed, with unique but unstable solutions. 
To overcome the ill-posed problem, regularization methods have been developed in mathematics. The core idea is to construct a nearby well-posed approximate problem that yields stable approximate solutions. These solutions converge to the true values as the input errors diminish.

In this work, we employ the standard Tikhonov regularization method, which is well known for its straightforward properties and particular friendliness towards beginners. 
The regularization parameter is determined using well-established and intuitive methods. 
The inverse problem approach is firmly grounded in rigorous mathematical principles.
Several toy models are examined to illustrate the key features of this approach. 
The results demonstrate that the solution precision can be systematically improved as input errors decrease, as well as by improving prior information and regularization methods.

This paper is organized as follows. In Sec.\,\ref{sec:dispersion_relation}, we give a brief introduction to the inverse problem of the dispersion relation. In Sec.\,\ref{sec:illposedness}, we prove the uniqueness and instability of the inverse problem. In Sec.\,\ref{sec:regularization}, the regularization method is introduced and used to overcome the ill-posedness. In Sec.\,\ref{sec:toymodel}, three toy models are discussed to demonstrate the effectiveness of the regularization method and illustrate the key features of the inverse problem approach. Finally, conclusions and perspectives are given in Sec.\,\ref{sec:conclusion}. 
For the convenience of readers, key mathematical concepts or proofs are presented in the appendix. 

%% file: 2-inverseproblem.tex
\section{Inverse Problem of Dispersion Relation}\label{sec:dispersion_relation}

%%%%%%%%%%%%%%%%%%%%%%%%%%%%%%%%%%%%%%%%%%%%%%%%%%%%%%%%%%%%%%%%%%%%%%%
%\subsection{Dispersion Relation}

The inverse problem approach for dispersion relations stems from the fact that they unify the perturbative and non-perturbative contributions within a single integral equation. 
In this section, we will introduce the inverse problem of dispersion-relation integral equations.

Dispersion relations have been widely used in particle physics and nuclear physics. 
The QCD sum rules approach is based on dispersion relations, which allow the calculation of hadronic properties using perturbative operator 
product expansion techniques \cite{Shifman:1978bx}. 
Dispersion relations also play a key role in hadron-hadron scattering, particularly in chiral perturbation theory \cite{Chew:1957zz,Mandelstam:1958xc,Dobado:1992ha}.
The nucleon charge radius can be determined via dispersion relations, which relates the full form factors to the integration of their imaginary parts \cite{Lin:2021xrc}.
Dispersion relations are also crucial in indirect searches for new physics beyond the Standard Model, such as analyzing charm-quark loop effects in $B\to K^*\ell^+\ell^-$ decays \cite{Khodjamirian:2010vf} and determining hadronic vacuum polarization contributions to muon g-2 through a data-driven approach \cite{Aoyama:2020ynm}.
The core principle of dispersion relations is to determine non-perturbative hadronic quantities by leveraging perturbative calculations or experimental measurements.

Dispersion relations arise from the causality and hence the analytic properties of quantum field theory (QFT). 
We outline the derivation of a general dispersion relation, which starts from a well-defined quantity in QFT, such as a correlation function
\begin{equation}
    \Pi(q^2) = i \int \mathrm{d}^4 x \, e^{iqx} \langle\Omega| T\{\mathcal{O}_1(x)\mathcal{O}_2(0) \}|\Omega\rangle \,.
\end{equation}
This two-point correlation function is simple and convenient for discussion, without the explicit expressions of Lorentz structures. 
It involves the time-ordered product ($T$) of two local operators at distinct four-dimensional spacetime points, $\mathcal{O}_1(x)$ and $\mathcal{O}_2(0)$, evaluated in the QCD vacuum state $|\Omega\rangle$. $\Pi(q^2)$ is defined in the momentum space. 

Then we consider all singularities of the correlation function in the complex plane. 
\begin{figure}    
\centering    
\includegraphics[width=0.5\linewidth]{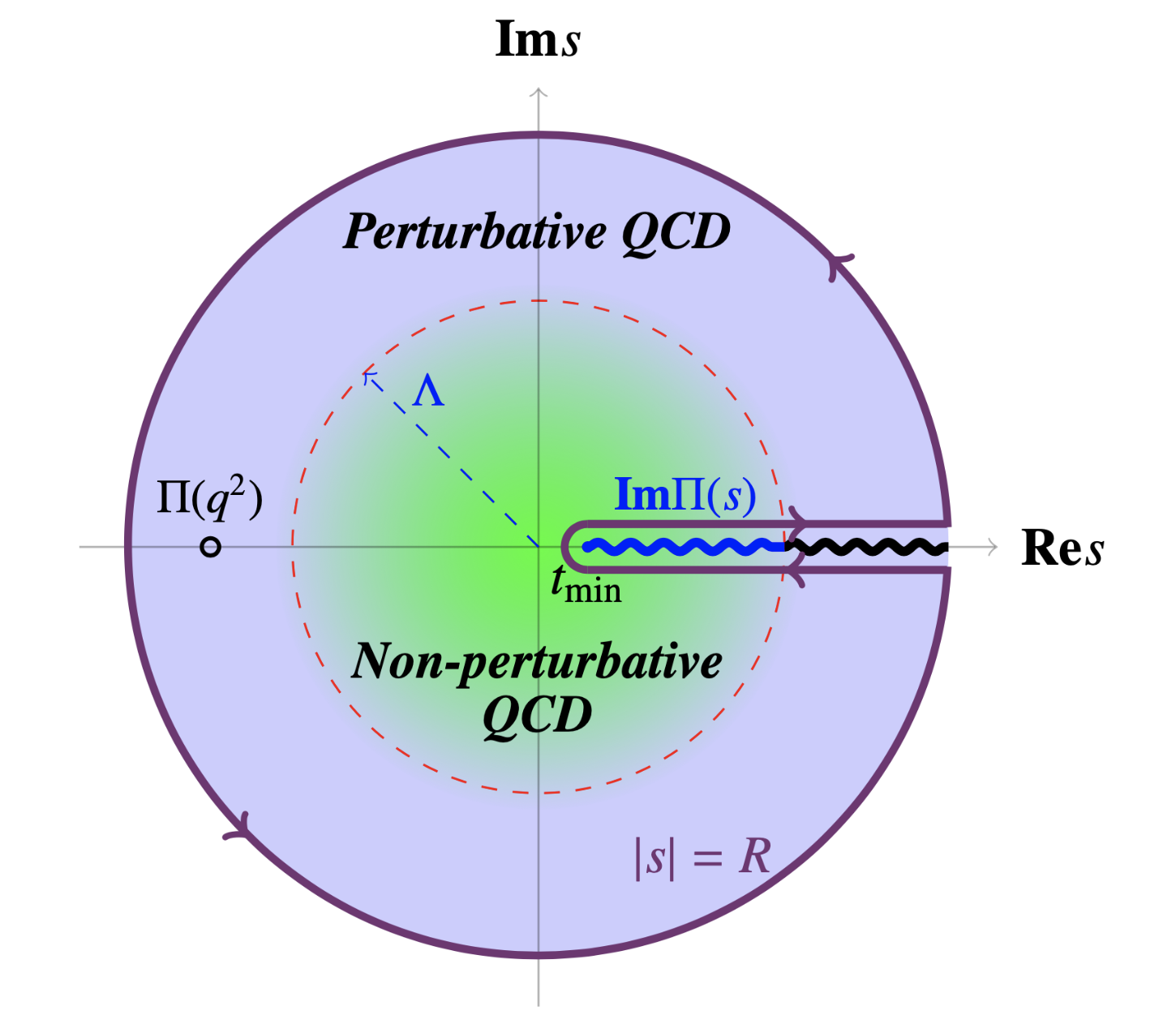}   
\caption{Contour in the complex $s$ plane of the correlation function. The physical branch cut starts at $t_{\rm min}$ and extends along the positive real axis, shown as wave lines. The contour runs just above and below the cut, and closes through the large circle $|s|=R$. The scale $\Lambda$ separates the perturbative and non-perturbative regimes of QCD. With $q^2<-\Lambda$, Eq.~(\ref{eq:InverseProblem-DispersionRelation}) serves as the integral equation for the inverse problem of dispersion relation, where non-perturbative quantities are determined from perturbative calculations. }    \label{fig:ComplexPlane}
\end{figure}
In QCD, the complex plane is free from singularities except along the positive real axis, as illustrated in Fig.\,\ref{fig:ComplexPlane}.
Within the region enclosed by the contour, the correlation function $\Pi(s)$ is analytic. 
Applying the Cauchy integral formula of complex functions, we derive the following dispersion relation,
\begin{equation}
    \begin{aligned}
         \Pi(q^2) &= \frac{1}{2\pi i} \oint_C \mathrm{d}s\,\frac{\Pi(s)}{s-q^2} \\
        &= \frac{1}{2\pi i} \oint_{| s |=R} \mathrm{d}s\,\frac{\Pi(s)}{s-q^2}
         + \frac{1}{2 \pi i} \int_{t_{\rm min}}^R \mathrm{d}s\,\frac{\Pi(s+i\epsilon)-\Pi(s-i\epsilon)}{s-q^2} \,,
    \end{aligned}\label{Complex dispersion relation}
\end{equation}
where $R$ is the radius of the large circle in Fig.\,\ref{fig:ComplexPlane}, taken to be infinitely large, and $t_{\mathrm{min}}$ denotes the threshold of the lowest singularities. Here, $\epsilon>0$ is an infinitesimal quantity introduced to shift the path slightly above or below the positive real axis.

The first term on the right-hand side of Eq.~(\ref{Complex dispersion relation}) vanishes if $\Pi(s)$ decays to zero as $s \to \infty$, leaving only the integral along the real $s$-axis. In many realistic physical scenarios, however, $\Pi(s)$ does not asymptotically vanish at infinity. In such cases, subtracted or derivative dispersion relations are required to eliminate the contribution from the large-circle contour integral.
Nevertheless, the proof of ill-posedness and the necessity of regularization depend solely on the smoothness and compactness of the integral kernel, rather than the specific form of the dispersion relation. Therefore, to avoid unnecessary algebraic complexity without loss of generality, we restrict our analysis in this work to the simplest case where $\Pi(s) \to 0$ as $s \to \infty$, which eliminates the integral term of the large-circle contour.

Applying the Schwarz reflection principle, which gives $\Pi(s + i\epsilon) - \Pi(s - i\epsilon) = 2i \operatorname{Im} \Pi(s)$ for $s > t_{\mathrm{min}}$, the dispersion relation becomes
\begin{equation}
    \begin{aligned}
        \Pi(q^2) = \frac{1}{\pi} \int_{t_{min}}^\infty \mathrm{d}s\,\frac{\operatorname{Im}\Pi(s)}{s-q^2}\,.
    \end{aligned} \label{Inf dispersion relation}
\end{equation}
With $q^2<0$, it is widely used in QCD or light-cone sum rules \cite{Shifman:1978bx,Colangelo:2000dp,Balitsky:1989ry,Braun:1988qv}, as well as in relating space-like and time-like form factors \cite{Lin:2021xrc,Khodjamirian:2010vf,Aoyama:2020ynm}. 
For $q^2>0$, it becomes a dispersion relation between the real and imaginary parts of correlation functions \cite{Li:2020xrz}, with the following discussions and conclusions remain applicable.  

%=============================================
%\subsection{Inverse Problem}
Using the dispersion relation in Eq.(\ref{Inf dispersion relation}) as a bridge between perturbative and non-perturbative physics, we formulate an inverse problem to calculate non-perturbative quantities from perturbative input.
In the deep Euclidean region of $q^2 \ll 0$, the correlation function $\Pi(q^2)$ is perturbatively calculable by operator-product expansions (OPE). 
Correspondingly, for the spectral function $\operatorname{Im} \Pi(s)$ with $s>0$, we identify a characteristic scale $\Lambda$ that separates perturbative and non-perturbative regimes.
The region $t_{\rm min} <s<\Lambda$ contains non-perturbative hadronic information, while for $s>\Lambda$, $\operatorname{Im} \Pi(s)$  can be reliably calculated using perturbation theory. 
In QCD, the transition between perturbative and non-perturbative regimes is not sharply defined. Hence $\Lambda$ can be chosen within a reasonable interval, with the physical results expected to be largely insensitive to variations within this range.
This separation of scales leads to a modified dispersion relation, which we formulate explicitly as an inverse problem,
\begin{equation}
    \begin{aligned}
        \int_{t_\mathrm{min}}^\Lambda \mathrm{d}s\,\frac{\operatorname{Im} \Pi(s)}{s-q^2} = \pi \,\Pi(q^2) -  \int_\Lambda^\infty \mathrm{d}s\,\frac{\operatorname{Im} \Pi(s)}{s-q^2}\,.
    \end{aligned} \label{eq:InverseProblem-DispersionRelation}
\end{equation}
In this form, the right-hand side serves as a known source term built from perturbative calculations. 
The task is to solve this equation for the unknown non-perturbative spectral density $\operatorname{Im} \Pi(s)$ in the low-energy region on the left-hand side. 
The core challenge is to extract the integrand function from information about its integral. 
Fig. \ref{fig:Inverse_Problem} provides a schematic illustration of the inverse problem of dispersion relation.

To simplify the subsequent discussion, we consider a reduced form of the modified dispersion relation,
\begin{equation}
    \begin{aligned}
        \int_{t_{\rm min}}^\Lambda \frac{f(x)}{x-y}\,\mathrm{d}x = g(y)\,, \quad \textrm{with} ~ y \in [q^2_{\rm low},q^2_{\rm up}]\,,
    \end{aligned}\label{eq:fxgy}
\end{equation}
where $\Lambda>t_{\rm min}\geqslant 0$ and $q^2_{\rm low}<q^2_{\rm up}\ll0$.
$f(x)$ is the non-perturbative spectral density $\operatorname{Im} \Pi(s)$, while $g(y)$ represents the perturbatively calculable right-hand side of Eq.\,(\ref{eq:InverseProblem-DispersionRelation}).
It is a Fredholm integral equation of the first kind. The corresponding inverse problem is to determine the unknown function $f(x)$ from the known input $g(y)$. 

This integral equation can be expressed as an operator equation of the form,
\begin{equation}
    \begin{aligned}
        Kf=g \,, \quad  \textrm{with} ~ f \in F \,, ~ g \in G\,,
    \end{aligned} \label{eq:Kf=g}
\end{equation}
where $K: F\to G$ is the linear integral operator and $F$ and $G$ are the function spaces  $F=L^2(t_{\rm min},\Lambda)$ and $G=L^2(q_{\rm low}^2,q_{\rm up}^2)$, respectively. 
A brief introduction to $L^2$ space is given in Appendix \ref{Hilbert space}. 
Consequently, reconstructing $f$ from the operator $K$ and the input data $g$ constitutes a linear inverse problem \cite{Kirsch-2011}.

%% file: 3-illposedness.tex
\section{Ill-posedness of the Inverse Problem}\label{sec:illposedness}

The integral equation in Eq. (\ref{eq:InverseProblem-DispersionRelation}) provides a framework for obtaining non-perturbative quantities, but solving it presents a substantial challenge. 
Intuitively, one might discretize Eqs. (\ref{eq:fxgy}) and (\ref{eq:Kf=g}) into the matrix formulation $K_{ij}f_j=g_i$, with the formal solution $f_j=(K^{-1})_{ji} g_i$. 
However, this naive approach is not guaranteed to yield a suitable solution, due to the potential unboundedness (or divergence) of $K^{-1}$.
It is evident from the fact that the inverse of a matrix is inversely proportional to its determinant $K^{-1}\propto 1/{\rm det}(K)$. 
To illustrate, for Eq. (\ref{eq:fxgy}) with $x\in [0,1]$ and $y\in [-2,-1]$, the determinant becomes extremely small upon discretization: ${\rm det}(K)\approx 10^{-88}$ for 10 points and ${\rm det}(K)\approx 10^{-244}$ for  20 points.
This implies that any small error in the input $g(y)$ can induce an arbitrarily large deviation in the solution.

The core difficulty is the ill-posed nature of the underlying inverse problem in Eq. (\ref{eq:fxgy}). 
In this work, it is demonstrated for the first time that while the solution is unique, it is inherently unstable. 
Consequently, obtaining a reliable numerical solution is highly non-trivial. 
We begin with a rigorous proof of this ill-posedness in the present section, followed by an examination of resolution strategies in the next section. 

An ill-posed problem is defined in contrast to a well-posed one. Mathematically, the notion of well-posedness in the sense of Hadamard is formulated in the following way. 

\begin{defn}[well-posedness]
Let $F$ and $G$ be normed spaces, $K:F \to G$ a mapping. The operator equation $Kf=g$ is called well-posed if the following holds \cite{Kirsch-2011}:

1. \textbf{Existence}: For every $g\in G$ there is (at least one) $f\in F$ such that $K f=g$;

2. \textbf{Uniqueness}: For every $g\in G$ there is at most one $f\in F$ with $K f=g$;

3. \textbf{Stability}: The solution $f$ depends continuously on $g$; that is, for every sequence $(f_{n}) \subset F$ with $K f_{n}\rightarrow K f \,(n\rightarrow \infty)$, it follows that $f_{n}\rightarrow f \,(n\rightarrow \infty)$\,.
\end{defn}
A problem is classified as ill-posed if it fails to satisfy any of the aforementioned conditions. 
In the following, we will prove that the inverse problem of dispersion relation is ill-posed due to its instability. 
While a complete mathematical proof of existence remains open, the presence of non-perturbative QCD dynamics strongly suggests that a solution exists. 
Before presenting detailed proofs of uniqueness and instability, we first establish the ill-posedness of Eq. (\ref{eq:fxgy}) through an operator-theoretic analysis, with the basic functional concepts seen in Appendix \ref{sec:appendix-functional}.

\begin{thm}\label{thm:compact-K}
The operator $K$ in Eqs.~(\ref{eq:fxgy}) and (\ref{eq:Kf=g}) is linear, bounded, and compact from $F$ to $G$.
\end{thm}
\begin{proof}
It is straightforward to verify that $K f_1+K f_2=K (f_1+f_2)$ and $\alpha K f=K (\alpha f)$ which imply that the operator $K: F \to G$ operator is linear. 

For any $f \in {L^2}(t_{\rm min},\Lambda)$, applying the Cauchy-Schwarz inequality yields
\begin{eqnarray}\label{ineq}
\begin{aligned}
\|K f\|_{L^2(q_{\rm low}^2,q_{\rm up}^2)} ^2 
    &= \int_{q_{\rm low}^2}^{q_{\rm up}^2} \left(\int_{t_{\rm min}}^\Lambda \frac{1}{x-y}f(x) \,\mathrm{d}x \right)^2 \mathrm{d}y \\
&	\leqslant \int_{q_{\rm low}^2}^{q_{\rm up}^2} \mathrm{d}y \int_{t_{\rm min}}^\Lambda \left(\frac{1}{x-y} \right)^2   \mathrm{d}x \int_{t_{\rm min}}^\Lambda |f(x)|^2 \,\mathrm{d}x   \\
& = \ln\left|\frac{(\Lambda-q_{\rm up}^2)(t_{\rm min}-q_{\rm low}^2)}{(\Lambda-q_{\rm low}^2)(t_{\rm min}-q_{\rm up}^2)}\right| \, \|f\|_{L^2(t_{\rm min},\Lambda)}^2 <  + \infty\,,
\end{aligned}
\end{eqnarray}
where $\ln\big|(\Lambda-q_{\rm up}^2)(t_{\rm min}-q_{\rm low}^2)/(\Lambda-q_{\rm low}^2)(t_{\rm min}-q_{\rm up}^2)\big|$ is a finite constant. 
From inequality (\ref{ineq}), it follows that $K$ is a bounded operator.

Define the kernel function
$k(y, x) := \frac{1}{x - y}, \quad (y, x) \in [q_{\rm low}^2, q_{\rm up}^2] \times [t_{\min}, \Lambda]$.
Under the disjointness with $q_{\rm up}^2<t_{\rm min}$, $k$ is continuous and bounded on this compact rectangle. In particular
\begin{equation}
\iint_{[q_{\rm low}^2, q_{\rm up}^2] \times [t_{\min}, \Lambda]} |k(y, x)|^2 \,\mathrm{d}x\,\mathrm{d}y = C < \infty.
\end{equation}
Thus $k \in L^2\big( [q_{\rm low}^2, q_{\rm up}^2] \times [t_{\min}, \Lambda] \big)$, which means that $K$ is a \emph{Hilbert--Schmidt operator} from $L^2(t_{\min}, \Lambda)$ to $L^2(q_{\rm low}^2, q_{\rm up}^2)$.
A fundamental result in functional analysis states that every Hilbert--Schmidt operator between Hilbert spaces is compact (Hilbert--Schmidt theorem) \cite{Kirsch-2011}. Therefore, $K$ is compact.
\end{proof}

\begin{thm}\label{thm:No-inverse-of-compact-operator}
    A compact linear operator $K:F\to G$ cannot have a bounded inverse unless $F$ is a finite dimensional space \cite{Kirsch-2011}.
\end{thm}
This theorem is essential in proving that the inverse problem is ill-posed. 
In Theorem~\ref{thm:No-inverse-of-compact-operator}, a fundamental result in functional analysis asserts that in an infinite-dimensional Hilbert space, a compact operator cannot have a bounded inverse. Indeed, its inverse is necessarily unbounded and therefore discontinuous.  
This property immediately implies that the inverse problem for the dispersion relation is ill-posed in the sense of Hadamard. Consequently, the compactness of $K$ lies at the heart of the well-posedness challenges addressed in the subsequent analysis.

In the following, we provide a rigorous proof of both the uniqueness and instability of the inverse problem associated with Eq. (\ref{eq:fxgy}). For simplicity, we assume the spectral function $f(x)$ is real-valued.
\begin{thm}\label{the theorem of uniqueness}
{\rm\bf (Uniqueness).} Suppose that $f_{1}(x)$, $f_{2}(x)\in L^{2}(t_{\rm min},\Lambda)$. If $Kf_{1}=Kf_{2}=g$ for all $y \in [q_{\rm low}^2,q_{\rm up}^2]$, then $f_{1}(x)=f_{2}(x)$, a. e. $x\in [t_{\rm min},\Lambda]$.
\end{thm}

\begin{proof}
Since $K$ is linear, $Kf_{1}-Kf_{2}=K(f_{1}-f_{2})=0$. Let $\bar f=f_{1}-f_{2}$. It suffices to show that $K \bar f = 0$ implies $\bar f(x)=0$, for almost every $x\in [t_{\rm min},\Lambda]$.

The condition $|q_{\rm up}^2| > \Lambda$ ensures that $\Pi(s)$ can be perturbatively calculated for $s\in [q_{\rm low}^2,q_{\rm up}^2]$.
For $x\in [t_{\rm min},\Lambda]$ and $y\in [q_{\rm low}^2,q_{\rm up}^2]$ with $\Lambda>t_{\rm min}\geqslant0$, $q_{\rm low}^2<q_{\rm up}^2<0$, and $|q_{\rm up}^2|>\Lambda$, we have $|x/y|\leqslant \Lambda/|q_{\rm up}^2|<1$.
Therefore, the geometric series expansion $\frac{1}{x-y}=-{1\over y}\sum_{k=0}^\infty \left({x\over y}\right)^k$ converges absolutely. 
Substituting into $K\bar f$ gives
\begin{equation}
    \begin{aligned}
        K \bar f=\int_{t_{\rm min}}^{\Lambda}\frac{1}{x-y}\bar f(x)\, \mathrm{d}x
        =-\frac{1}{y}\int_{t_{\rm min}}^{\Lambda}\sum_{k=0}^{\infty}\left(\frac{x}{y}\right)^{k}\bar f(x)\,\mathrm{d}x\,.
    \end{aligned}
\end{equation}
Since $\left|\sum_{k=0}^{\infty}\left(\frac{x}{y}\right)^{k}\bar f(x)\right|\leqslant \sum_{k=0}^{\infty}\left|\frac{\Lambda}{q_{\rm up}^2}\right|^{k}|\bar f(x)|=\frac{1}{1-\Lambda/|q_{\rm up}^2|}|\bar f(x)|$ and $\int_{t_{\rm min}}^{\Lambda}|\bar f(x)|\,\mathrm{d}x<+\infty$ (because $\bar f\in L^2 \subset L^1$ on a bounded interval), the dominated convergence theorem allows exchanging the sum and integral. 
Hence,
\begin{equation}
-y\int_{t_{\rm min}}^{\Lambda}\frac{1}{x-y}\bar f(x)\, \mathrm{d}x=\sum_{k=0}^{\infty}\frac{1}{y^{k}}\int_{t_{\rm min}}^{\Lambda}x^{k}\bar f(x)\, \mathrm{d}x.
\end{equation}
Using $K\bar f=0$, we have
\begin{equation}\label{eq:sum_moment=0}
\sum_{k=0}^{\infty}\frac{1}{y^{k}}\int_{t_{\rm min}}^{\Lambda}x^{k}\bar f(x)\, \mathrm{d}x=0, \quad y\in [q_{\rm low}^2,q_{\rm up}^2]\,.
\end{equation}

The above equation is an infinite sum.
As a necessary step toward the conclusion $\bar f(x)=0$, we should firstly establish that all moments vanish,
\begin{equation}
    m_k:=\int_{t_{\rm min}}^{\Lambda}x^{k}\bar f(x)\, \mathrm{d}x=0,~~~~k=0,1,2,\cdots
\end{equation} 
The proof proceeds by considering two cases, $q_{\rm low}^2$ being negative infinity or finite.
\begin{itemize}
\item {Case 1: \( q_{\rm low}^2 = -\infty \).}  
From Eq.\, (\ref{eq:sum_moment=0}), we have
\begin{equation}
    \sum_{k=0}^{\infty} {m_k\over y^k}=0,~~~y\in(-\infty,q_{\rm up}^2).
\end{equation}
Taking the limit $y\rightarrow -\infty$ gives the first vanishing moment, $m_0=0$. 
Multiplying by $y$ and taking $y\rightarrow -\infty$ again yields the second $m_1=0$. 
Proceeding inductively, we obtain $m_k=0$ for all $k$. 

\item \noindent {Case 2: $q_{\rm low}^2$ is finite, i.e., \( q_{\rm low}^2 > -\infty \).}  
Let $z\in \mathbb{C}$ with $|z|\geqslant |q_{\rm up}^2|$. Then
\begin{equation}
    \left|{m_k\over z^k}\right|
    \leqslant{|m_k|\over |q_{\rm up}^2|^k}
    \leqslant \left({\Lambda \over |q_{\rm up}^2|}\right)^k\|\bar f\|_{L^1}.
\end{equation}
By the Weierstrass M-test, the series $\sum_{k=0}^\infty m_k/z^k$ converges absolutely and uniformly on $D=\{z: |z|\geqslant |q_{\rm up}^2|\}$. 
Since each term $m_k/z^{k}$ is analytic on $D$, the sum $\sum_{k=0}^\infty m_k/z^{k}$ is also analytic on $D$ by the Weierstrass convergence theorem. 
From Eq. (\ref{eq:sum_moment=0}), $\sum_{k=0}^\infty m_k/y^{k}=0$ holds for $y\in [q_{\rm low}^2, q_{\rm up}^2]$. 
By analytic continuation, it holds for all $y\in (-\infty, q_{\rm up}^2]$. 
The limiting argument from Case 1 then gives $m_k=0$ for all $k$.
\end{itemize}

In both cases, all moments of $\bar f(x)$ vanish. Since $C[t_{\rm min},\Lambda]$ is dense in $L^{2}(t_{\rm min},\Lambda)$, for any $\epsilon>0$, there exists $\tilde{f}(x) \in C[t_{\rm min},\Lambda]$ such that $\|\bar f(x)-\tilde{f}(x)\|_{L^{2}}\leqslant\epsilon$. 
By the Weierstrass approximate theorem, there exists a polynomial $Q_{n}(x)$ such that $\|\tilde{f}(x)-Q_{n}(x)\|_{C[t_{\rm min},\Lambda]}\leqslant\epsilon$. Hence,
\begin{equation}
    \begin{aligned}
        \|\bar f(x)-Q_{n}(x)\|_{L^{2}}
        &\leqslant \|\bar f(x)-\tilde{f}(x)\|_{L^{2}}+\|\tilde{f}(x)-Q_{n}(x)\|_{L^{2}} \\
        &\leqslant \epsilon+\sqrt{\Lambda-t_{\rm min}} \, \|\tilde{f}(x)-Q_{n}(x)\|_{C[t_{\rm min},\Lambda]}\\
        &\leqslant\epsilon+\epsilon\sqrt{\Lambda-t_{\rm min}}\,.
    \end{aligned}
\end{equation}
Since all moments vanish, $\int_{t_{\rm min}}^{\Lambda}\bar f(x)Q_{n}(x)\,\mathrm{d}x=0$. Applying the Cauchy-Schwarz inequality,
\begin{equation}
    \begin{aligned}
        \|\bar f(x)\|_{L^{2}}^{2}&=\int_{t_{\rm min}}^{\Lambda}|\bar f(x)|^2\,\mathrm{d}x
        =\int_{t_{\rm min}}^{\Lambda}\left(|\bar f(x)|^2-\bar f(x)Q_{n}(x)\right)\,\mathrm{d}x\\
        %&\leqslant \int_{t_{\rm min}}^{\Lambda}\left|\bar f(x)\right|\times\left|\bar f(x)-Q_{n}(x)\right|\,\mathrm{d}x \\
        &\leqslant \left(\int_{t_{\rm min}}^{\Lambda}|\bar f(x)|^2\,\mathrm{d}x\right)^{\frac{1}{2}}\left(\int_{t_{\rm min}}^{\Lambda}\left|\bar f(x)-Q_{n}(x)\right|^{2}\,\mathrm{d}x\right)^{\frac{1}{2}}\\
        &=\|\bar f(x)\|_{L^{2}} \cdot \|\bar f(x)-Q_{n}(x)\|_{L^{2}} \\
        &\leqslant \left(\epsilon+\epsilon\sqrt{\Lambda-t_{\rm min}}\right)  \|\bar f(x)\|_{L^{2}}\,.
    \end{aligned}
\end{equation}
Thus, $\|\bar f(x)\|_{L^{2}}\leqslant \epsilon+\epsilon\sqrt{\Lambda-t_{\rm min}}$. Letting $\epsilon\rightarrow 0$, we obtain $\|\bar f(x)\|_{L^{2}}=0$, i. e. $\bar f(x)=0$, a.e., $x\in [t_{\rm min},\Lambda]$.  Therefore, $f_1(x)=f_2(x)$ almost everywhere, proving uniqueness. 

The above proof relies on the geometric series expansion and the moment argument for $y\in [q_{\rm low}^2, q_{\rm up}^2]$ with $|q_{\rm up}^2|>\Lambda$. 
Remarkably, the uniqueness result remains valid even when $|q_{\rm up}^2| < \Lambda$, provided that the interval $[q_{\rm low}^2, q_{\rm up}^2]$ is non-degenerate. 
This is a consequence of the analyticity of the integral $\int_{t_{\min}}^{\Lambda} \frac{\bar f(x)}{x - y}\,dx$ as a function of $y$, where $y$ is a complex number not located in the interval $[t_{\min}, \Lambda]$ on the real axis. 
If this function vanishes on any subinterval of $(-\infty, 0)$, then by the identity theorem for analytic functions, it vanishes identically on $(-\infty, t_{\min})$. 
Hence, $\bar f(x) = 0$, and therefore $f_1(x)=f_2(x)$ almost everywhere in $[t_{\rm min},\Lambda]$. 
Note that the above argument regarding analyticity of the integral also ensures the uniqueness of the inverse problem for dispersion relation when $y\in [q_{\rm low}^2, q_{\rm up}^2]$ with $0<\Lambda<q_{\rm low}^2<q_{\rm up}^2$ \cite{Li:2020xrz}.

This completes the proof.
\end{proof}

\begin{thm}\label{The theorem proving the instability of the solution}
{\bf (Instability.)} 
Consider the inverse problem in Eq. (\ref{eq:fxgy}) with $0 \leqslant t_{\rm min} < \Lambda < +\infty$ and $0 > q_{\rm up}^2 > q_{\rm low}^2 > -\infty$. 
Then the problem is unstable: there exists a sequence of noisy data $g_n \to g$ in $L^2(q_{\rm low}^2, q_{\rm up}^2)$, but the corresponding solutions $f_n$ satisfy $\|f_n - f\|_{L^2(t_{\rm min},\Lambda)} \to \infty$.
\end{thm}

\begin{proof}
We construct a specific counterexample.
Let $t_{\rm min}=0$ and $\Lambda = 2$, and define the noisy solution
\begin{equation}
f_n(x) = f(x) + \sqrt{n} \cos(n\pi x), \quad x \in [0,2].
\end{equation}
Let $g_n(y)$ be the data corresponding to $f_n$, i.e.,
\begin{equation}
g_n(y) - g(y) = \int_0^2 \frac{\sqrt{n} \cos(n\pi x)}{x - y}\,dx.
\end{equation}
Since $y \leqslant q_{\rm up}^2 < 0$ and $x \in [0,2]$, we have $x - y \geqslant -y \geqslant -q_{\rm up}^2 =: d > 0$. Thus the kernel is smooth and bounded.
Integrating it by parts,
\begin{equation}
\begin{aligned}
g_n(y) - g(y) 
&= \frac{\sqrt{n}}{n\pi} \int_0^2 \frac{1}{x - y}\, d\big(\sin(n\pi x)\big) 
= \frac{1}{\sqrt{n}\pi} \left[ \frac{\sin(n\pi x)}{x - y} \Big|_{x=0}^{x=2} + \int_0^2 \frac{\sin(n\pi x)}{(x - y)^2}\,dx \right].
\end{aligned}
\end{equation}
Since $\sin(2n\pi) = \sin(0) = 0$, the boundary term vanishes. Using $|\sin(n\pi x)| \leqslant 1$ and $|x - y| \geqslant |q_{\rm up}^2|$, we obtain
$\left|g_n(y) - g(y)\right| \leqslant \frac{1}{\sqrt{n}\pi} \frac{2}{|q_{\rm up}^2|^2}$.
Therefore,
\begin{equation}
\|g_n - g\|_{L^2(q_{\rm low}^2, q_{\rm up}^2)} 
\leqslant \frac{2}{\sqrt{n}\pi |q_{\rm up}^2|^2} \sqrt{q_{\rm up}^2 - q_{\rm low}^2}
=: \frac{M}{\sqrt{n}} \xrightarrow[n\to\infty]{} 0,
\end{equation}
where $M = 2\sqrt{q_{\rm up}^2 - q_{\rm low}^2}/(\pi |q_{\rm up}^2|^2)$ is finite and independent of $n$.

On the other hand, the error in the solution is
\begin{equation}
\|f_n - f\|_{L^2(0,2)} = \left( \int_0^2 n \cos^2(n\pi x)\,dx \right)^{1/2}
= \sqrt{n} \left( \int_0^2 \cos^2(n\pi x)\,dx \right)^{1/2}= \sqrt{n} \xrightarrow[n\to\infty]{} \infty
\end{equation}
since $\int_0^2 \cos^2(n\pi x)\,dx = 1$ for all integers $n \geqslant 1$.

Thus, an arbitrarily small error in the data can produce an arbitrarily large uncertainty in the solution, proving the instability of the inverse problem.
\end{proof}

\begin{figure}
    \centering
    \includegraphics[width=0.4\linewidth]{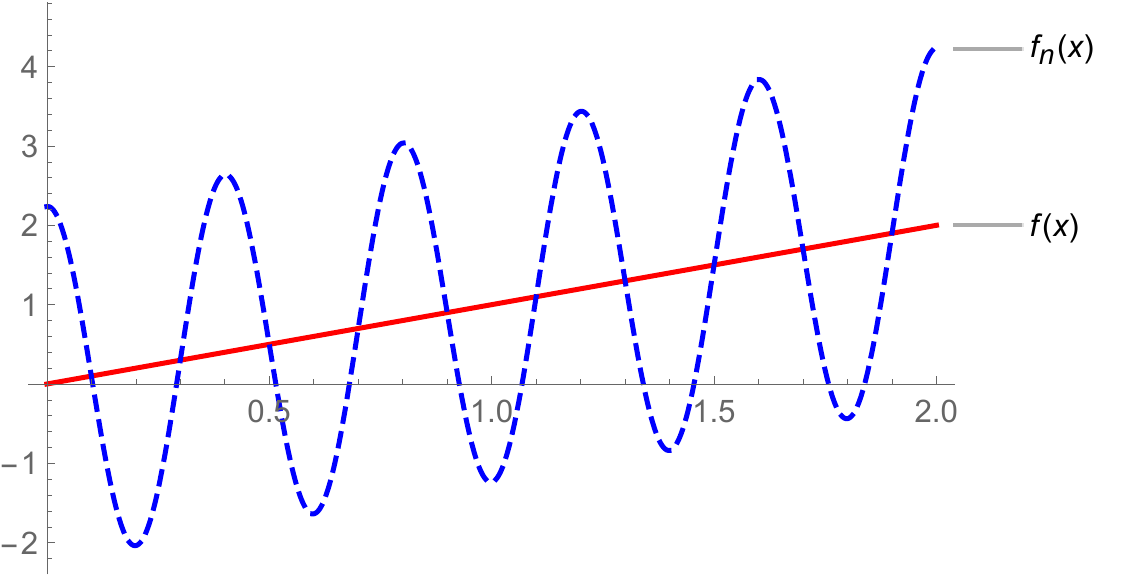}
    \hspace{1cm}
    \includegraphics[width=0.4\linewidth]{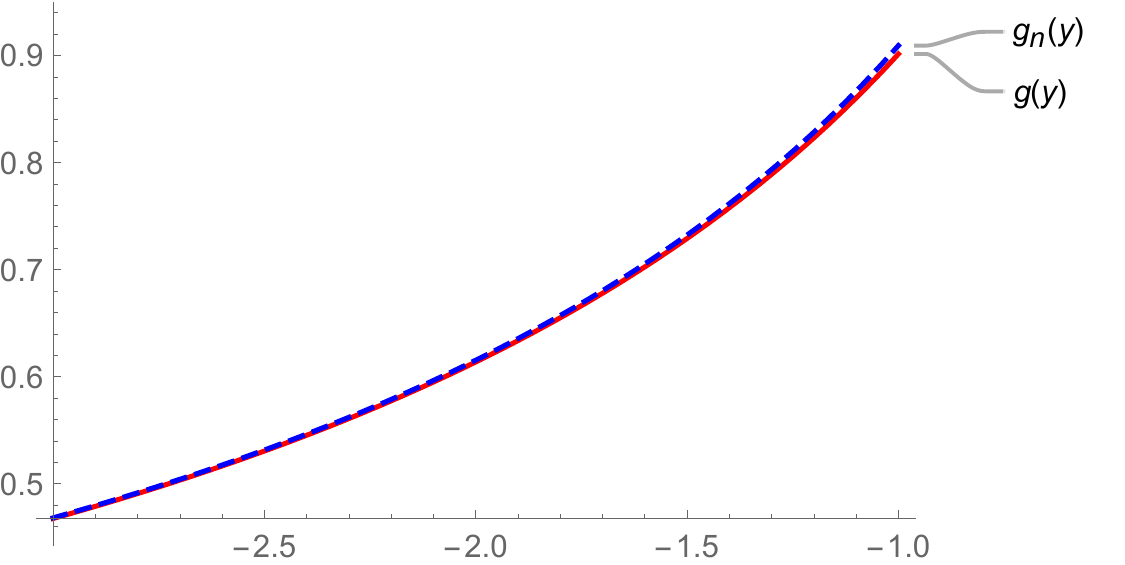}
    \caption{Illustration of instability in the inverse problem. The left panel shows the function $f(x)=x$ and a noisy function $f_{n=5}(x)=x+\sqrt{5}\cos(5\pi x)$ in the solution space. The right panel shows the corresponding data $g(y)$ and $g_{n=5}(y)$, which are nearly indistinguishable due to the smoothing effect of the integral kernel. This contrast demonstrates the instability of the inverse problem. }
    \label{fig:illposed}
\end{figure}

In summary, we have rigorously established that the inverse problem associated with the dispersion relation is ill-posed. Although the solution is unique, it lacks stability. 
This ill-posedness is illustrated in Fig. \ref{fig:illposed}. 
Uniqueness ensures a one-to-one mapping between any perturbative input and its corresponding non-perturbative solution. However, the absence of stability poses a fundamental challenge, that arbitrarily small errors in the input data can lead to large deviations in the reconstructed solution. 
Critically, because perturbation theory is only asymptotically convergent, such input errors are inevitable in practice. Consequently, regularization techniques are essential to stabilize the inversion procedure.

%% file: 4-regularization.tex
\section{Regularization Methods}\label{sec:regularization}
\begin{figure}
    \centering
    \includegraphics[width=1\linewidth]{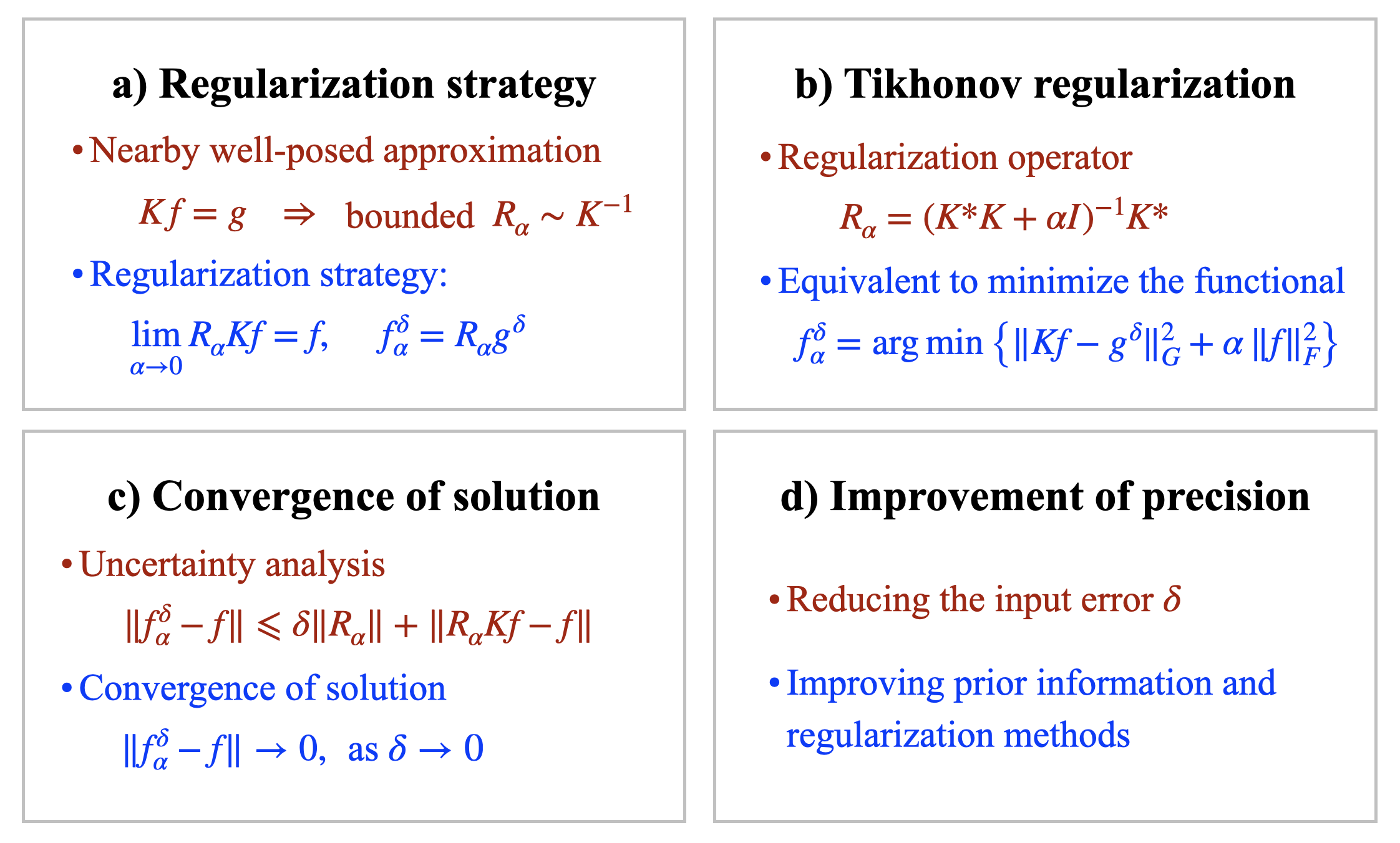}
    \caption{Illustration of the regularization method.}
    \label{fig:Illustration-Regularization}
\end{figure}

In this section, we discuss the application of regularization methods to solve ill-posed inverse problems. 
Regularization theory has been rigorously developed in mathematics since the mid-20th century and is extensively covered in numerous texts \cite{Oldest1,Oldest2,Kirsch-2011,Regularization_tools}. 
Here, we provide a brief introduction to ensure the logical flow of the article and to make the material accessible for readers with a physics background. 
The basic idea of the regularization method is illustrated in Fig. \ref{fig:Illustration-Regularization}.

%%%%%%%%%%%%%%%%%%%%%%%%%%%%%%%%%%%%%%%%%%%%%%%%%%%%%%%%%%%%%%%%%%%%
\subsection{General Regularization Theory}
The basic idea of regularization methods is to approximate an ill-posed problem by a well-posed problem that is close to the original one, ensuring that the approximate solution converges to the true solution as the input error diminishes.

\begin{defn}[Regularization strategy]
A regularization strategy is a family of linear and bounded operators $R_\alpha: G \rightarrow F$, $\alpha>0$, such that
\begin{equation}\label{eq:RegularizationOperator}
    \lim \limits_{\alpha \rightarrow 0} R_{\alpha} K f=f \quad \text{for all}~  f \in F,
\end{equation}
where $\alpha$ is called the regularization parameter \cite{Kirsch-2011}. 
\end{defn}

From this definition, it follows that the regularization operators possess two fundamental properties. 
Firstly, the operator $R_\alpha$ is bounded for any fixed $\alpha > 0$. This ensures that the approximate solution $f_\alpha = R_\alpha g$ is continuously dependent on the data $g$, rendering the approximate problem well-posed and numerically solvable even in the presence of data errors. 
Secondly, as $\alpha \to 0$, the composite operators $R_\alpha K$ converge pointwise to the identity operator on $F$, implying $R_\alpha g \to K^{-1}g$. The limit establishes both convergence and consistency.

Regularization theory for ill-posed problems can be intuitively understood through a linear algebra analogy. 
Consider a square matrix $A$ in a finite-dimensional space, which is positive semidefinite and singular (non-invertible). 
A common regularization method approximates $A$ by $A+\alpha I$, where $I$ is the identity matrix, and $\alpha>0$.
This approximation is consistent, since $A+\alpha I\to A$ as $\alpha\to 0$.
In this context, the regularization operator corresponds to  
\begin{equation}\label{eq:SqureMatrix}
    R_\alpha=(A+\alpha I)^{-1}.
\end{equation}
In the following, we demonstrate that $R_\alpha$ is well-defined and bounded. 
%As discussed in the previous section and Theorem \ref{thm:No-inverse-of-compact-operator}, ill-posedness arises because the compact linear operator $K$ lacks a bounded inverse. 
For matrices, non-invertibility is equivalent to a vanishing determinant, i.e., $\det(A)=0$. 
Since the determinant equals the product of the eigenvalues $\lambda_i$
(i.e., $\det(A)=\prod_i\lambda_i$), $\det(A)=0$ implies that at least one eigenvalue $\lambda_i$ is zero.
Regarding the eigenvalue equation $Ax=\lambda x$, we observe that $(A+\alpha I)x=(\lambda+\alpha)x$. 
Thus, $\lambda_i+\alpha$ are the eigenvalues of $A+\alpha I$. 
Since $A$ is positive semidefinite, $\lambda_i \geqslant 0$. 
For any $\alpha>0$, the shifted eigenvalues satisfy $\lambda_i+\alpha\neq0$.
Consequently, the determinant becomes $\det(A+\alpha I)=\prod_i(\lambda_i+\alpha)\neq0$. 
This implies that $A+\alpha I$ is invertible.
Therefore, the regularization operator $R_\alpha=(A+\alpha I)^{-1}$ exists and is bounded.
This illustrates that adding the term $\alpha I$ removes the singularity and makes the problem well-posed. 
The explicit formula of the regularization operator will be presented in detail in the next subsection on Tikhonov regularization. 

In the following, we discuss the error estimation. In practice, input data $g$ is typically observed as noisy data $g^\delta\in G$ with $\|g^\delta-g\|\leqslant\delta$. We define 
\begin{equation}
    f_\alpha^\delta := R_\alpha g^\delta
\end{equation}
as an approximation of the true solution of $Kf=g$. The error of solution is quantified by the norm of the discrepancy $\| f_\alpha^\delta - f\|$. 
By invoking the triangle inequality, this error can be decomposed into two parts,
\begin{align}
    \| f_\alpha^\delta - f\| &\leqslant \| f_\alpha^\delta - f_\alpha \| + \| f_\alpha - f\|
    \nonumber\\
    &= \| R_\alpha g^\delta - R_\alpha g \| + \| R_\alpha g - f\|
     \nonumber\\
    &\leqslant \| R_\alpha\| ~\| g^\delta -g\| + \| R_\alpha K f - f\|.
\end{align}
Here, the final step relies on the inequality property of the operator norm, whose definition and theorem are provided in Definition \ref{defn:operator-norm} and Theorem \ref{thm:operator-norm-inequality} of Appendix \ref{sec:appendix-functional}, respectively.
Given the error of noisy input data $\| g^\delta -g\|\leqslant \delta$, we obtain the key error bound
\begin{equation}\label{eq:error-bound}
    \| f_\alpha^\delta - f\| \leqslant  
    \delta \,\| R_\alpha\| + \| R_\alpha K f - f\| .
\end{equation}
This relation is central to our uncertainty analysis and convergence proof.  

\begin{figure}
    \centering
    \includegraphics[width=0.5\linewidth]{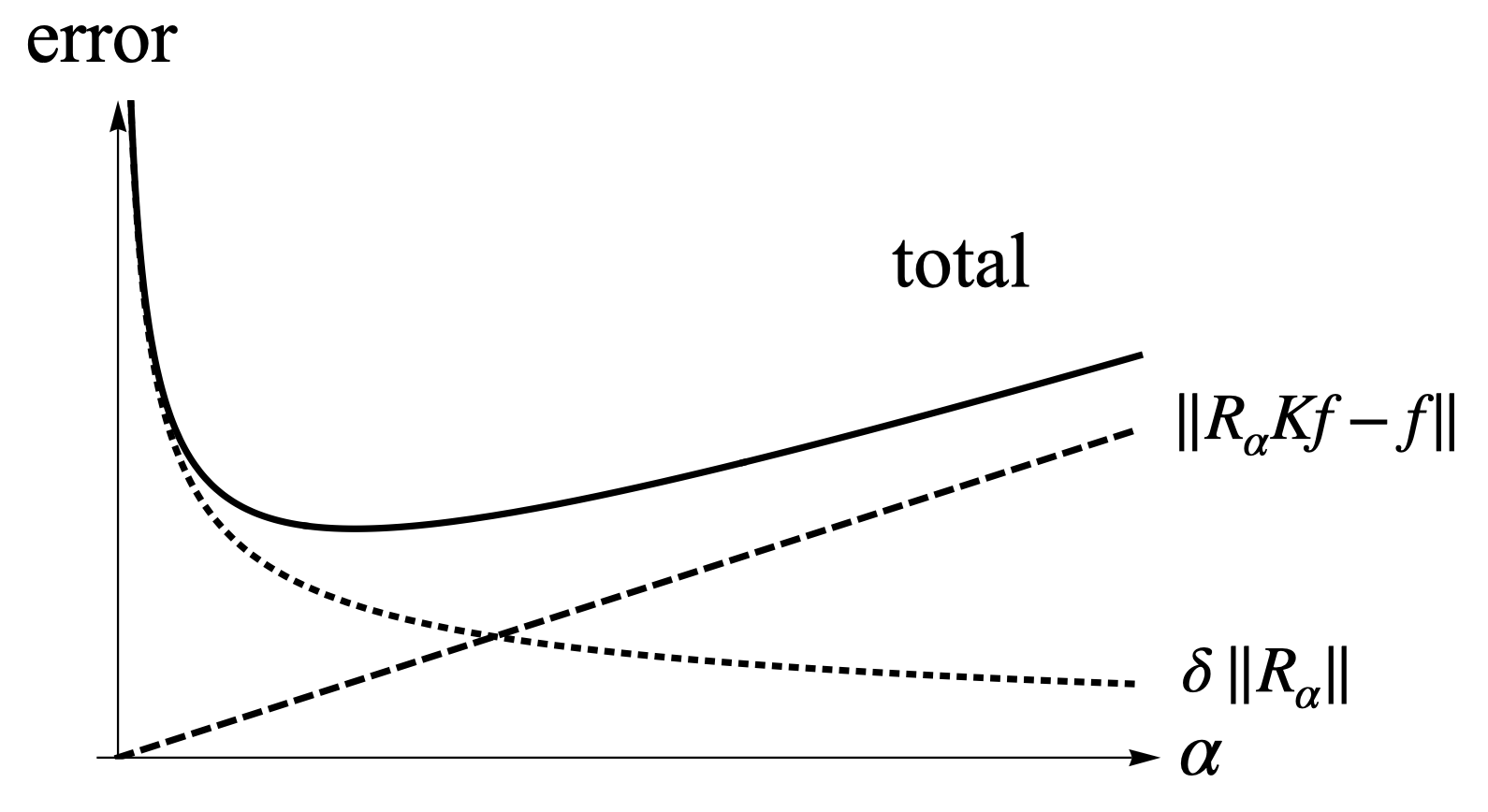}
    \caption{Illustration of the total error and its decomposition into two components of $\| R_\alpha K f - f\|$ and $\delta \,\| R_\alpha\|$. }
    \label{fig:error-illustration}
\end{figure}

The two terms in Eq.~(\ref{eq:error-bound}) are analyzed below, with their impact on the total error depicted in Fig.~\ref{fig:error-illustration}.
The first term,  $\delta ~\| R_\alpha\|$, represents the propagation of the data error into the solution error, scaled by the operator norm $\| R_\alpha\|$. For ill-posed problems where $K^{-1}$ is unbounded, the regularized operator $R_\alpha$ approaches $K^{-1}$ as $\alpha\to 0$, causing the operator norm $\| R_\alpha\|$ to blow up to infinity. In this sense, $\| R_\alpha\|$ increases as $\alpha$ decreases (increasing ill-conditioning) and decreases as $\alpha$ increases (improving stability).
The second term, $\| R_\alpha K f - f\|$, quantifies the error due to the regularization approximation. According to the definition of the regularization operator in Eq.~(\ref{eq:RegularizationOperator}), this term vanishes as $\alpha\to 0$ (i.e. $\| R_\alpha K f - f\| \to 0$). Thus, the approximation error decreases as $\alpha$ diminishes and increases as $\alpha$ grows. 

As illustrated in Fig. \ref{fig:error-illustration}, the regularization parameter $\alpha$ cannot be chosen arbitrarily large or small. Rather, it must strike a balance between two competing requirements.
On the one hand,  ensuring the convergence of the regularized solution to the true solution, $\| (R_\alpha - K^{-1})g \|\to 0$, calls for a sufficiently small $\alpha$. 
On the other hand, maintaining the stability of the solution requires that the operator norm $\| R_\alpha \|$ remain bounded, which prohibits $\alpha$ from being too small. 
Consequently, the choice of $\alpha$ reflects an inherent trade-off between approximation accuracy and numerical stability.

Therefore, the fundamental principle of regularization for solving ill-posed problems is to establish a strategy for selecting the parameter $\alpha = \alpha(\delta)$. 
This strategy aims to balance the approximation error and the stability term by effectively minimizing the following total error bound, $\delta \,\| R_\alpha \| + \| R_\alpha K f - f \|$, with respect to $\alpha$.
Crucially, as the noise level $\delta \to 0$, the choice of parameters must ensure that this total error converges to zero. This property constitutes a core principle of regularization theory that regularized solutions must converge to the true solution as data accuracy improves.

\begin{defn}[Admissible regularization strategy]\label{defn:convergence-solution}
A regularization strategy $\alpha=\alpha(\delta)$ is called {\it admissible} if, as the noise level $\delta\to 0$, the parameter satisfies $\alpha(\delta)\to 0$ and the regularized solution converges to the true solution, $f_\alpha^\delta \to f$, or equivalently, $\| f_\alpha^\delta -f \|\to 0$ \cite{Kirsch-2011}. 
\end{defn}

An intuitive interpretation of this definition involves two key conditions. First, the regularization parameter must satisfy $\alpha(\delta) \to 0$ as $\delta \to 0$, which is easily achieved by choosing $\alpha$ as a decaying function of the noise level. Second, the total error bound $\delta\,\| R_\alpha \| + \| R_\alpha K f - f \|$ must vanish in this limit. The crucial challenge lies in the stability term $\delta\,\| R_\alpha \|$. Since $\| R_\alpha \|$ typically diverges as $\alpha \to 0$ (an inverse relationship), the parameter choice strategy must ensure that this divergence is sufficiently slow. Specifically, $\| R_{\alpha(\delta)} \|$ must grow strictly slower than $1/\delta$; only then will the product $\delta \,\| R_\alpha \|$ converge to zero, guaranteeing the convergence of the regularized solution.
In the following subsection, we will demonstrate the convergence of approximate solutions for Tikhonov regularization. 

It is important to clarify that our discussion does not imply the notion of ``fine-tuning'' often encountered in physics. 
While the approximate solution formally depends on the parameter $\alpha$, the toy models presented in the next section demonstrate that the solutions remain insensitive to $\alpha$ over a broad range of values. 
This behavior aligns with physical expectations that a valid regularized solution should exhibit a stable plateau with respect to variations in $\alpha$.

%%%%%%%%%%%%%%%%%%%%%%%%%%%%%%%%%%%%%%%%%%%%%%%%%%%%%%%%%%%%%%%%%%%%%%%%%
\subsection{Tikhonov Regularization}
Before introducing a specific regularization method, we recall that such methods are designed to satisfy two fundamental properties: (1) restoring well-posedness by ensuring the existence, uniqueness, and stability of a solution; and (2) guaranteeing convergence, meaning that the regularized solution approaches the true solution as the  error of the input data tends to zero.

In this work, we employ the Tikhonov regularization method, one of the most prevalent and well-established techniques in the field.
Originally introduced by A. N. Tikhonov in the 1940s~\cite{Oldest1,Oldest2}, this approach is now central to the development of methodologies for a wide range of ill-posed problems.
Distinguished by its conceptual simplicity, the method offers an elegant framework that is both easy to implement and theoretically rigorous. 
In this subsection, we first introduce the Tikhonov regularization operator, then demonstrate the convergence of its solution, and finally present the minimization of the Tikhonov functional for practical applications.

\paragraph{Tikhonov Regularization Operator\\}
Analogous to the matrix example in the previous subsection and Eq. (\ref{eq:SqureMatrix}), Tikhonov regularization involves adding a term $\alpha I$. However, unlike the case of a square matrix, the operator $K: F\to G$ in the equation $Kf=g$ maps between two generally distinct Hilbert spaces. Directly adding $\alpha I$ to $K$ is not well defined, since the identity operator maps $F$ onto itself, resulting in a dimensional mismatch. 
To resolve this, applying the adjoint operator $K^*$ (see Definition \ref{defn:adjoint-operator}) to both sides of $Kf=g$ yields the normal equation $K^*Kf=K^*g$. The composite operator $K^*K: F\to F$ is self-adjoint, making it analogous to a square matrix. 
The original ill-posed problem, formally corresponding to $f=(K^*K)^{-1}K^*g$, is transformed into a well-posed approximate problem by introducing the regularization term $\alpha I$.
The regularized solution is given by 
\begin{equation*}
    f_\alpha=(K^*K+\alpha I)^{-1}K^*g. 
\end{equation*}
This expression defines the Tikhonov regularization operator 
\begin{equation}\label{eq:TikhonovOperator}
    R_\alpha=(K^*K+\alpha I)^{-1}K^*.
\end{equation}
For noisy data $g^\delta$, the regularized solution is computed by $f_\alpha^\delta= (K^*K+\alpha I)^{-1}K^* g^\delta$.
The operator $R_\alpha$ in Eq. (\ref{eq:TikhonovOperator}) provides a stable and bounded approximation of the unbounded inverse of $K$, and satisfies the essential convergence property in Eq. (\ref{eq:RegularizationOperator}). Hence, $R_\alpha$ constitutes a valid regularization operator.  

\paragraph{Convergence of solution\\}
In the following, we discuss the convergence of the solution under Tikhonov regularization. 
According to Definition~\ref{defn:convergence-solution}, the convergence of solutions implies that as $\delta \to 0$, we have $\alpha(\delta) \to 0$ and $\|f_\alpha^\delta - f\| \to 0$. 

Note that this should not be confused with the convergence of regularization operators, which refers to the fact that as $\alpha \to 0$, $\| (R_\alpha - K^{-1}) g \| \to 0$, i.e., $R_\alpha \to K^{-1}$. In this case, the ill-posedness is recovered.
What is meaningful is that the regularization operator consistently works as a regularization. By first letting $\delta \to 0$ and consequently $\alpha(\delta) \to 0$, we ensure $\| f_\alpha^\delta - f \| \to 0$. This is precisely the convergence of solutions.

The convergence of solutions followed directly from the error bound given in Eq.~(\ref{eq:error-bound}),
\begin{equation*}
    \| f_\alpha^\delta - f\| \leqslant 
    \delta \| R_\alpha \| + \| R_\alpha K f - f \|.
\end{equation*}
We derive upper bounds for $\delta \| R_\alpha \|$ and $\| R_\alpha K f - f \|$, respectively. It will be demonstrated that the sum of these terms vanishes as $\delta \to 0$, provided that the regularization parameter $\alpha(\delta)$ is chosen appropriately.

First, we analyze the approximation error $\| R_\alpha K f - f \|$. 
Consider a specific \textit{a priori} source condition that $f = K^*K z$ for some $z \in F$ with $\|z\|_F \leqslant E$. 
This condition imposes both boundedness and smoothness on the true solution $f$. 
The boundedness requirement is naturally satisfied in physical contexts, as all physical quantities are inherently finite. 
Regarding smoothness, the representation $f = K^*K z$ implies that $f$ lies in the range of the composite operator $K^*K$. Since compact operators (and their adjoints) possess a smoothing effect, applying $K^*K$ to $z$ yields a solution $f$ that is smoother than the source element $z$. 
%Such smoothness assumptions are standard in most inverse problems. 
In the following, we investigate the behavior of the approximation error $\| R_\alpha K f - f \|$ as a function of the regularization parameter $\alpha$.

Using the definition in Eq.~(\ref{eq:TikhonovOperator}) and letting $T = K^*K$, the error term simplifies as follows:
\begin{equation*}
    \begin{aligned}
        R_\alpha K f - f 
        &= (T + \alpha I)^{-1} T f - f \\
        &= (T + \alpha I)^{-1} \big( T f - (T + \alpha I) f \big) \\
        &= -\alpha (T + \alpha I)^{-1} f \\
        &= -\alpha (T + \alpha I)^{-1} T z,
    \end{aligned}
\end{equation*}
where the last step utilizes the source condition $f = Tz$. 
Taking the norm of both sides and applying the compatibility property of the operator norm (see Theorem \ref{thm:operator-norm-inequality}), we obtain the following estimate
\begin{equation}\label{eq:norm-regularization-bias}
    \begin{aligned}
    \| R_\alpha K f - f \| 
    & = \alpha \, \| (T + \alpha I)^{-1} T z \|    \\
    & \leqslant \alpha \, \| (T + \alpha I)^{-1} T \| \cdot \| z \|_F    \\
    & \leqslant \alpha \,E\, \| (T + \alpha I)^{-1} T \|,
    \end{aligned}
\end{equation}
where we have employed the \textit{a priori} bound $\|z\|_F \leqslant E$.

Let $A = (T+\alpha I)^{-1}T$. We can rewrite $A$ as follows $A = (T+\alpha I)^{-1}(T+\alpha I - \alpha I) = I - \alpha (T+\alpha I)^{-1}$. 
Since $T = K^*K$ is a self-adjoint operator, the resolvent $(T+\alpha I)^{-1}$ is also self-adjoint; consequently, $A$ is self-adjoint. According to the definition of the norm for a self-adjoint operator in Eq.~(\ref{eq:norm-self-adjoint-operator}), we have $\| A \| = \sup_{\| f \| = 1} \left| (Af, f)_F \right|$.
For any unit vector $f$ (i.e., $\| f \| = 1$), the inner product expands to
\begin{equation}\label{eq:A-norm}
    \begin{aligned}
        (Af, f) &= (f, f) - \alpha \left( (T+\alpha I)^{-1} f, f \right) \\
        &= 1 - \alpha \left( (T+\alpha I)^{-1} f, f \right).
    \end{aligned}
\end{equation}
Observe that $(Tf, f) = (K^*Kf, f) = (Kf, Kf) = \|Kf\|^2 \geqslant 0$. Thus, $T$ is a positive semi-definite operator ($T \geqslant 0$). It follows that $T + \alpha I \geqslant \alpha I$. By the monotonicity property of the inverse for positive definite operators, we obtain $(T+\alpha I)^{-1} \leqslant (\alpha I)^{-1} = \frac{1}{\alpha} I$.
This implies that for any unit vector $f$, $0 \leqslant \left( (T+\alpha I)^{-1} f, f \right) \leqslant \frac{1}{\alpha} (f, f) = \frac{1}{\alpha} \|f\|^2 = \frac{1}{\alpha}$.
Multiplying by $\alpha$, we get $0 \leqslant \alpha \left( (T+\alpha I)^{-1} f, f \right) \leqslant 1$.
Substituting this bound into Eq.~(\ref{eq:A-norm}) yields $0 \leqslant (Af, f) \leqslant 1$.
Therefore, the operator norm satisfies $\|A\| \leqslant 1$. Applying this result to Eq.~(\ref{eq:norm-regularization-bias}), we conclude
\begin{equation}\label{eq:error-bound-regularization}
    \| R_\alpha Kf - f \| \leqslant \alpha \,E.
\end{equation}

Under different a priori source conditions, the results may differ slightly. For instance, consider the source condition $f = K^*z$ with $z \in G$ and $\|z\|_G \leqslant E$. In this case, the error bound becomes $\| R_\alpha Kf - f \| \leqslant \frac{\sqrt{\alpha}}{2} E$.
Compared to the case where $f = K^*Kz$, the regularity (smoothness) of $f$ is weaker. Consequently, the convergence rate of $\| R_\alpha Kf - f \|$ with respect to $\alpha$ is slower, specifically $\mathcal{O}(\sqrt{\alpha})$ versus $\mathcal{O}(\alpha)$. In the following discussion, we focus exclusively on the case where $f = K^*Kz$.

Next, we analyze the term $\delta \,\| R_\alpha \|$. Since the regularized solution is given by $f_\alpha^\delta = R_\alpha g^\delta$, we invoke Definition \ref{defn:norm-operator-ratio} for the operator norm: $\|R_\alpha \| = \sup_{g^\delta \neq 0} \frac{\|R_\alpha g^\delta\|_F}{\|g^\delta\|_G}$.  Estimating $\| R_\alpha \|$ is equivalent to find an upper bound for the ratio of ${\| R_\alpha g^\delta \|}$ to ${\| g^\delta \|}$.

The regularized solution satisfies the normal equation $(K^*K+\alpha I)f_\alpha^\delta = K^* g^\delta$. Taking the inner product of both sides with $f_\alpha^\delta$ yields $\left((K^*K+\alpha I)f_\alpha^\delta, f_\alpha^\delta \right) = (K^* g^\delta, f_\alpha^\delta)$, which expands to $\| K f_\alpha^\delta \|^2 + \alpha \| f_\alpha^\delta \|^2 = (g^\delta, K f_\alpha^\delta)$. By the Cauchy-Schwarz inequality, the right-hand side is bounded by $(g^\delta, K f_\alpha^\delta) \leqslant \| g^\delta \| \cdot \| K f_\alpha^\delta \|$. Meanwhile, applying the Arithmetic Mean-Geometric Mean (AM-GM) inequality to the left-hand side gives $\| K f_\alpha^\delta \|^2 + \alpha \| f_\alpha^\delta \|^2 \geqslant 2 \sqrt{\alpha} \| K f_\alpha^\delta \| \cdot \| f_\alpha^\delta \|$. Combining these results leads to $2 \sqrt{\alpha} \| K f_\alpha^\delta \| \cdot \| f_\alpha^\delta \| \leqslant \| g^\delta \| \cdot \| K f_\alpha^\delta \|$. Assuming $\| K f_\alpha^\delta \| \neq 0$ (the case where it equals zero holds trivially), we divide by $\| K f_\alpha^\delta \|$ to obtain $\frac{\| f_\alpha^\delta \|}{\| g^\delta \|} \leqslant \frac{1}{2\sqrt{\alpha}}$. Since $f_\alpha^\delta = R_\alpha g^\delta$, taking the supremum over all $g^\delta \neq 0$ allows us to conclude that the operator norm satisfies
\begin{equation}
    \|R_\alpha\| \leqslant \frac{1}{2\sqrt{\alpha}}.
\end{equation}
This result aligns with the discussion in the previous subsection. As $\alpha$ decreases, $\| R_\alpha \|$ increases and may even diverge; conversely, as $\alpha$ increases, $\| R_\alpha \|$ decreases, as depicted in Fig.~\ref{fig:error-illustration}. Consequently, we establish the final bound for the noise propagation term
\begin{equation}\label{eq:error-bound-noise}
    \delta\,\|R_\alpha\| \leqslant \frac{\delta}{2\sqrt{\alpha}}.
\end{equation}

The error bounds in Eqs.~(\ref{eq:error-bound-regularization}) and (\ref{eq:error-bound-noise}) can also be derived using the spectral theorem and singular value decomposition. However, these derivations are omitted here for brevity.

By combining the error bounds from Eqs.~(\ref{eq:error-bound-regularization}) and (\ref{eq:error-bound-noise}), we establish the convergence of the solution by minimizing the upper bound
\begin{equation}
    \| f_\alpha^\delta -f \| \leqslant \alpha\,E + {\delta \over 2\sqrt{\alpha}}.
\end{equation}
The minimum is achieved by choosing $\alpha = (\delta/E)^{2/3}$, which yields
\begin{equation}\label{eq:convergence-of-solution}
    \| f_\alpha^\delta -f \| 
    \leqslant  {3\over2}\, \delta^{2\over 3}\,E^{1\over3}.
\end{equation}
It is evident that, regardless of the magnitude of $E$, the solution error converges to zero, 
\begin{equation}
\| f_\alpha^\delta -f \| \to 0,\quad \delta\to 0.
\end{equation}
This demonstrates the convergence of the solution. Under Tikhonov regularization, the approximate regularized solution converges to the true solution in the limit of vanishing noise $\delta\to 0$.

The convergence of the solution is the most critical property of regularization methods. It implies that the inverse problem approach can be systematically improved as the input error decreases. 
Conversely, even when the noise level $\delta$ is not infinitesimally small, Eq.~(\ref{eq:convergence-of-solution}) demonstrates that the uncertainty of the solution possesses an upper bound. This indicates that the uncertainty remains under control, thereby confirming that regularization ensures stability. 
Furthermore, in establishing solution convergence, the choice of the regularization parameter $\alpha$ is not arbitrary. It must depend on $\delta$. For a fixed $\delta$, $\alpha$ must be chosen carefully. It cannot be too large nor too small. This aligns with the discussion in the previous subsection. 
Finally, although the \textit{a priori} bound $E$ may be difficult to estimate in practice, this does not affect the convergence of the solution. Regardless of the specific method used to select $\alpha$, the convergence of the solution is guaranteed.

\paragraph{Minimization of the Tikhonov functional\\}
The discussions thus far have been grounded on the regularization operator $R_\alpha$ and the corresponding operator equation $f_\alpha^\delta = R_\alpha g^\delta$. 
Alternatively, this can be viewed through the minimization of the Tikhonov functional, an equivalent scheme that is often preferred in practical applications. 
From the variational perspective, regularization redefines the solution of ill-posed problems via incorporating problem-specific prior information and introduces stabilizing functionals to establish stable solutions that approximate the original ones.

\begin{thm}\label{thm:minimization-functional}
    Let $K:F \to G$ be a linear and bounded operator between Hilbert spaces and $\alpha>0$. Then the Tikhonov functional $J_\alpha$,
    \begin{equation}\label{eq:Tikhonov-functional}
        J_\alpha(f):=\|Kf-g^\delta\|_G^2 + \alpha \,\|f\|_F^2 ~~~~~\text{for}~~f\in F,
    \end{equation}
    has a unique minimizer 
    \begin{equation}\label{eq:minimizer-functional}
        f_\alpha^\delta = \mathop {\arg \min }\limits_{f \in F} J_\alpha(f).
    \end{equation} 
    This minimizer $f_\alpha^\delta$ is the unique solution of the normal equation  \cite{Kirsch-2011}
    \begin{equation*}
        \alpha f_\alpha^\delta + K^*Kf_\alpha^\delta = K^* g^\delta.
    \end{equation*}
\end{thm}
The proof of the existence of the minimizer and its equivalence to the solution of the normal equation are shown in the appendix \ref{sec:appendix-proof-regularization}.

The minimization of the Tikhonov functional offers an intuitive perspective on regularization. 
The first term, $\|Kf-g^\delta\|_G^2$, represents the residual of the data, analogous to $\chi^2$ in the fitting procedures. 
While minimizing the residual term seeks a solution to $Kf=g^\delta$, driving it too close to zero often leads to overfitting the noise in $g^\delta$.
This instability arises because $K$ is typically a continuous smoothing operator. 
In a discretized setting, this implies that a solution $f$ with rapidly alternating signs (where adjacent components $f_i$ and $f_{i+1}$ cancel each other out) can still yield a product $Kf$ extremely close to the observed data $g^\delta$. 
Consequently, minimizing the residual term alone fails to penalize such high-frequency oscillations, resulting in an unstable and physically meaningless solution.

To counteract this, the second term, $\alpha \|f\|_F^2$, serves as a regularization or penalty term that encodes prior information about the smoothness or boundedness of the solution.
Mathematically, since high-frequency oscillations typically correspond to large norms in the function space $F$, penalizing $\|f\|_F^2$ effectively suppresses these unstable components that the smoothing operator $K$ overlooks.
The regularization parameter $\alpha > 0$ acts as a crucial balance factor, mediating the trade-off between data fitting and solution stability. 
By preventing the norm from growing excessively, a properly chosen $\alpha$ mitigates the ill-posedness of the original problem. 
As established in Theorem~\ref{thm:minimization-functional}, this combined functional is strictly convex in the $L^2$ or $H^1$ spaces, guaranteeing a unique and stable regularized solution $f_\alpha^\delta$ that converges to the true solution as the noise level vanishes.

The variational principle offers several key advantages. First, it enhances the flexibility of numerical algorithms with functional minimization. Finding extrema is generally more feasible than directly solving equations. Algorithmically, this formulation can be interpreted as an optimization problem that achieves a good balance between fitting noisy data and satisfying prior constraints. It can also address non-differentiable or constrained problems, such as boundary conditions and non-negativity constraints.

Second, a key strength of the Tikhonov functional is its adaptability to a priori physical conditions. In particular, the penalty term can be customized to incorporate the known properties of the solution. For example, while the standard $L^2$ norm ($\alpha \|f\|_{L^2}^2$) penalizes the overall magnitude of the solution, it can be replaced by the $H^1$ norm ($\alpha \|f\|_{H^1}^2$) if smoothness is expected. As detailed in Appendix~\ref{sec:appendix-functional}, the $H^1$ norm includes derivative information to explicitly enforce smoothness, and Section~\ref{sec:toymodel} demonstrates the effectiveness of this modification using several toy models.

During the past half-century, numerous advanced regularization techniques have evolved, such as iterative regularization, discrete regularization, Bayesian approach, and so on \cite{Kirsch-2011,Regularization_tools,Wei_Ting,Yan_Xiong_Bin}. 
However, in this work, which establishes the initial mathematical foundation for an inverse problem approach to non-perturbative QCD, we focus exclusively on the Tikhonov framework. As the earliest and most fundamental method for solving ill-posed problems, Tikhonov regularization is distinguished by its conceptual clarity and computational efficiency, making it an ideal starting point. Future studies will build on this basis by exploring more sophisticated methods to address complex physical scenarios.

%%%%%%%%%%%%%%%%%%%%%%%%%%%%%%%%%%%%%%%%%%%%%%%%%%%%%%%%%%%%%%%%%%%
\subsection{Choices of the Regularization Parameter}

As discussed in the preceding subsections, the Tikhonov regularization ensures that the regularized solution converges to the true solution asymptotically as $\delta \to 0$. However, in practical applications, the ideal limit of vanishing noise ($\delta \to 0$) is rarely attainable. Consequently, we must determine an appropriate value of $\alpha$ for a finite noise level $\delta$.
This requires a balance that $\alpha$ must be chosen neither too large nor too small. If $\alpha$ is too small, it fails to provide adequate regularization, and the solution retains the ill-posed characteristics of the original problem. For instance, the solution may become unstable and highly oscillatory. Conversely, if $\alpha$ is too large, the approximation error increases significantly. In this case, the penalty term dominates the Tikhonov functional, leading to an oversmoothed solution that fails to capture the primary features of the input data. Hence, a rigorous and systematic selection of $\alpha$ is essential.

Formally, strategies for selecting the regularization parameter $\alpha$ are generally classified into two broad categories, a priori and a posteriori methods. In the a priori approach, $\alpha$ is determined prior to minimizing the Tikhonov functional, relying on the knowledge of the true solution. This approach offers a theoretical guarantee of convergence. For instance, under the source condition $f = K^* K z$ with $\|z\|_{F} \leqslant E$, setting $\alpha=(\delta/E)^{2/3}$ yields the error estimate $\|f_\alpha^\delta-f\| \leqslant \frac{3}{2}\delta^{2/3}E^{1/3}$, which converges to zero as $\delta \to 0$.
This result relies on two a priori assumptions, the boundedness of $\|z\|_F$ by $E$ and the smoothness of the true solution $f$ inherited from the smoothing operators $K$ and $K^*$.
Despite its theoretical soundness, the a priori strategy is often impractical in real physical applications. This is because essential constants, such as $E$, are rarely known. This limitation motivates the use of a posteriori methods, which determine $\alpha$ based solely on the noisy data $g^\delta$ and/or an estimate of the noise level $\delta$.

In the a posteriori approach, the regularization parameter $\alpha$ is varied over a wide range of values. For each candidate value $\alpha_i$, a regularized solution $f_{\alpha_i}^\delta$ is obtained by Tikhonov regularization, generating a family of solutions $\{f_{\alpha_i}^\delta\}$. Subsequently, a specific selection criterion is applied to identify the optimal parameter $\alpha$ and its corresponding solution $f_\alpha^\delta$ from this set. Widely used and effective criteria include the discrepancy principle, the quasi-optimality criterion, etc.

Before introducing specific criteria, it is important to note that we expect the physical solution to be robust and not overly sensitive to the specific value of $\alpha$. Ideally, there should exist a ``plateau'' in the parameter space where the solution remains stable. 
We adopt a dual-criterion strategy to select $\alpha$ and the corresponding solution. Specifically, the choice of $\alpha$ is determined by the rigorous mathematical principles introduced below, while simultaneously requiring the existence of a stability plateau. This dual requirement ensures robustness, making the solution more reliable. This expectation will be demonstrated through the toy model analyzes presented in the following section.
At this stage, if such a plateau is not observed, we must either modify the regularization method or revise the selection criteria. Furthermore, we expect the regularization method and the selection criteria to remain consistent within the same class of physical problems. However, it may be necessary to adapt the selection criteria or even change the regularization method entirely if the nature of the problem changes, for instance, transitioning from continuous to discrete problems, or from single-peak to multi-peak structures.

\paragraph{(1) Discrepancy Principle\\}
Regularization methods essentially seek a balance between data fitting and solution stability. Given a noise level $\delta$, the residual $\|Kf_\alpha^\delta - g^\delta\|$ should be neither excessively large nor small.
If $\|Kf_\alpha^\delta - g^\delta\| \ll \delta$, the model fits the observed data more closely than the precision of the data allows. This indicates an overfitting of the noise, which implies that the regularization parameter is too small.
In contrast, if $\|Kf_\alpha^\delta - g^\delta\| \gg \delta$, the fit of the data is poor, failing to capture the dominant features of the signal. This represents an underfit, indicating that the regularization parameter is too large.
The discrepancy principle aims to find a value of $\alpha$ such that $\|Kf_\alpha^\delta - g^\delta\| \approx \delta$. This ensures an appropriate balance between suppressing noise and retaining effective information.

The discrepancy principle is conceptually analogous to the criterion of a $\chi^2$ fit. In statistical estimation, a model is generally considered well-fitted when the reduced chi-squared statistic, $\chi^2_{\rm min}/\text{d.o.f.}$, is approximately unity. A value significantly smaller than one suggests overfitting, whereas a value substantially larger than one indicates a poor fit. Consequently, the optimal fit corresponds to $\chi^2_{\rm min}/\text{d.o.f.} \approx 1$. In the context of Tikhonov regularization, this principle motivates the selection of the regularization parameter $\alpha$ such that the residual norm matches the noise level, i.e., $ \|K f_\alpha^\delta - g^\delta\| = \delta$.

\begin{thm}[Discrepancy Principle]
    Let $K: F \to G$ be linear, compact and one-to-one with dense range in $G$. Let $Kf=g$ with $f\in F, g\in G, g^\delta \in G$ such that $\| g-g^\delta \|\leqslant \delta < \|g^\delta \|$. Let the Tikhonov solution $f^\delta_\alpha$ satisfy 
    \begin{equation}
        \left\| K f^\delta_\alpha -g^\delta \right\| = \delta.
    \end{equation} 
    Then $f^\delta_\alpha \to f$ for $\delta \to 0$. That is, the discrepancy principle is admissible \cite{Kirsch-2011}. 
    \label{thm:discrepancy-principle}
\end{thm}
Its proof is provided in Appendix \ref{sec:appendix-proof-regularization}.

The most significant advantage of the discrepancy principle lies in its rigorous theoretical foundation. Specifically, it possesses a strict proof of convergence, guaranteeing that the regularized solution converges to the true solution as the noise level $\delta$ approaches zero. In contrast, establishing convergence for a posteriori methods is often a formidable challenge. As will be demonstrated subsequently, widely used techniques such as the quasi-optimality criterion lack such proofs of convergence. 

On the other hand, the limitation of the discrepancy principle is evident. The criterion $\|Kf_\alpha^\delta-g^\delta\|=\delta$ requires precise knowledge of the noise level $\delta$, a condition that is rarely met in practice. Since $\delta$ is typically unknown, any variation in its estimation directly affects the regularization parameter $\alpha$, thus altering the solution. The discrepancy principle is primarily applicable when $\delta$ is known with certainty, such as in experimental data or lattice QCD where rigorous error analysis is available.

\paragraph{(2) Quasi-Optimality Criterion\\}
When the noise error $\delta$ is unknown, the quasi-optimality criterion can be employed to determine the value of $\alpha$ via \cite{Kirsch-2011}
\begin{equation}
    \begin{aligned}
        \alpha_{\rm opt}=\arg\min _{\alpha>0}\left\{\left\| \alpha \frac{\mathrm{~d} f^\delta_\alpha}{\mathrm{d} \alpha} \right\|\right\}\,.
    \end{aligned}
\end{equation}
The underlying rationale is as follows. First, one seeks a plateau of $\alpha$ by minimizing $\text{d}f_\alpha^\delta /\text{d}\alpha$. Second, preference is given to smaller values of $\alpha$, which correspond to a smaller residual (i.e., a smaller regularization approximation error).
The quasi-optimality criterion minimizes the difference between adjacent regularized solutions. This approach ensures the robustness and insensitivity of the physical results to the specific value of $\alpha$. Thus, it is widely used in practical applications.

The quasi-optimality criterion offers the distinct advantage of not requiring an exact value for the noise level $\delta$. When the solution landscape exhibits a single plateau with respect to $\alpha$, the criterion effectively identifies it. In cases with multiple plateaus, the method inherently favors the one corresponding to a smaller $\alpha$. 
%This approach aims to achieve higher resolution, representing a strategy that prioritizes fidelity. 

When input errors are significant, complications arise.  If the quasi-optimality criterion selects an excessively small $\alpha$, the constraining power of the penalty term becomes too weak. The regularization process is then dominated by the residue term. Under such conditions, a model with a weak restricting effect fails to distinguish between the true signal and random noise. The resulting solution often exhibits spurious high-frequency oscillations, a phenomenon known as under-regularization. Therefore, in practice, this criterion sometimes requires a combination with other methods or manual judgment to ensure robust results.

%In practice, various other principles and criteria are also widely employed. A full discussion is beyond the scope of this work. However, the L-curve criterion is presented in Appendix \ref{sec:appendix-L-curve} as an example.

%%%%%%%%%%%%%%%%%%%%%%%%%%%%%%%%%%%%%%%%%%%%%%%%%%%%%%%%%%%%%%%%
\subsection{Discretization for Tikhonov Regularization}
The Tikhonov regularization method seeks the minimizer of the functional defined in Eq.~(\ref{eq:minimizer-functional}). However, obtaining an analytical expression for this minimizer is generally difficult.
%since the functional is defined over the continuous space $F$ in Eq.~(\ref{eq:Kf=g}), such as $F=L^2(t_{\rm min},\Lambda)$ or $H^1(t_{\rm min},\Lambda)$. 
Therefore, it is necessary to discretize the functional to compute a numerical solution. This can be achieved by building a finite-dimensional subspace $X_n = \operatorname{span} \{ \varphi_0, \varphi_1, \cdots, \varphi_n \}$ of $F$ and solving the problem within this subspace. Finally, we demonstrate that the solution in the discretized space converges to the true solution as the space dimension approaches infinity.

To discretize the Tikhonov functional, a finite element method based on piecewise linear basis functions is employed. The finite element method offers broad applicability, making it capable of handling complex geometries. The interval $[t_{\rm min},\Lambda]$ is divided into uniform sub-intervals $n$ by grid points defined as
\begin{equation}\label{eq:discretize-step}
    \begin{aligned}
        x_i = t_{\rm min} + ih\,, \quad i = 0,1,2, \cdots ,n\,,
    \end{aligned}
\end{equation}
where the step size is $h = (\Lambda - t_{\rm min})/n$. On this grid,  the  piecewise linear hat functions are defined in the following
\begin{equation}\label{eq:finite-element}
\begin{aligned}
\varphi_i(x) &= 
\begin{cases} 
\frac{x - x_{i-1}}{h}, & x \in [x_{i-1}, x_i]\,, \\
-\frac{x - x_{i+1}}{h}, & x \in [x_i, x_{i+1}]\,,\\
0, & \text{otherwise}\,,
\end{cases}
&
\varphi_0(x) &= 
\begin{cases} 
-\frac{x - x_1}{h}, & x \in [x_0, x_1]\,, \\
0, & \text{otherwise}\,,
\end{cases}
&
\varphi_n(x) &= 
\begin{cases} 
\frac{x - x_{n-1}}{h}, & x \in [x_{n-1}, x_n]\,, \\
0, & \text{otherwise}\,,
\end{cases}
\end{aligned}
\end{equation}
for $i=1,\cdots,n-1$. These functions form a basis of the subspace $X_n$. The continuous minimization problem is then restricted to $X_n$, leading to the discrete problem
\begin{equation}\label{eq:minimize-discrete-functional}
f_{\alpha,n} ^\delta  = \mathop {\arg \min }\limits_{f \in X_n} J_\alpha(f)=\| Kf - g^\delta \|_{G}^2 + \alpha \| f \|_{F}^2\,.
\end{equation}
The approximate solution is then expressed as
\begin{equation}\label{eq:discrete-expansion}
    \begin{aligned}
        f_{\alpha,n} ^\delta(x) =\sum\limits_{i = 0}^n  c_i \varphi _i(x)\,,
    \end{aligned}
\end{equation}
with the coefficients $c_i$ to be determined. 
Substituting it into the functional $J_\alpha(f)$ leads to
\begin{equation}
    \begin{aligned}
         J_\alpha(f_{\alpha,n} ^\delta)
         =& \left\| \sum\limits_{i = 0}^n c_i K\varphi _i-g^\delta\right\|_{G}^2 + \alpha \left\|\sum\limits_{i = 0}^n c_i \varphi _i \right\|_{F}^2\\
	     =& \sum\limits_{i,j = 0}^n c_i c_j(K\varphi_i, K\varphi_j)_{G} - 2\sum\limits_{i = 0}^n c_i (K\varphi _i,g^\delta )_{G}+ (g^\delta , g^\delta )_{G} 
        + \alpha \sum\limits_{i,j = 0}^n c_ic_j(\varphi _i,\varphi _j)_{F} \,.
    \end{aligned}
\end{equation}

In order to find the extreme values of this multivariate function, we set the partial derivative of $J_\alpha(f_{\alpha,n} ^\delta)$ with respect to each coefficient $c_i~(i=0,1,\cdots,n)$ equal to zero. This results in the following linear system:
\begin{equation}\label{eq:eqation-ABCD}
(A+\alpha B)\,C=D\,,
\end{equation}
where 
\begin{equation}\label{eq:ABCD}
    A_{ij} = (K\varphi _i,K\varphi _j)_{G}\,, ~~~~ B_{ij}= (\varphi _i, \varphi _j)_{F}\,, ~~~~ C_j=(c_0,c_1, \cdots , c_n)^T\,,~~~~
         D_i  = (K\varphi _i, g^\delta)_{G}^T\, .
\end{equation}
The coefficient vector $C$ is obtained by solving the linear system, which gives the numerical minimizer of the Tikhonov functional $f_{\alpha ,n}^\delta (x)$ in Eq.~(\ref{eq:discrete-expansion}). 
Note that, due to the continuity of the finite element basis functions in Eq.~(\ref{eq:finite-element}), the inner products in Eq.~(\ref{eq:ABCD}) are still defined in the continuous spaces $F$ and $G$.

%Alternatively, the above linear system is equivalently obtained by projecting the regularized normal equation, $(K^*K+\alpha I) f_\alpha^\delta = K^* g^\delta$, onto the basis functions $\varphi_i$ via the inner product,
%\begin{equation}\left(\varphi_i,(K^*K+\alpha I)\sum_{j} c_j\varphi_j\right)_F=\left(\varphi_i, K^*g^\delta\right)_F\,.\end{equation}
%Then we obtain 
%\begin{equation}\sum_j c_j(K\varphi_i,K\varphi_j)_G + \sum_j c_j \alpha (\varphi_i,\varphi_j)_F = (K \varphi_i,g^\delta)_G,\end{equation}
%which is equivalent to Eq.~(\ref{eq:eqation-ABCD}).

While discretization inherently introduces approximation errors, the convergence and reliability of this numerical solution are guaranteed by Theorem~\ref{thm:discrete}. Specifically, as $n \rightarrow \infty$, the numerical solution $f^\delta_{\alpha,n}$ converges to the theoretical solution $f^\delta_\alpha$, ensuring that numerical errors can be controlled through mesh refinement.

\begin{thm} \label{thm:discrete}
For fixed noise level $\delta$ and regularization parameter $\alpha>0$, we have 
\begin{equation}
\|f_{\alpha ,n}^\delta  -f^\delta_\alpha\|_F\rightarrow 0, \quad n\rightarrow \infty.
\end{equation}
\end{thm}
The proof of this theorem is given in Appendix \ref{sec:appendix-proof-regularization}.

\subsection{Short summary of Regularization}

In summary, this section has presented a comprehensive overview of general regularization theory for ill-posed inverse problems, with a particular focus on the Tikhonov regularization method and the associated strategies for selecting the optimal value of the regularization parameter $\alpha$. A rigorous theoretical foundation has been established. It guarantees a stable and unique regularized solution that converges to the true solution as the input error tends to zero.   
Importantly, the overall procedure is based on rigorous mathematical principles and introduces no additional model-dependent assumptions beyond those intrinsic to the chosen regularization scheme. 

From the error bound given in Eq.~(\ref{eq:error-bound}), $\|f_\alpha^\delta-f\|\leqslant \|R_\alpha Kf-f\|+\delta\|R_\alpha\|$, it is straightforward to analyze the uncertainties of the solution. The bound consists of two parts,  the error from the input data amplified by $\|R_\alpha\|$ and the error due to the regularization approximation. The convergence, $\|f_\alpha^\delta-f\|\to 0$ as $\delta\to0$, indicates that the solution can be systematically improved by reducing the input error $\delta$. We will present the numerical results and the corresponding error analysis in the next section, using three toy models. 

%We may interpret regularization from an alternative perspective. Instead of regarding errors as undesirable noise to be removed, we treat them as inherent features of practical systems and introduce regularization terms to maintain balance. By converting ill-posed problems into well-posed approximate counterparts, regularization ensures the existence, uniqueness, and stability of the approximate solutions. The core concern is whether such solutions can converge to the true physical solution, which has been strictly verified for Tikhonov regularization.

During the development of this work, the inverse problem associated with dispersion relations has been widely addressed using polynomial expansion and the inverse matrix method. Applications include studies of hadron spectra \cite{HnLi-spectrum,Mutuk:2025lak,Zhao:2024drr}, light-cone distribution amplitudes \cite{HnLi-LCDA}, neutral meson mixings \cite{HnLi-mixing}, determination of Standard Model parameters \cite{HnLi-SMparameters}, and investigations of new physics such as a fourth generation \cite{HnLi-NP}. The inverse matrix method is physically intuitive, yet its mathematical foundation remains unclear and warrants further investigation. The theoretical framework developed in this work, such as Tikhonov regularization, can be applied to all of the above problems.

Furthermore, inverse problems are widespread throughout physics. Although the proposed scheme is motivated by dispersion relations, its regularization framework is sufficiently general to be applied to a broad range of inverse problems in different disciplines.
For instance, Tikhonov regularization has been used to establish the mathematical properties of the inverse Fourier transformation in lattice QCD within the large-momentum effective theory \cite{Xiong:2025obq,Ling:2025olz}, while also validating the corresponding extrapolation method. It is also applied to determine the source functions in hadron-hadron momentum correlations \cite{Xiong:2025bmd}. 
Other typical applications include extracting non-perturbative parameters such as parton distribution functions from experimental measurements \cite{Candido:2024hjt}, and reconstructing  spectral functions from lattice QCD data \cite{Liang:2019frk}. Common approaches used in such tasks include the Backus-Gilbert method \cite{Backus:1968svk}, the maximum entropy method \cite{Asakawa:2000tr}, the Bayesian method \cite{Burnier:2013nla}, and so on. All these inverse problems can be addressed within the standard regularization theory, which facilitates systematic error analysis.

%% file: 5-toymodel.tex
\section{Tests of Toy Models}\label{sec:toymodel}

In the previous section, the mathematical framework of the inverse problem approach is introduced. Before tackling detailed physical problems, we can test some toy models to gain an intuitive understanding of how the regularization mechanism affects the solution of the inverse problem for the dispersion relation.

%%%%%%%%%%%%%%%%%%%%%%%%%%%%%%%%%%%%%%%%%%%%%%%%%%%%%%%
\subsection{Three Toy Models}\label{Three toy model}
To evaluate the effectiveness of the Tikhonov regularization method, we employ three representative toy models:
\begin{equation}\label{eq:toy-model}
\begin{aligned}
& \textbf{Model 1 (monotonic)}: f(x)=a_1 2\log(x+0.5)+{a_2\over x+3} ,\\
& \textbf{Model 2 (simple non-monotonic)}: f(x)=a_1 4 x e^{-a_2  x} ,\\
& \textbf{Model 3 (resonance)}: f(x)={1\over \pi}{ a_1 m\Gamma\over (x-m^2)^2+(a_1 m\Gamma)^2}+a_2 {x\over 5} .
\end{aligned}
\end{equation}
These models are designed to serve as benchmarks for a broad class of physical problems, with $a_1=a_2=1.0$. For instance, Models 1 and 2 are analogous to continuum spectral functions, which often exhibit simple monotonic or non-monotonic behaviors, as well as parton distribution functions and distribution amplitudes. 
Model 2 vanishes as $x\to \infty$, whereas model 1 does not. In addition, models 1 and 2 capture the typical behaviors of exponential and logarithmic functions.  
Model 3 represents resonance phenomena, with $m=0.8$ and $\Gamma=0.5$. 
The primary objective is to assess the performance of the regularization method, especially in scenarios where the input data are subject to errors represented by the parameters $a_{1,2}$.

\begin{figure}
    \centering
    \includegraphics[scale=0.3]{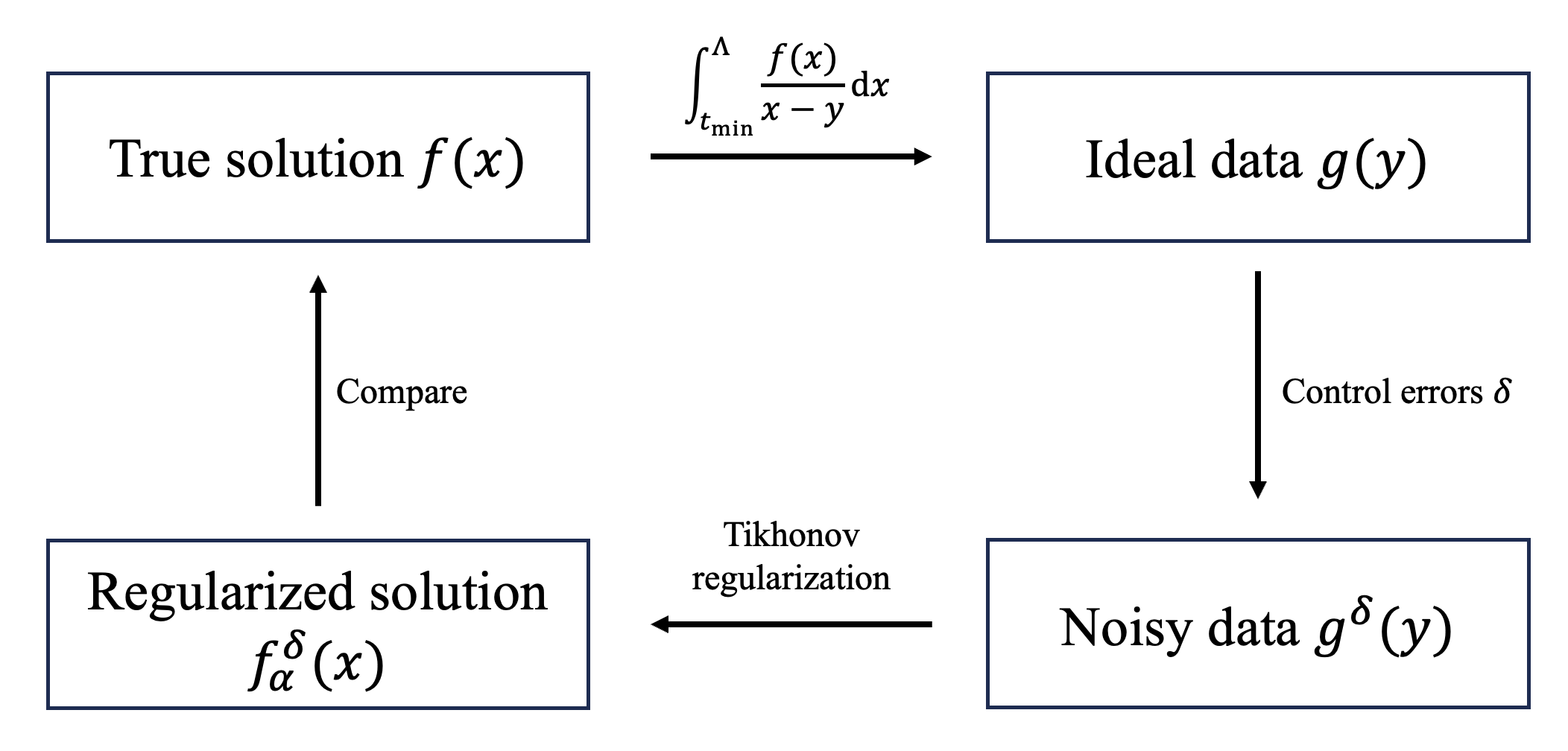}
    \caption{Numerical procedure for evaluating the effectiveness of the inverse problem approach.}
    \label{fig:flow-toy-model}
\end{figure}

The numerical procedure is illustrated in Fig.\,\ref{fig:flow-toy-model}. For each toy model, the true solution $f(x)$ is transformed by the dispersion integral to obtain the exact $g(y)$. Controlled errors are then added to simulate real physics, resulting in noisy data $g^\delta(y)$. Tikhonov regularization is applied to invert $g^\delta(y)$, producing the regularized solution $f^\delta_\alpha(x)$, which is then compared to the original $f(x)$ to assess the effectiveness of the method.

There are two types of error in $g^\delta(y)$. The first approach adds random noise to each discrete point of $g(y_i)$, which closely mimics the experimental data. However, in the present inverse problem approach, the input data is always obtained from perturbative calculations, and thus correlated between discrete points. Therefore, a second approach is adopted, where errors enter through analytic functions with uncertain input parameters (e.g., the energy scale or vacuum condensates). The resulting error formulas are expressed analytically and this type of input error is tested.

Specifically, the input data are given by
$g^\delta(y,a_1, a_2)=\int_{t_{\rm min}}^{\Lambda}{f(x,a_1,a_2)\over x-y}dx$,
where $a_{1,2}$ are explicitly expressed from $f(x)$ in Eq.~(\ref{eq:toy-model}).
The values of $a_{1,2}$ are randomly sampled from Gaussian distributions with mean $\mu_i$ and standard deviation $\sigma_i$ ($i=1,2$). Here, $\mu_{1,2}=1.0$, and $\sigma_{1,2}$ take the values $0.3$, $0.1$, and $0.01$, corresponding to the input error levels of $30\%$, $10\%$, and $1\%$, respectively.
The parameters in Eq.~(\ref{eq:fxgy}) are taken as $t_{\rm min}=0$, $\Lambda=2$, $q_{\rm low}^2=-100$, $q_{\rm up}^2=-5$, with the discretization step sizes $h_x=0.01$ in Eq. (\ref{eq:discretize-step}) and $h_y=0.01$.

%%%%%%%%%%%%%%%%%%%%%%%%%%%%%%%%%%%%%%%%%%%%%%%%%%%%%%%%%%
\subsection{Numerical Results Under Regularization}\label{sec:numerical-regularization}

The efficacy of the regularization schemes is validated using toy models. The boundary conditions are introduced prior to the detailed methodology.
Since the original integral equation in Eq.~(\ref{eq:InverseProblem-DispersionRelation}) separates perturbative and non-perturbative contributions, the continuity of the imaginary part of the correlation function serves as a boundary condition for the inverse problem. More prior knowledge leads to better inversion results. Therefore, the boundary conditions on $f(x)$ should be exploited to improve the inversion in all toy models.
Let the boundary conditions be $f(t_{\rm min})=M$ and $f(\Lambda)=N$. The function $f(x)$ can then be decomposed as $f(x)=u(x)+v(x)$, where $v(x)$ is linear and satisfies the same boundary conditions, i.e., $v(x)=M\frac{(x-\Lambda)}{t_{\rm min}-\Lambda}+N\frac{(x-t_{\rm min})}{\Lambda-t_{\rm min}}$. The unknown function $u(x)$ is then solved from the modified integral equation:
\begin{equation*}
    \int_{t_{\rm min}}^\Lambda \frac{u(x)}{x-y}dx=g^{\delta}-\int_{t_{\rm min}}^\Lambda \frac{v(x)}{x-y} dx:=G^{\delta}
\end{equation*}
with homogeneous boundary conditions $u(t_{\rm min})=u(\Lambda)=0$. Applying a regularization method yields the regularized solution $u^{\delta}_\alpha$, and the final solution is reconstructed as $f^{\delta}_\alpha(x)=u^{\delta}_\alpha(x)+v(x)$. 

\paragraph{Without Regularization\\}
The necessity of regularization for the ill-posed inverse problem is manifested by the absence of regularization. To this end, the penalty term is neglected in the Tikhonov functional in Eq.~(\ref{eq:Tikhonov-functional}), retaining only the residual term, equivalent to the conventional $\chi^2$ fitting approach. In minimizing the functional, $Kf_\alpha^\delta$ becomes infinitely close to $g^\delta$.

\begin{figure}
    \centering
    \includegraphics[scale=0.5]{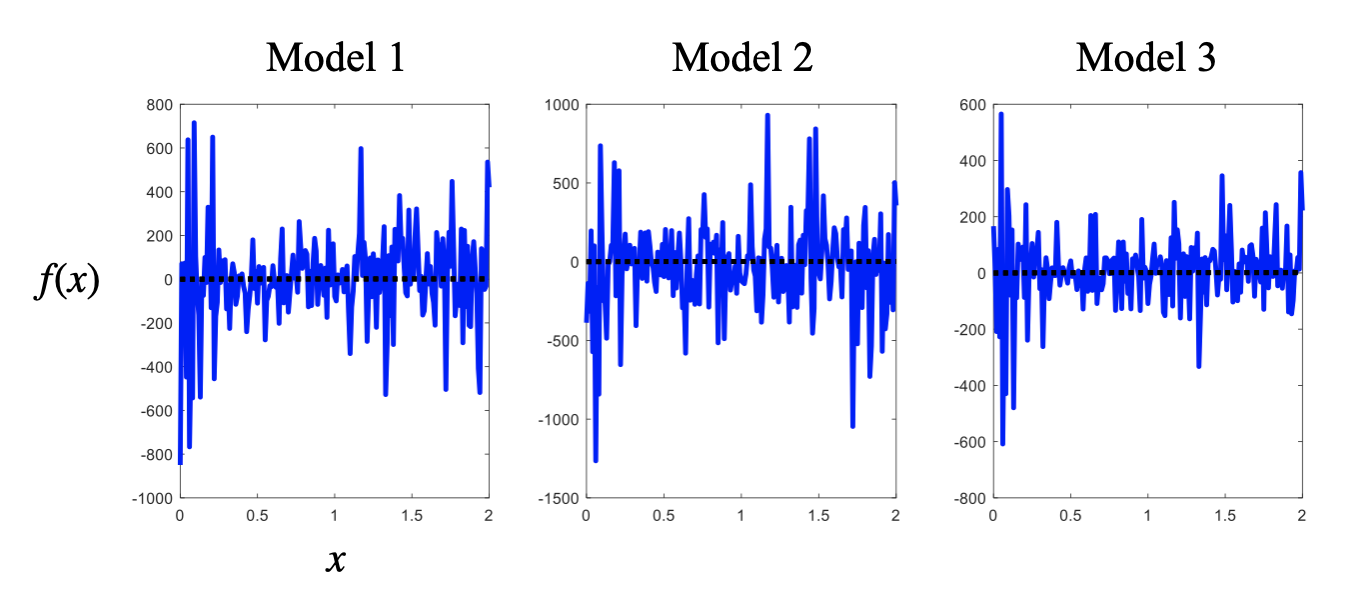}
    \caption{Solutions without regularization (blue lines) compared to the exact solution (black dotted lines).}
    \label{fig:noRegularization}
\end{figure}

The results without regularization are shown in Fig.\,\ref{fig:noRegularization} for the three toy models. The numerical solutions (blue lines) exhibit high-frequency oscillations and deviate significantly from the exact solutions (black lines). This behavior reveals the instability characteristic of the underlying ill-posed problem. Without regularization, stable and physically meaningful solutions cannot be reliably recovered.

\begin{figure}
\centering
\includegraphics[scale=0.35]{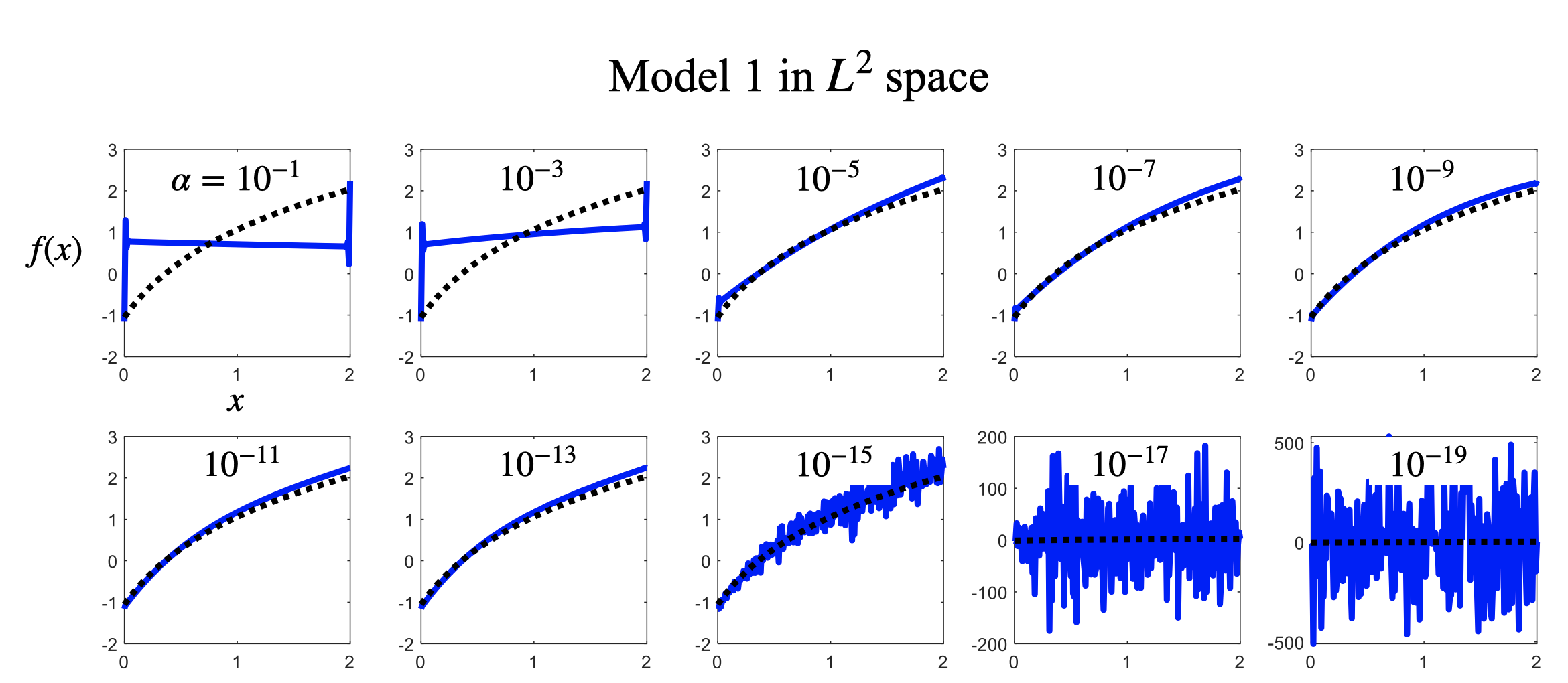}
\includegraphics[scale=0.35]{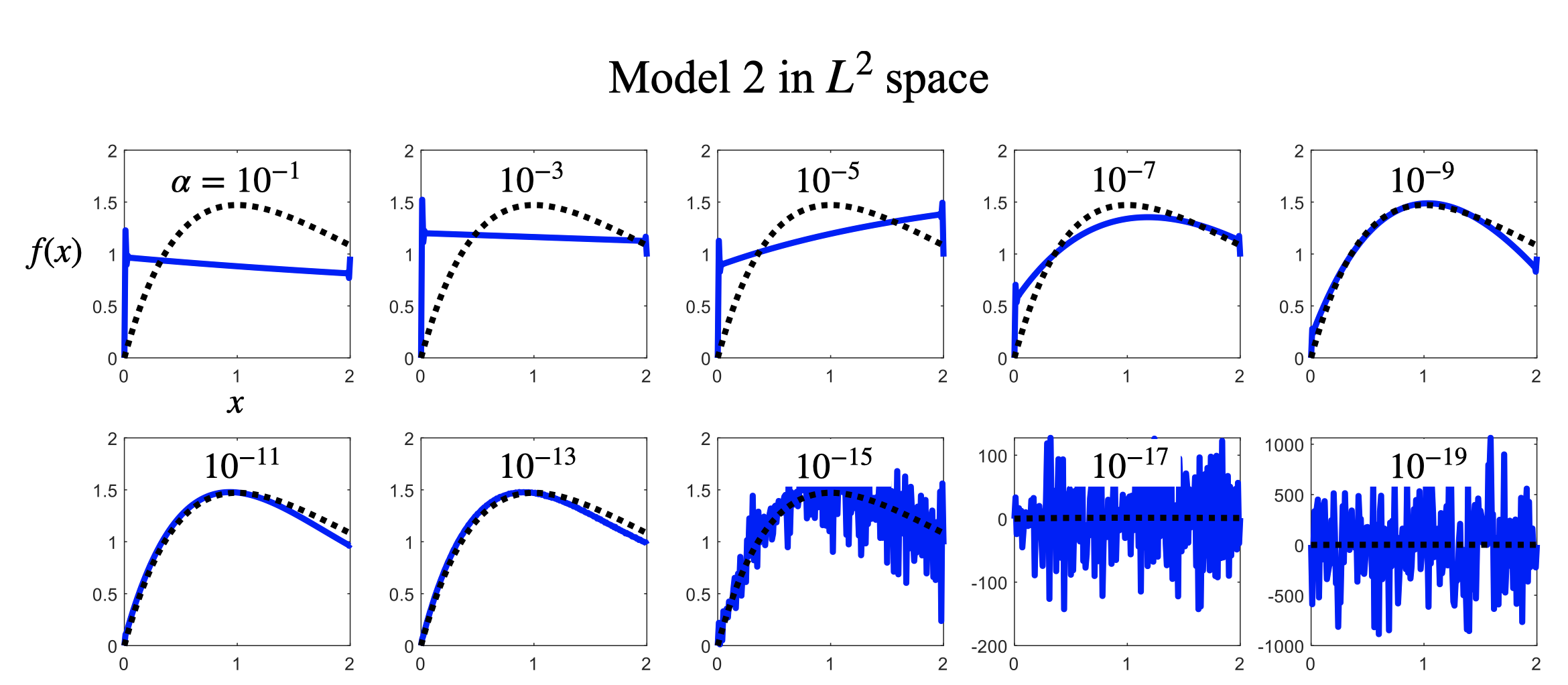}
\includegraphics[scale=0.35]{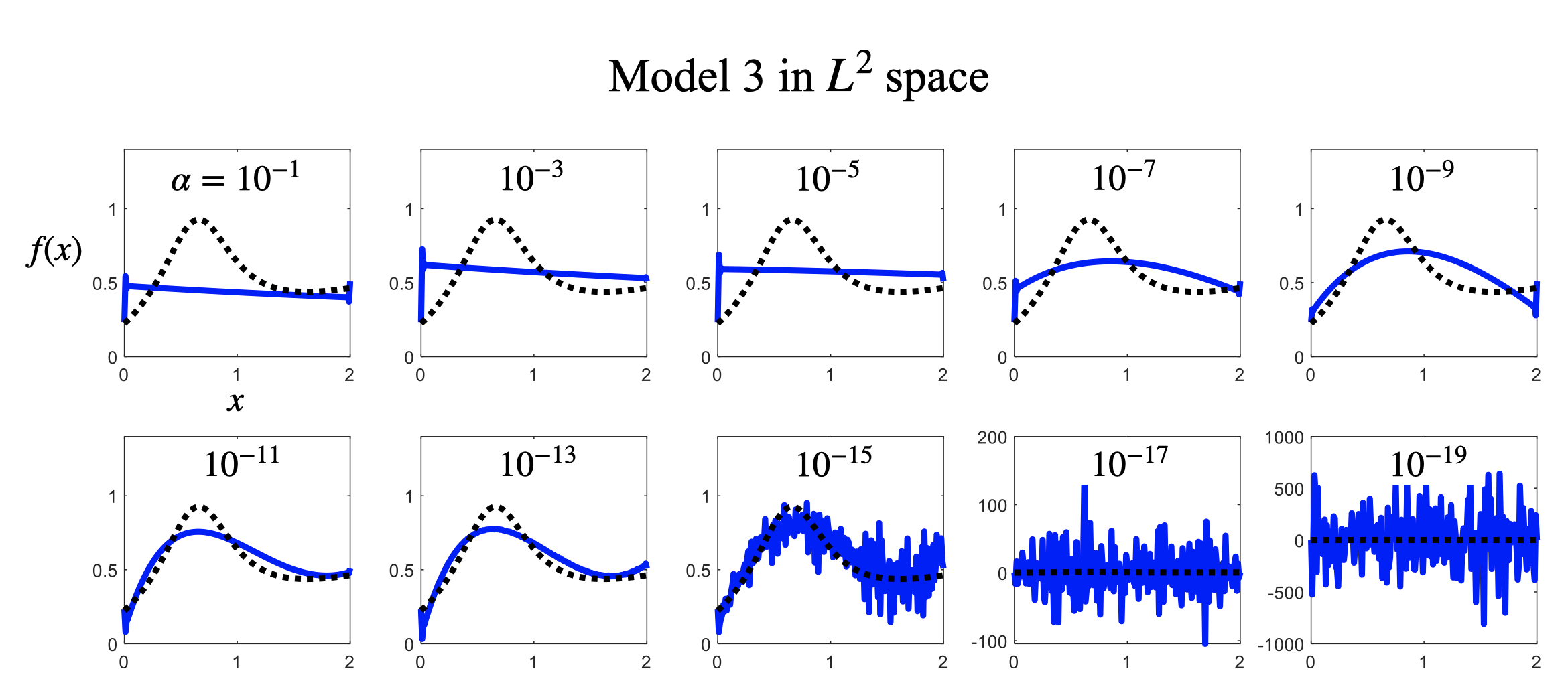}
\caption{Numerical solutions for the three toy models using Tikhonov regularization with the solution space $F=L^2(t_{\rm min}, \Lambda)$. Each figure displays the regularized solutions $f_\alpha^\delta(x)$ (blue solid) and the exact solutions $f(x)$ (black dashed), with the corresponding $\alpha$ value at the top. A $10\%$ input error is adopted throughout all cases. }
\label{fig:numerical-toy-models-L2}
\end{figure}

\paragraph{Results under Tikhonov Regularization in the $L^2$ space\\}
The numerical results obtained from Tikhonov regularization are presented using Eq.~(\ref{eq:minimizer-functional}) with the solution space $F$ taken as the $L^2$ space. The regularization parameter $\alpha$ is scanned from $1\times10^{-1}$ to $1\times10^{-19}$ to examine its influence on the solution, and the input error is set to a $10\%$ deviation ($a_1 = a_2 = 1.1$). The results are shown in Fig.~\ref{fig:numerical-toy-models-L2}.

We find that regularization yields considerably improved solutions compared to the highly oscillatory ones shown in Fig.~\ref{fig:noRegularization}. With an appropriate choice of $\alpha$, the solution approaches the true value, confirming both the necessity and the effectiveness of regularization for ill-posed problems.

It is observed that $\alpha$ must be chosen within an appropriate range. A value of $\alpha$ that is too large leads to poor approximation and significant deviations, while a value that is too small renders regularization ineffective, gradually restoring ill-posedness and causing oscillations and instability. Satisfactory results are achieved for intermediate $\alpha$ values, which is completely consistent with the expectations discussed in the preceding section.

Notably, within this intermediate regime, the results are not particularly sensitive to $\alpha$. Values spanning several orders of magnitude yield comparable outcomes. This is a highly favorable feature from a physical perspective and will be discussed in detail in the following section.

Approximate solutions approach the true solution for two reasons. First, the true solution lies within a constrained subspace rather than being randomly distributed over the full solution space. Second, physical signals are intrinsically bounded, whereas unphysical oscillations triggered by noise drive the solution norm to diverge. Accordingly, least-norm solutions inherently converge toward genuine physical states.
This accounts for the exclusion of highly oscillatory solutions caused by minor input errors, which do not belong to the true solution. Instead, such undesirable behaviors stem from the unbounded inverse of smoothing operators. Regularization eliminates these unphysical components via prior information. Low-frequency parts dominated by valid signals are preserved, whereas noise-dominated high-frequency parts are suppressed. Eliminating these irrelevant components helps to recover the true physical solution.

\paragraph{Results under Tikhonov Regularization in the $H^1$ space\\}

From Fig.~\ref{fig:numerical-toy-models-L2}, the solutions in the end-point region exhibit oscillations for some large values of $\alpha$ in each model. This is due to the use of boundary conditions and the fact that the $L^2$ space requires only square integrability, not smoothness. In many physical problems, solution functions are smooth, analogous to the three toy models. Therefore, we can enforce a prior smoothness condition by changing the solution space from $L^2$ to $H^1$ (see Appendix~\ref{sec:appendix-functional} for definitions).
In the $H^1$ space, the solution function is first-order differentiable, which will improve the solution to our problem.

\begin{figure}
\centering
\includegraphics[scale=0.35]{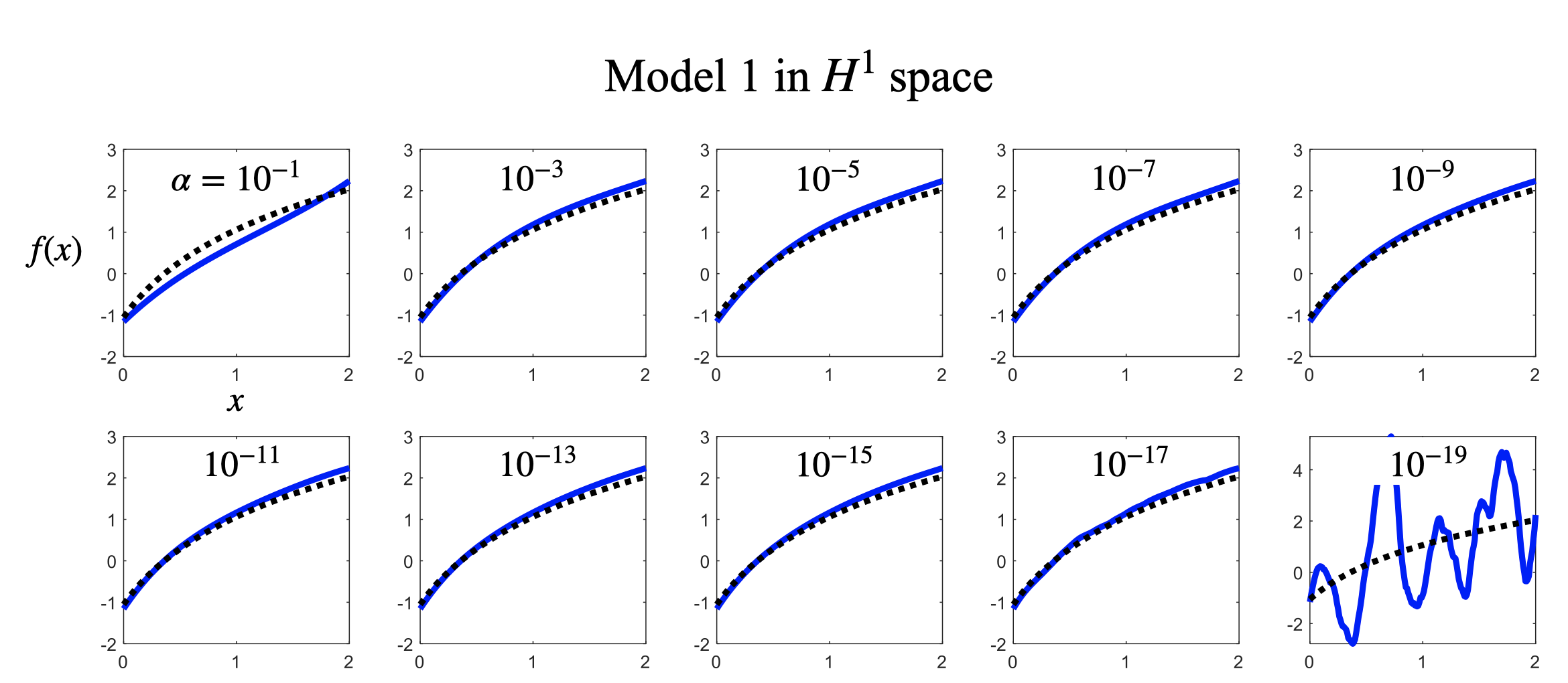}
\includegraphics[scale=0.35]{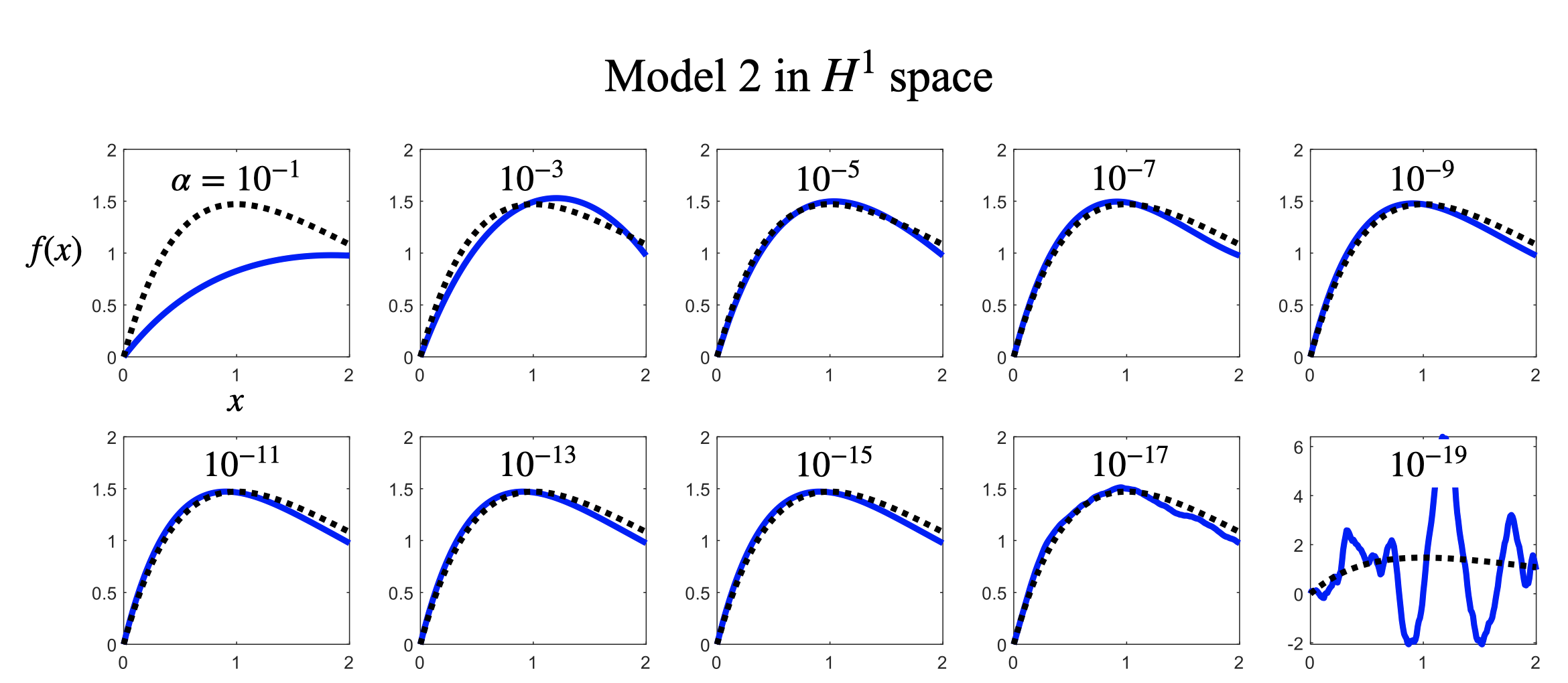}
\includegraphics[scale=0.35]{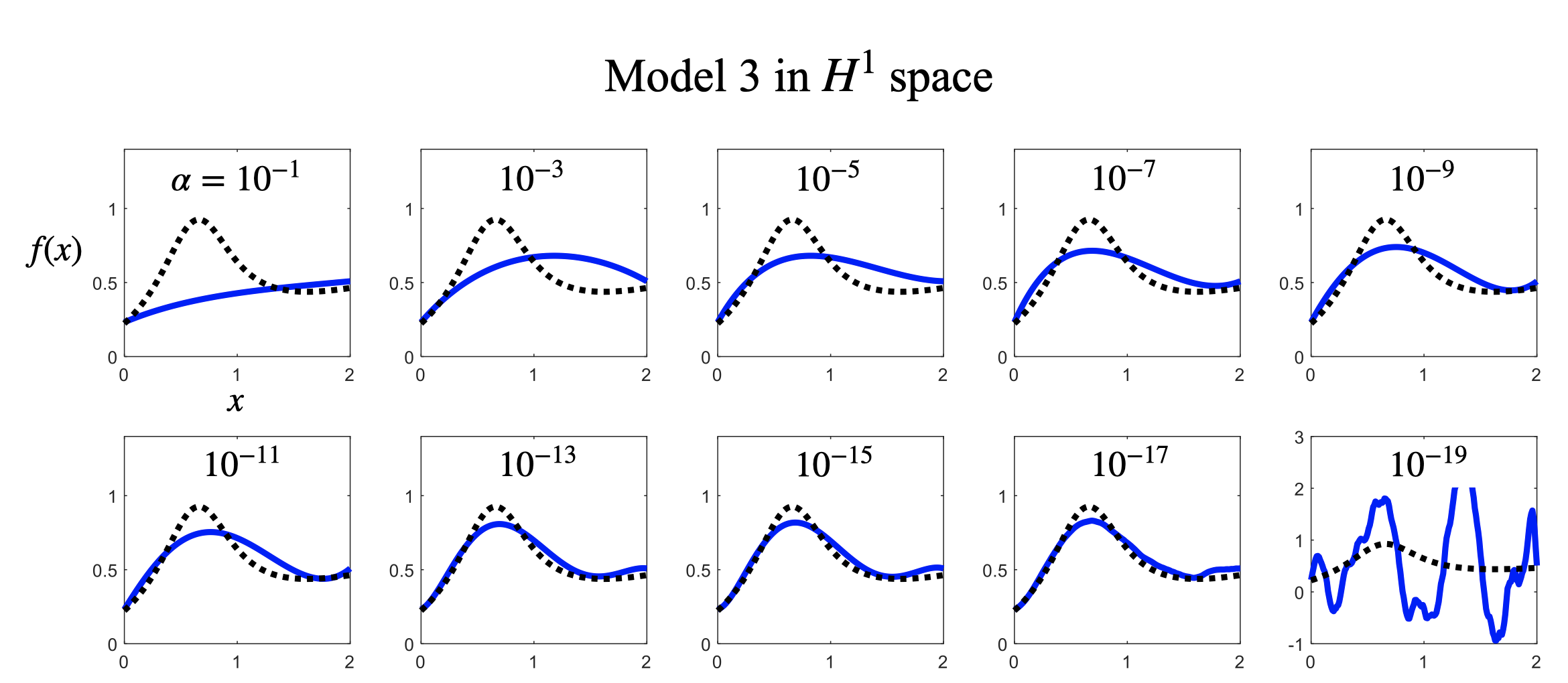}
\caption{Same as Fig.~\ref{fig:numerical-toy-models-L2} but for the solution space $F=H^1(t_{\rm min}, \Lambda)$.}
\label{fig:numerical-toy-models-H1}
\end{figure}

The numerical results in the $H^1$ space are presented in Fig.~\ref{fig:numerical-toy-models-H1}. 
Compared to the results of the $L^2$ space in Fig.~\ref{fig:numerical-toy-models-L2}, all three models show improvement, exhibiting no oscillations in the end-point regions. Models 1 and 2 display a wider plateau with respect to $\alpha$. For Model 3, improved solutions are obtained at $\alpha = 10^{-13}$ and $10^{-15}$, which are consistent with the true solution given an uncertainty of $10\%$.

As smoothness is a common a priori condition in physical applications, we adopt the $H^1$ space in the subsequent discussion, though the $L^2$ space would be used in practical cases where smoothness is not satisfied.

\paragraph{Stability Plateaus of Parameters\\}

\begin{figure}
    \centering
    \includegraphics[scale=0.45]{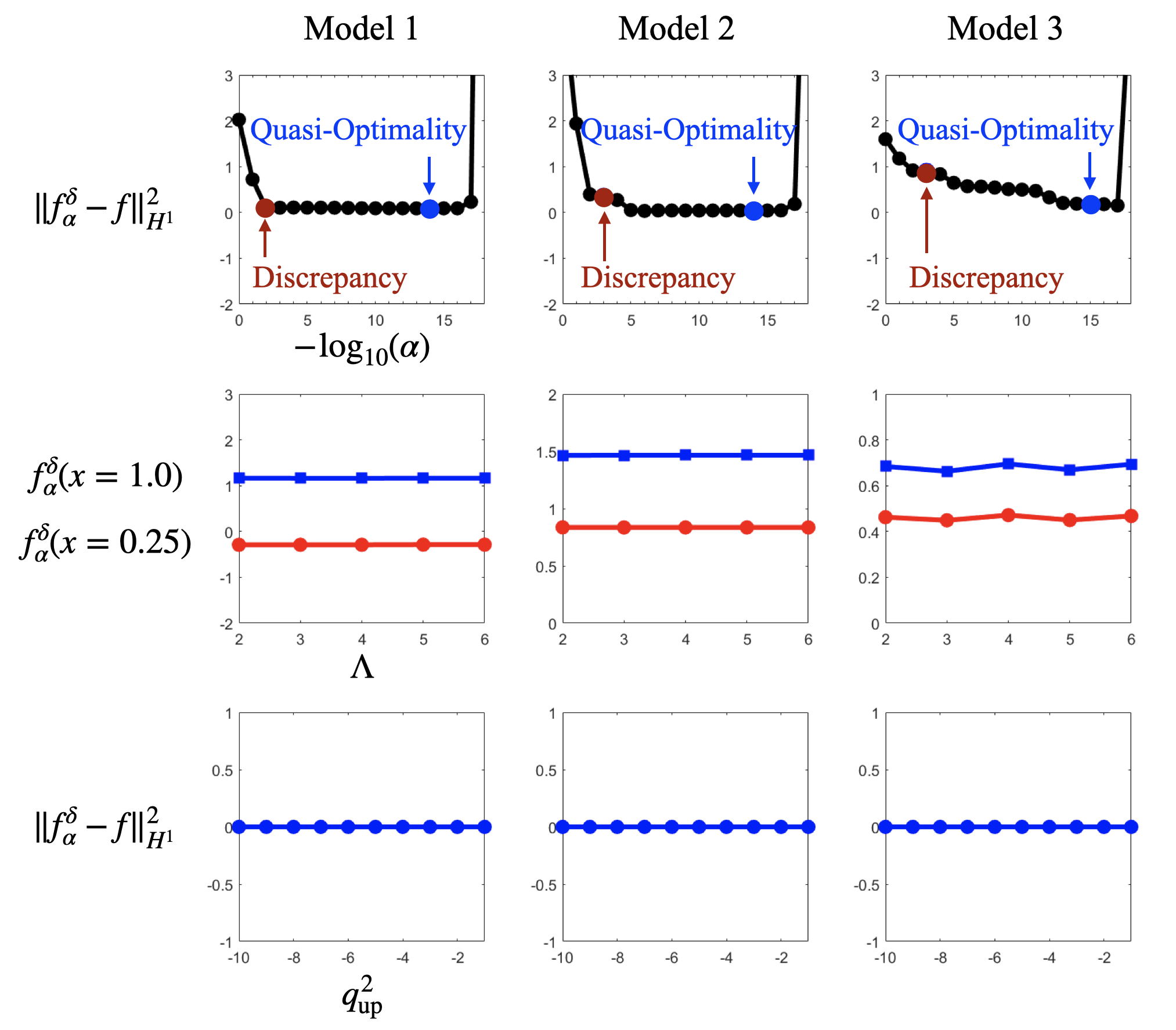}
    \caption{The dependence of numerical results on the regularization parameter $\alpha$ (top row), the perturbative-nonperturbative boundary $\Lambda$ (center row), and the upper bound of the input range $q_{\rm up}^2$ (bottom row) are shown for the three toy models (left to right). The full-range reconstruction errors $\|f_\alpha^\delta-f\|_{H^1}^2$ are presented for $\alpha$ and $q_{\rm up}^2$, while solutions at $x=0.25$ and $x=1.0$ are provided as illustrative examples for $\Lambda$.} 
    \label{fig:plateau-parameters}
\end{figure}

The sensitivity of the results to all parameters in the inverse problem approach is examined here. In the regularization method, the primary parameter is $\alpha$. As observed in Fig.~\ref{fig:numerical-toy-models-H1}, the results exhibit a broad stability plateau with respect to $\alpha$. This behavior is investigated in detail below. Additionally, Eq.~(\ref{eq:fxgy}) involves the scale parameter $\Lambda$, which separates the perturbative and non-perturbative regimes, together with the upper and lower bounds of the integration interval $q_{\rm up}^2$ and $q_{\rm low}^2$. The dependence of the results on these parameters is also analyzed. 
The numerical results of the tests are shown in Fig.~\ref{fig:plateau-parameters}.

We first examine the dependence of the results on $\alpha$, using the norm of the difference between the regularized solution and the exact solution, $\|f_\alpha^\delta - f\|_{H^1}^2$.  Clear plateaus spanning several orders of magnitude are observed, as shown at the top of  Fig.~\ref{fig:plateau-parameters}. 
Models~1 and~2 exhibit relatively wide plateaus, ranging from $\alpha = 10^{-2}$ to $10^{-16}$ for model~1 and from $10^{-5}$ to $10^{-16}$ for model~2, while model~3 shows a narrow plateau from $10^{-13}$ to $10^{-17}$.
Although $\alpha$ cannot be extremely large or small, a very wide range of intermediate values is acceptable, ensuring that the results remain insensitive to $\alpha$.
For simple problems such as model~1, different selection criteria choose values of $\alpha$ within the same plateau, demonstrating that the results are highly robust to the choice of $\alpha$.

Furthermore, the values of $\alpha$ determined by the quasi-optimality criterion are always significantly lower than those obtained from the discrepancy principle. The former typically yields the best solutions for all three models. 
Although the discrepancy principle possesses the favorable mathematical property of proven convergence, it is generally more appropriate for scenarios with minimal input errors (e.g., smaller than $0.1\%$), as discussed in the next subsection. 
Given the substantial input errors ($1\% - 30\%$), the quasi-optimality criterion is preferred to determine $\alpha$, and will be used in the following discussions unless otherwise specified. 

The dependence of the results on $\Lambda$ is shown in the center row of Fig.~\ref{fig:plateau-parameters}, using the solutions at $x=0.25$ and $x=1.0$ as examples. 
Plateaus are observed in $\Lambda\in[2,6]$, which is consistent with our expectations, since the transition from perturbative to non-perturbative regimes is continuous and smooth. 
In practical physical problems, the plateau of $\Lambda$ might not be as wide as those in toy models. A $\Lambda$ that is too large increases the amount of unknown information to be determined, thereby complicating the problem. In contrast, a $\Lambda$ that is too small introduces larger non-perturbative errors in the input term of Eq.~(\ref{eq:InverseProblem-DispersionRelation}), leading to deteriorated results. 
Note that the values of $\Lambda$ shown here are dimensionless in the toy models and do not correspond to the actual energy scales, which have dimensions of GeV$^2$. 

The influence of the input range is shown at the bottom of Fig.~\ref{fig:plateau-parameters}, where $q_{\rm up}^2\in [-10,-1]$ with $q_{\rm low}^2=-100$. The resulting plateaus are broad and stable. Although not shown, fixing $q_{\rm up}^2$ and varying $q_{\rm low}^2$ also yield excellent plateaus. Provided that the input data are sufficiently accurate, the results remain relatively insensitive to the choice of input range.

Finally, we discuss model 3. The plateau of $\alpha$ in model 3 is narrower than those in models 1 and 2. Furthermore, the dependence on $\Lambda$ in model 3 exhibits slight fluctuations, in contrast to the very flat behavior observed in models 1 and 2. 
We have also examined that reducing the width of the Breit-Wigner peak in model 3 leads to further deterioration of the results.
In the case of two or more peaks, the reconstruction  problem becomes substantially more difficult. 
These observations indicate that fine structures, such as Breit-Wigner peaks, are inherently more difficult to resolve under Tikhonov regularization in the $L^2$ and $H^1$ space. 
Other regularization methods may better handle such cases. 
Since the present work is intended as a foundational introduction rather than a realistic spectroscopy calculation, a full numerical demonstration is not necessary. Therefore, we maintain our focus on Tikhonov regularization in the $H^1$ space in the subsequent analysis of this work.

This finding indicates that no single regularization method is universally optimal for all inverse problems. Under a given regularization scheme, certain problems cannot be solved satisfactorily. Without further refinement, some problems may remain inherently unsolvable, especially with finite input errors.
By contrast, incorporating additional prior information and selecting a suitable regularization method can substantially improve the accuracy and reliability of the solution, as will be discussed in the following subsection.

%%%%%%%%%%%%%%%%%%%%%%%%%%%%%%%%%%%%%%%%%%%%%%%%%
\subsection{Ability to Analyze Uncertainties and Improve Precision}

Rigorous uncertainty analysis is essential for theory-experiment comparisons and the search for new physics. The ability to systematically improve precision is also a prerequisite for first-principles calculations. For instance, perturbative QCD can achieve higher precision through higher-order corrections, while lattice QCD can do so by increasing computational resources. 
In this subsection, we demonstrate that the inverse problem approach enables the rigorous analysis of both statistical and systematic uncertainties, as well as the controllable improvement of precision. This improvement is reflected in two aspects: (1) reducing the input error $\delta$, and (2) improving the prior information and regularization methods, as illustrated in Fig.~\ref{fig:Illustration-Regularization}.
%
%\begin{figure}
%    \centering
%    \includegraphics[width=0.8\linewidth]{Figure/fig-uncertainty-improve-precision.png}
%    \caption{Illustration of the ability to analyze statistical and systematic uncertainties and to improve precision systematically.}
%    \label{fig:precision-systematically-improvement}
%\end{figure}

\paragraph{Uncertainty Analysis\\}
From Eq.~(\ref{eq:error-bound}), the uncertainty of solution consists of two parts:
\begin{equation*}
    \| f_\alpha^\delta - f\| \leqslant  
    \delta \,\| R_\alpha\| + \| R_\alpha K f - f\|.
\end{equation*}
The first term $\delta\,\|R_\alpha\|$ represents the uncertainty generated from the input error $\delta$. It can be understood as statistical uncertainty. To analyze this uncertainty in detail, $\alpha$ is first fixed by solving the inverse problem for the central curve of $g^\delta(y)$. Then, all possible values of the input parameters (e.g., $a_{1}, a_{2}$) within the error bound are considered. For each of such $g_i^\delta$, the corresponding solution $f_{\alpha,i}^\delta(x)$ is computed. Repeating this process for all different inputs $g_i^\delta$ yields an ensemble of solutions, from which the statistical mean and the associated standard deviation are computed at each point $x$. 
In practice, such uncertainties originate from input parameters in perturbative calculations, including the renormalization-scale dependence, quark masses, condensates of OPE operators, and so on. 

The second term $\|R_\alpha Kf-f\|$ is the uncertainty due to the regularization approximation, i.e., the difference between the regularized solution $R_\alpha Kf$ and the exact solution $f$. This term can be interpreted as the systematic uncertainty. 
Since the exact solution $f$ is typically unknown, it can be replaced in practice by the regularized solution $f_\alpha^\delta$. Note that $\delta$ and $\alpha$ are fixed here. We then examine the difference between $R_\alpha K f_\alpha^\delta$ and $f_\alpha^\delta$.

The systematic uncertainties also include variations of parameters in the inverse problem approach, such as $\Lambda$, $q_{\rm max}^2$, $q_{\rm min}^2$, and $\alpha$, while respecting the plateaus within the range of each parameter. Since the plateau behaviors of these parameters in the toy models are perfect, as shown in Fig.~\ref{fig:plateau-parameters}, we do not consider such uncertainties in this work. However, they should be carefully accounted for in practical studies of non-perturbative QCD.

A key advantage of the inverse problem approach is its ability to rigorously quantify systematic uncertainty. Yet model-dependent methods typically cannot do this precisely. 
Theoretically, an admissible regularization requires $\alpha(\delta)\to 0$ as $\delta\to 0$ (see Definition~\ref{defn:convergence-solution}).  Hence, a smaller $\delta$ leads to a smaller $\alpha$ and consequently to a smaller systematic uncertainty. This chain of reasoning shows how the inverse problem approach can systematically improve precision, as illustrated in the following.

\paragraph{Improvement of Precision (I): Reducing Input Errors\\}
Having established a reasonable method for uncertainty analysis, we now examine how the solution behaves with input errors. 
\begin{figure}
\centering
\includegraphics[scale=0.35]{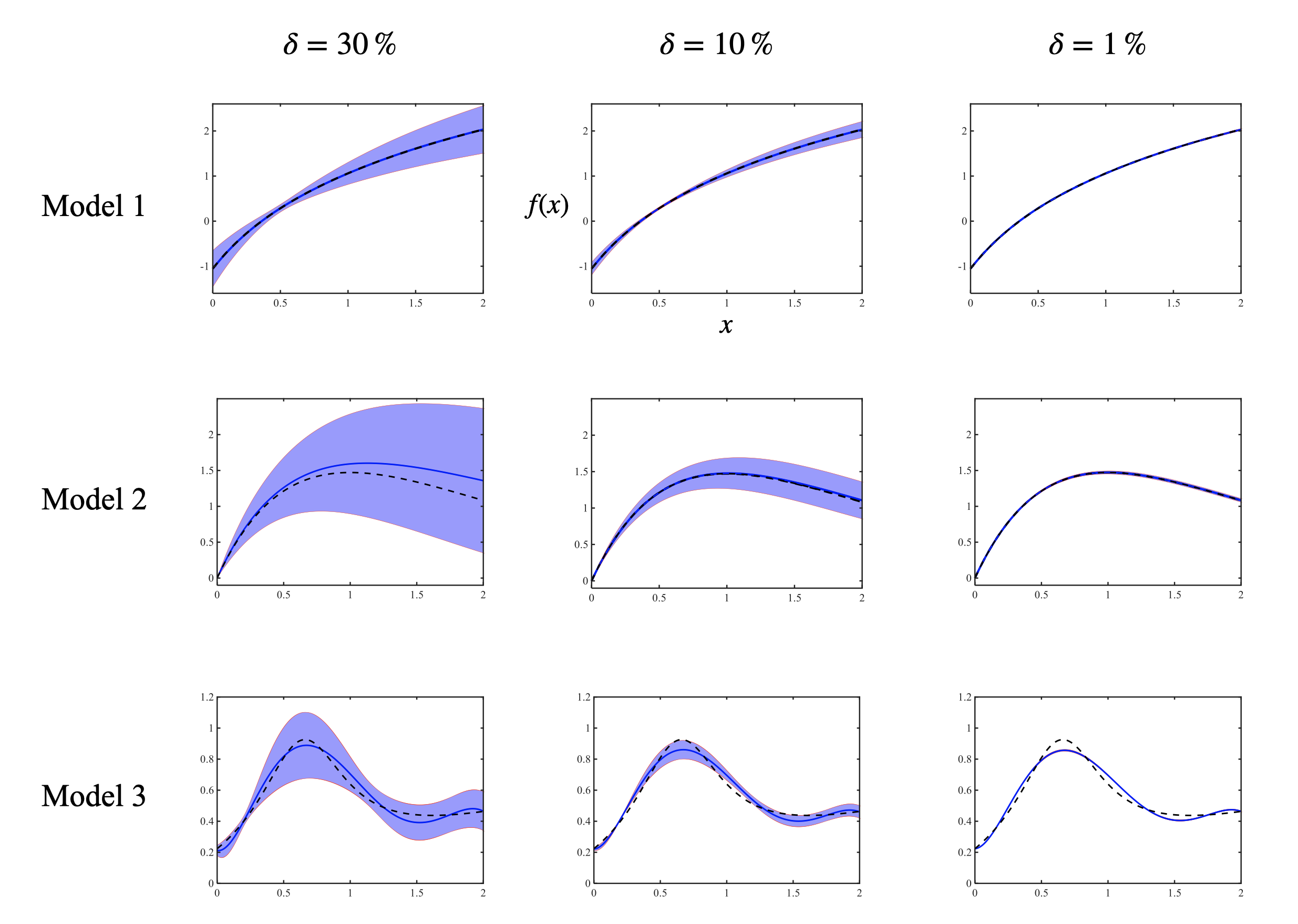}
\caption{Numerical results under three levels of input errors. The central values $f^{\delta}_{\alpha}$ (blue solid curve) with statistical uncertainties (purple band) and systematic uncertainties (red band, lying outside the purple band) are compared against the true solutions (black dashed curve). Rows: models 1, 2, 3 (top to bottom). Columns: input errors $30\%$, $10\%$, $1\%$ (left to right).}\label{fig:improve-precision-input-error}
\end{figure}
Three levels of input error ($30\%$, $10\%$, and $1\%$) are examined. The numerical results are shown in Fig.~\ref{fig:improve-precision-input-error}. 
The central values $f^{\delta}_{\alpha}$ represented by solid blue curves, together with their statistical uncertainties indicated by purple bands and systematic uncertainties shown as red bands outside the purple bands, are compared against the true solutions represented by black dashed curves.

The statistical uncertainties of the solutions are approximately $30\%$, $10\%$, and $1\%$, respectively, corresponding to the three levels of input errors. It is evident that the statistical uncertainties diminish with decreasing input errors, as expected. 
This concretely demonstrates that the solution precision improves as the input errors decrease.
In addition, a clear convergence of the solutions is observed in models 1 and 2. In particular, for the input error of $1\%$, the approximate solutions converge almost entirely to the true solutions.

%For model 3 with large input errors, the true solution lies within the uncertainty region. In contrast, when the input error is reduced to $1\%$, the uncertainty narrows but exhibits a noticeable deviation from the true solution. This once again verifies the limited power of Tikhonov regularization in the $L^2$ and $H^1$ spaces in addressing fine-structure problems like model 3. In this case, alternative regularization methods are required. 

The systematic uncertainties are too small to be clearly visible in the figures, because $\alpha$ is determined to be $10^{-14}$ or $10^{-15}$ by the quasi-optimality criterion. 
The numerical results show that the systematic uncertainties are of the order of $10^{-3}$ or $10^{-4}$, which is much lower than the statistical uncertainties of $30\%$–$1\%$.
The overall uncertainties are thus dominated by the statistical ones.

In addition, the values of $\alpha$ determined by the quasi-optimality criterion are almost unchanged when the input errors vary, i.e., they do not satisfy convergence. This is not a problem for input errors as large as $30\%-1\%$, where the systematic uncertainties are much smaller than the statistical ones.
When input errors are further reduced, e.g., to $10^{-3}$–$10^{-5}$, the discrepancy principle applies. However, since such low error levels lack a clear physical correspondence, we do not explicitly show the numerical results in this work.
The values of $\alpha$ from the discrepancy principle decrease with the input errors, leading to a reduction in systematic uncertainties. Consequently, both statistical and systematic uncertainties diminish as input errors decrease, indicating convergence of solutions, as proved in Theorem \ref{thm:discrepancy-principle}. Thus, the precision of solutions in the inverse problem approach can be systematically improved by reducing input errors.

In practice, physical input errors cannot be arbitrarily small, and the truncation error in perturbative QCD cannot be reduced infinitesimally by including more orders, since the series is asymptotic and diverges at high orders owing to renormalon effects. Nevertheless, our toy model tests show that even with input errors at the level of $1\%–10\%$, the precision of the solutions can still be improved as the input error decreases, without requiring extremely small input errors. This is illustrated in Fig. \ref{fig:improve-precision-input-error}.

\paragraph{Improvement of Precision (II): Improving Prior Information and Regularization Method\\}

Solving an ill-posed inverse problem can be taken by minimizing a regularized functional, which can be defined as \cite{Regularization_tools}
\begin{equation}
J_\alpha(f)=\|Kf-g^\delta\|_G^2 + \alpha \,\Omega[f],
\end{equation}
where $\Omega[f]$ is an appropriate norm that encodes prior information about the solution. 
For example, $\Omega[f]=\|f\|_F^2$ in Tikhonov regularization where the norm is defined on the space $F$ to characterize the properties of the solution $f$ and to control its stability.
Refining the prior information allows us to improve the definition of the functional in the regularization method, consequently leading to a more accurate solution.

The most intuitive improvement is the transition from the $L^2$ space to the $H^1$ space discussed in Sec.~\ref{sec:numerical-regularization}. When $F = L^2(t_{\rm min},\Lambda)$, only the boundedness of the solution is imposed, without any smoothness requirement. The resulting solutions are shown in Fig.~\ref{fig:numerical-toy-models-L2}, where oscillations appear at the left and right endpoints, or over the entire domain, for certain values of $\alpha$. When the smoothness prior is added, i.e., $F = H^1(t_{\rm min},\Lambda)$, the solutions are significantly improved, as shown in Fig.~\ref{fig:numerical-toy-models-H1}. In this case, the oscillations disappear and the plateau of $\alpha$ becomes substantially wider.

As a counterexample, Fig.~\ref{fig:improve-precision-input-error} shows that for model 3 at the $1\%$ input error, the obtained solution does not fully match the true solution. This is because Tikhonov regularization in the $L^2$ or $H^1$ norm space penalizes the squared magnitude of the solution or its derivatives integrated over the domain, where the peak values and steep slopes are amplified by the square and thus preferentially suppressed. Such regularization inherently favors smooth and flat solutions and is therefore less effective for problems with fine details. Alternative methods, such as $L^1$ norm, total variation regularization, iterative regularization, or discrete regularization \cite{Kirsch-2011,YanfeiYang}, may be more suitable, but their exploration is beyond the scope of this work and is left for future research.

%Another type of prior information originates from external constraints, such as precise experimental measurements or accurate calculations from methods like lattice QCD. For instance, when accurate experimental or lattice results are available at a specific energy point within the non-perturbative region, incorporating these constraints into the inverse problem can enhance the accuracy of the solution across the entire region. Similarly, in spectral problems, if the masses and widths of ground-state or excited-state particles have been experimentally measured, such prior information can be leveraged to refine the solution. Therefore, inverse problem methodologies are capable of working in synergy with experiments or lattice QCD to jointly advance the resolution of non-perturbative problems.

Finally, we discuss the influence of prior information on regularization methods and the resulting solution. In general, mathematical techniques alone cannot solve all problems. They cannot compensate for a lack of information, especially when the input data have errors. This is because the information is conserved in the sense that the solution cannot contain more information than is provided by the data and the prior combined.
Consequently, having more and correct prior information typically improves the quality of the solution. However, highly detailed prior information is not necessary. It often suffices to know qualitative properties of the solution, such as whether it is smooth or contains fine-scale structures like resonances. Then a regularization method that properly encodes such information should be chosen.

%% file: 6-summary.tex
\section{Conclusion and Outlook}\label{sec:conclusion}

In this work, we develop a novel theoretical framework, the inverse problem approach, to calculate non-perturbative QCD quantities, supported by rigorous mathematical theorems and proofs. The main ideas are illustrated in Fig.~\ref{fig:Inverse_Problem}. It begins with the QFT dispersion relation, separates the high-energy and low-energy scales, and uses known perturbative theories to determine unknown non-perturbative quantities by solving an inverse problem. Generally, we prove that the inverse problem of dispersion relation is ill-posed. It admits a unique but unstable solution. To obtain stable approximate solutions, regularization methods must be utilized. The idea behind regularization is to solve a nearby well-posed problem instead and use its solution as an approximation to the original ill-posed one.
We employ the Tikhonov regularization in this work, and present three toy models to illustrate the main features of the inverse problem approach. 

We summarize the advantages of the inverse problem approach:
\begin{itemize}
\item The framework is based on rigorous mathematical principles. The dispersion-relation inverse problem is proved to admit a unique but unstable solution. The well-established Tikhonov regularization is employed, ensuring that the regularized solution converges to the true solution as the input error decreases. 

\item It enables one to analyze statistical and systematic uncertainties. The precision of solution can be  systematically improved by reducing the input errors and by refining prior information and regularization methods. 

\item The framework treats the entire non-perturbative region simultaneously. It is beneficial for studying excited states and the continuum spectrum. However, resolving narrow resonant peaks requires incorporating prior information or refining the corresponding regularization methods. 

\item Numerical calculations can be completed with only modest computational resources. For illustration, all the tests of toy models in this work were run on a standard laptop, requiring just a few seconds to obtain a solution as in Figs.~\ref{fig:numerical-toy-models-L2} and \ref{fig:numerical-toy-models-H1}, and only a few minutes for uncertainties as shown in  Fig.~\ref{fig:improve-precision-input-error}. Provided that the perturbative input is available, the numerical solution of the inverse problem can be obtained very efficiently.

\end{itemize}

Due to the limited scope of this work, the discussion herein is restricted primarily to the mathematical level and does not address specific physical problems. Nevertheless, it is foreseeable that the inverse problem method has broad application prospects in high-energy physics. We have already employed this approach to study several physical quantities, such as the decay constants and light-cone distribution amplitudes of pion and $B$ meson. Our results will be released publicly shortly.

In addition, the regularization theory and methods presented in this work can be extended to a wider range of inverse problems in particle physics. The regularization formalism elaborated in Sec.~\ref{sec:regularization} does not rely on the dispersion relation. It is general enough to be applied to any ill-posed problem. Based on this framework, comprehensive investigations into various physical inverse problems become possible, such as extracting diverse non-perturbative quantities from experimental or lattice QCD data. Although the integral kernels in such problems may differ from those derived from dispersion relations, the regularization approaches remain universally applicable. Moreover, the regularization methods introduced in this work guarantee convergence of the approximate solution to the true solution and enable a reliable estimation of systematic uncertainties.

As experimental data continue to improve in precision, the demand for high-precision calculations is steadily increasing. Theoretically, achieving such high-precision predictions requires both high-order corrections in perturbative expansions and high-precision inputs for non-perturbative quantities. The inverse problem approach may provide a complementary method for determining non-perturbative physical quantities. 

Further improvements to the inverse problem framework based on dispersion relations are still required to fully achieve the above research objectives. In particular, it is crucial to enhance the accuracy of the perturbative input in dispersion integrals. Moreover, as inverse problem theory continues to develop as a branch of mathematics, more robust and efficient regularization schemes can further advance the field. Such methodologies can help explore the intrinsic properties of solutions and provide deeper insights into non-perturbative QCD.

%%%%%%%%%%%%%%%%%%%%%%%%%%%%%%%%%%%%%%%%
\section*{ACKNOWLEDGMENTS}
We are very grateful to Hsiang-nan Li, Hiroyuki Umeeda, and Fanrong Xu for collaboration on the pioneering work \cite{Li:2020xrz}, and especially to Hsiang-nan Li for many fruitful discussions throughout this project; to Kuang-Ta Chao, Xu Feng, and Yan-Qing Ma for their encouragement over the long-term development of this work; to Jun Hua, Xiao Huang, Dong-Hao Li, Sreeraj Nair, Ji Xu, and Jun Zeng for generous help with the writing; to Jian Liang for the analysis of statistical and systematic uncertainties; and to Jin Cheng and Xiong-Bin Yan for discussions on the mathematical aspects of the inverse problem. This work is supported by the Scientific Research Innovation Capability Support Project for Young Faculty under Grant No. ZYGXQNJSKYCXNLZCXM-P2,  by the Fundamental Research Funds for the Central Universities under No. lzujbky-2023-stlt01, lzujbky-2024-oy02 and lzujbky-2025-eyt01, and the National Natural Science Foundation of China under Grant No.12571455, and No.12335003.

%% file: Appendix.tex
\appendix

\section{Basic Functional Analysis}\label{sec:appendix-functional}
This appendix provides a summary of fundamental definitions and theorems from functional analysis \cite{Kirsch-2011}. 

\paragraph{Normed spaces ($L^2$ and $H^1$ spaces and so on)\\} \label{Hilbert space}
The physical problems here are all formulated within the framework of quantum field theory. Therefore, the relevant quantities are naturally defined in a Hilbert space, which is endowed with an inner product structure. 
In particular, we work within the $L^2$ and $H^1$ spaces throughout our analysis, as they provide the appropriate functional framework for describing the physics in its most general formulation. 

$L^2$ space (square-integrable function space) consists of all measurable functions whose squared absolute value has finite integral: 
\begin{equation}
 \int_a^b |f(x)|^2\, \mathrm{d}x < \infty .
\end{equation}
$H^1$ space (Sobolev space) consists of all functions in $L^2$ whose first weak derivatives also belong to $L^2$: 
\begin{equation}
\int_a^b |f(x)|^2 + |f'(x)|^2 \, \mathrm{d}x < \infty ,
\end{equation}
where $f'(x)$ is the first weak derivative of $f(x)$. 

The norms and inner products associated with these spaces are defined as follows. For the $L^2(a,b)$ space:
\begin{equation}
\begin{aligned}
& \|f\|_{L^2(a,b)} = \left( \int_a^b |f(x)|^2 \, \mathrm{d}x \right)^{1/2}\,, \\
& (f_1, f_2)_{L^2(a,b)} = \int_a^b f_1(x) f_2(x) \, \mathrm{d}x\,.
\end{aligned}
\end{equation}
For the $H^1(a,b)$ space:
\begin{equation}
\begin{aligned}
& \|f\|_{H^1(a,b)} = \left( \int_a^b \left( |f(x)|^2 + |f'(x)|^2 \right) \, \mathrm{d}x \right)^{1/2}\,, \\
& (f_1, f_2)_{H^1(a,b)} = \int_a^b \left( f_1(x) f_2(x) + f'_1(x) f'_2(x) \right) \, \mathrm{d}x\,.
\end{aligned}
\end{equation}

In addition, the function space $C[a,b]$ and $L^1[a,b]$ are used in the proof of uniqueness for the inverse problem of dispersion relation. 

$C[a,b]$ is the space of all real-valued (or complex-valued) continuous functions defined on the closed interval $[a,b]$. 
The standard norm on $C[a,b]$ is $\|f\|_{C[a,b]}=\max_{x\in[a,b]}|f(x)|$. 
An important property of $C[a,b]$ is that for $f_n,f\in C[a,b]$, $\|f_n- f\|_{C[a,b]}\to 0$ as $n\to \infty$, is equivalent to $f_n$ converging to $f$ uniformly on $[a,b]$. 

$L^1[a,b]$ is the space of all functions whose absolute value has a finite integral, i.e. $\|f\|_{L^1[a,b]}=\int_a^b\,|f(x)|\,dx<\infty$.

\paragraph{Linear Bounded and Compact Operators\\} \label{compact operators}

\begin{defn} [Linear Operator]
An operator $K: F \to G$ is called linear if
\begin{equation}
    K(\alpha f_1 + \beta f_2)=\alpha K f_1 + \beta K f_2\,, \quad \forall f_1, f_2 \in F, \alpha, \beta \in \mathbb{R}~ (~or~ \mathbb{C})\,.
\end{equation}
\end{defn}

\begin{defn} [Bounded Operator]
A linear operator $K: F \to G$ is called bounded if there exists a finite positive number $C$ such that
\begin{equation}
    \|Kf\|_{G} \leqslant C \, \|f\|_{F}\,,
\end{equation}
for all $f\in F$. Each number $C$ for which this inequality holds is called a bound for the operator $K$.
\end{defn}

\begin{thm}
    For a linear operator $K:F \to G$ mapping a normed space $F$ into normed space $G$ the following properties are equivalent: (a) $K$ is bounded. (b) $K$ is continuous at $f=0$, i.e., $f_j\to0$ implies that $Kf_j \to 0$. (c) $K$ is continuous for every $f\in F$.
\end{thm}

\begin{defn} [Operator Norm]\label{defn:operator-norm}
    Let $K:F \to G$ be a bounded linear operator. The operator norm of $K$ is defined as the supremum (least upper bound) of $\|Kf\|_G$ over all $f$ in the unit ball: 
    \begin{equation}
        \|K\|:=\sup\nolimits_{\|f\|_F\leqslant 1}\|Kf\|_G.
    \end{equation}
\end{defn}
If $f\neq0$, the operator norm can be equivalently defined as 
    \begin{equation}\label{defn:norm-operator-ratio}
        \|K\|:=\sup_{f\neq0}{\|Kf\|_G\over\|f\|_F}.
    \end{equation}
    This is a normalized perspective. For any non-zero input, we calculate the relative ratio by which the output is amplified, and then take the supremum of this ratio.

\begin{thm}\label{thm:operator-norm-inequality}
    For a bounded linear operator $K:F \to G$ and every $f\in F$, and by the definition of the operator norm, the following inequality holds: 
    \begin{equation}
        \|Kf\|_G\leqslant \|K\|\cdot \|f\|_F.
    \end{equation}
\end{thm}

\begin{defn}[Adjoint Operator]\label{defn:adjoint-operator}
    Let $F$ and $G$ be Hilbert spaces. The adjoint of a bounded linear operator $K:F \to G$ is the unique operator $K^*: G\to F$ satisfying the relation
    \begin{equation}
        ( Kf,g)_G = ( f,K^*g)_F,
    \end{equation}
    for all $f\in F$ and $g\in G$.
\end{defn}

\begin{defn}[Self-Adjoint Operator]
    A bounded linear operator $A: F\to F$ is called self-adjoint if $A = A^*$. Equivalently, it satisfies $( Af, f)_F = ( f, Af)_F$ for all $f\in F$.
\end{defn}

For a self-adjoint operator $A$, its norm admits an equivalent characterization via the quadratic form:
\begin{equation}\label{eq:norm-self-adjoint-operator}
    \|A\| = \sup_{\|f\| = 1} |( Af, f)_F|.
\end{equation}

\begin{defn}[Compact Operator]
    A linear operator $K:F \to G$ is called compact if it maps each bounded set in $F$ into a relatively compact set in $G$.
\end{defn}

%%%%%%%%%%%%%%%%%%%%%%%%%%%%%%%%%%%%%%%%%%%%%%
\section{Proofs of theorems in Section~\ref{sec:regularization}}\label{sec:appendix-proof-regularization}

In this appendix, we complete the proofs of results in the section of regularization, Sec.~\ref{sec:regularization}.

\noindent\textbf{Theorem 4.3.}
   \textit{ Let $K:F \to G$ be a bounded linear operator between Hilbert spaces and $\alpha>0$. Then the Tikhonov functional $J_\alpha$,
    \begin{equation}
        J_\alpha(f):=\|Kf-g\|_G^2 + \alpha \,\|f\|_F^2 ~~~~~\text{for}~~f\in F,
    \end{equation}
    has a unique minimizer $f_\alpha = \arg\min_{f\in F} J_\alpha(f)$. This minimizer $f_\alpha$ is the unique solution of the normal equation \cite{Kirsch-2011}
    \begin{equation}
        \alpha\, f_\alpha + K^*Kf_\alpha = K^* g.
    \end{equation}}

\begin{proof}
\text{Firstly, existence and uniqueness are proven.}

Let $\{f_n\}\subset F$ be a minimizing sequence such that $J_\alpha(f_n)\to J_0=\inf_{f\in F}J_\alpha(f)$ as $n\to \infty$. By expanding the squared norms, we have
\begin{align*}
       J_\alpha(f_n) + J_\alpha(f_m) 
       &= \|Kf_n - g\|^2 + \|Kf_m - g\|^2 + \alpha\left(\|f_n\|^2 + \|f_m\|^2\right) \\
       &= 2\left\|K\left(\frac{f_n+f_m}{2}\right) - g\right\|^2 + \frac{1}{2}\|K(f_n - f_m)\|^2 \\
       &\quad + 2\alpha\left\|\frac{f_n+f_m}{2}\right\|^2 + \frac{\alpha}{2}\|f_n - f_m\|^2 \\
       &= 2J_\alpha\left(\frac{f_n+f_m}{2}\right) + \frac{1}{2}\|K(f_n - f_m)\|^2 + \frac{\alpha}{2}\|f_n - f_m\|^2.
\end{align*}
Since $J_\alpha\left(\frac{f_n+f_m}{2}\right) \geqslant J_0$ and $\|K(f_n - f_m)\|^2 \geqslant 0$, it follows that
\begin{equation*}
    J_\alpha(f_n) + J_\alpha(f_m) \geqslant 2J_0 + \frac{\alpha}{2}\|f_n - f_m\|^2.
\end{equation*}
Taking the limit as $n, m \to \infty$, the left-hand side converges to $2J_0$. Consequently, $2J_0 \geqslant 2J_0 + \frac{\alpha}{2} \lim_{n,m \to \infty} \|f_n - f_m\|^2$,
which implies $\lim_{n,m \to \infty} \|f_n - f_m\| = 0$ because $\alpha > 0$. Thus, $\{f_n\}$ is a Cauchy sequence in $F$. Since $F$ is a Hilbert space (and thus complete), there exists an element $f_\alpha \in F$ such that $f_n \to f_\alpha$.

The functional $J_\alpha$ is continuous (as it is composed of continuous linear operators and norms). Therefore,
    \begin{equation*}
        J_\alpha(f_\alpha) = \lim_{n \to \infty} J_\alpha(f_n) = J_0.
    \end{equation*}
This establishes the existence of a minimizer. The uniqueness follows from the strict convexity of $J_\alpha$. Specifically, the term $\alpha\|f\|^2$ with $\alpha > 0$ is strictly convex, and $\|Kf-g\|^2$ is convex; the sum of a strictly convex function and a convex function is strictly convex, ensuring at most one minimizer. Alternatively, uniqueness is immediate from the fact that any two minimizers would generate a minimizing sequence with distance zero as shown above.

\text{Secondly, equivalence with the normal equation is demonstrated.}

($\Rightarrow$) Suppose $f_\alpha$ is the unique minimizer of $J_\alpha$. Then for any arbitrary element $f \in F$, we have $J_\alpha(f) \geqslant J_\alpha(f_\alpha)$.
The difference $J_\alpha(f) - J_\alpha(f_\alpha)$ becomes
    \begin{align*}
        J_\alpha(f) - J_\alpha(f_\alpha) 
        =& \left( \|K f - g\|^2 + \alpha \|f\|^2 \right) - \left( \|Kf_\alpha - g\|^2 + \alpha \|f_\alpha\|^2 \right) \\
        =& (Kf,Kf)-(Kf_\alpha,Kf_\alpha)-2\text{Re}(Kf,g) + 2\text{Re}(Kf_\alpha,g)+\alpha (f,f)-\alpha (f_\alpha,f_\alpha) \\
        =& (K^*Kf,f) - (K^*Kf_\alpha, f_\alpha) - 2\text{Re}(K^*g,f) + 2\text{Re}(K^*g,f_\alpha)+\alpha (f,f)-\alpha (f_\alpha,f_\alpha)\\
        =& 2\text{Re}(K^*Kf_\alpha + \alpha f_\alpha - K^*g,f) 
          -2\text{Re}(K^*Kf_\alpha + \alpha f_\alpha - K^*g,f_\alpha)\\
          &- 2\text{Re}(K^*Kf_\alpha, f) + (K^*Kf_\alpha,f_\alpha) + (K^*Kf,f) 
          + \alpha (f_\alpha,f_\alpha) + \alpha (f,f) - 2\text{Re}(f_\alpha,f)\\
        =& 2\,\text{Re}\left(K^*K f_\alpha+\alpha f_\alpha - K^*g,\, f-f_\alpha \right) + \|K(f-f_\alpha)\|^2 + \alpha\,\|f-f_\alpha\|^2.
    \end{align*}
Since $f_\alpha$ minimizes $J_\alpha(f)$, we substitute $f=f_\alpha+tz$ for any $t>0$ and $z\in F$, and arrive at 
\begin{equation*}
    2t\,\text{Re}\left(K^*K f_\alpha+\alpha f_\alpha - K^*g,\, z \right) + t^2\|Kz\|^2 + \alpha\,t^2\|z\|^2 \geqslant0.
\end{equation*}
Division by $t>0$ and taking the limit $t\to0$ yields
\begin{equation*}
    \text{Re}\left(K^*K f_\alpha+\alpha f_\alpha - K^*g,\, z \right) 
    \geqslant 0,
\end{equation*}
for all $z\in F$. Since $z$ is arbitrary, we can replace $z$ with $-z$, which reverses the sign of the inner product term while keeping the inequality valid. This implies the term must be zero. Thus, we have $K^*K f_\alpha+\alpha f_\alpha - K^*g=0$. That is, $f_\alpha$ solves the normal equation.

($\Leftarrow$) Conversely, suppose $f_\alpha$ satisfies the normal equation $(\alpha I + K^*K)f_\alpha = K^*g$. 
The difference formula reduces to:
\begin{equation*}
    J_\alpha(f) - J_\alpha(f_\alpha) = \|K(f-f_\alpha)\|^2 + \alpha\|f-f_\alpha\|^2 \geqslant 0,
\end{equation*}
which confirms that $f_\alpha$ is the global minimizer.
\end{proof}

\noindent\textbf{Theorem 4.4} $\,$(Discrepancy Principle)\textbf{.}
    \textit{Let $K: F \to G$ be linear, compact and one-to-one with dense range in $G$. Let $Kf=g$ with $f\in F, g\in G, g^\delta \in G$ such that $\| g-g^\delta \|\leqslant \delta < \|g^\delta \|$. Let the Tikhonov solution $f^\delta_\alpha$ satisfy 
    \begin{equation}
        \left\| K f^\delta_\alpha -g^\delta \right\| = \delta.
    \end{equation} 
    Then $f^\delta_\alpha \to f$ for $\delta \to 0$ \cite{Kirsch-2011}. 
}
\begin{proof}
We first prove weak convergence. For any $z\in G$, 
\begin{equation*}
\begin{aligned}
\left|(Kf_\alpha^\delta-g,z)\right| 
&\leqslant \left\| Kf_\alpha^\delta-g \right\|\, \left\| z \right\| \\
&\leqslant \left(\left\| Kf_\alpha^\delta - g^\delta \right\| + \left\| g^\delta - g \right\|\right)\,\left\|z\right\| \\
&\leqslant 2\,\delta \,\left\|z\right\|.
\end{aligned}
\end{equation*}
This implies that $Kf_\alpha^\delta \rightharpoonup g$, or equivalently $f_\alpha^\delta \rightharpoonup f$, as $\delta\to 0$.

The proof of strong convergence proceeds as follows. The minimization property of the Tikhonov functional implies the inequality $\|Kf_\alpha^\delta-g^\delta\|^2+\alpha \|f_\alpha^\delta\|^2 \leqslant \|Kf-g^\delta\|^2+\alpha \|f\|^2$ for any $f \in F$. From this, the following chain of inequalities is obtained
\begin{equation*}
    \begin{aligned}
        \delta^2 + \alpha \|f_\alpha^\delta\|^2 
        &= \| Kf_\alpha^\delta - g^\delta \|^2 + \alpha \|f_\alpha^\delta\|^2 \\
        &\leqslant \| Kf-g^\delta \|^2 + \alpha \|f\|^2 \\
        &= \| g-g^\delta \|^2 + \alpha \|f\|^2 \\
        &\leqslant \delta^2 +  \alpha \|f\|^2.
    \end{aligned}
\end{equation*}
Subtracting $\delta^2$ from both sides and dividing by $\alpha > 0$ yields
\begin{equation*}
\|f_\alpha^\delta\|\leqslant \|f\|.
\end{equation*}
Finally, an analysis of the norm of the difference gives:
\begin{equation*}
    \begin{aligned}
        \| f_\alpha^\delta -f \|^2
        &= \|f_\alpha^\delta\|^2-2\,\text{Re}(f_\alpha^\delta, f) + \|f\|^2 \\
        &\leqslant 2\|f\|^2 -2\,\text{Re}(f_\alpha^\delta, f) \\
        &= 2\,\text{Re}(f-f_\alpha^\delta, f).
    \end{aligned}
\end{equation*}
Due to the weak convergence established earlier, the right-hand side converges to 0 as $\delta\to 0$. Thus, $\| f_\alpha^\delta -f \|\to 0$, which proves the strong convergence $f_\alpha^\delta \to f$ as $\delta\to0$.
\end{proof}

\noindent\textbf{Theorem 4.5}$\,$ (Convergence of the finite element method)\textbf{.}
\textit{
For fixed noise level $\delta$ and regularization parameter $\alpha>0$, we have 
\begin{equation}
\|f_{\alpha ,n}^\delta  -f^\delta_\alpha\|_F\rightarrow 0, \quad n\rightarrow \infty.
\end{equation}}

\begin{proof}
We begin by proving weak convergence, and subsequently demonstrate strong convergence. 

\emph{Weak convergence.}
Let $P_n: F\to X_n$ denote the orthogonal projection onto the finite-dimensional subspace $X_n\subset F$, where $F$ is either $L^2{(t_{\rm min},\Lambda)}$ or $H^1{(t_{\rm min},\Lambda)}$. By definition, for any $u\in F$ and all $v\in X_n$, it satisfies $(u-P_n u, v)_F=0$.
Standard finite element approximation results ensure that $\|u-P_n u\|_F\rightarrow 0$ as $n \rightarrow \infty$, for $u\in H^1(t_{\rm min},\Lambda)$. 
Since $f^\delta_\alpha$ solves $\alpha f^\delta_\alpha+K^*K f^\delta_\alpha=K^*g^\delta$, 
and $K^*$ maps $G$ into $C^\infty[t_{\rm min},\Lambda]$ due to the smoothing property of the integral kernel, it follows that $f^\delta_\alpha\in H^1(t_{\rm min},\Lambda)$. Consequently, 
\begin{equation}\label{eq:Pnfalpha}
    \left\|f^\delta_\alpha-P_n f^\delta_\alpha \right\|_{F}\rightarrow 0,\quad n\to \infty.
\end{equation}

Now $f_\alpha^\delta$ is the global minimizer of the functional in $F$ from Eq.~(\ref{eq:minimizer-functional}), while $f_{\alpha,n}^\delta$ minimizes the same functional restricted to $X_n$ as in Eq.~(\ref{eq:minimize-discrete-functional}). Hence, $J_\alpha(f^\delta_\alpha) \leqslant J_\alpha(f_{\alpha,n}^\delta) \leqslant J_\alpha(P_n f^\delta_\alpha)$. Then,
\begin{equation*}
    \begin{aligned}
        0 \leqslant & J_\alpha(P_n f^\delta_\alpha)-J_\alpha(f^\delta_\alpha) \\
        =& 
        \left\| K(P_n f^\delta_\alpha-f^\delta_\alpha)\right\|_{G}^2 + 2\left(K(P_n f^\delta_\alpha-f^\delta_\alpha),Kf^\delta_\alpha-g^\delta\right)_{G} \\
        & + \alpha \left\| P_n f^\delta_\alpha-f^\delta_\alpha \right\|_{F}^2 + 2\alpha \left(P_n f^\delta_\alpha-f^\delta_\alpha,f^\delta_\alpha \right)_{F} \\
        \leqslant &
        \| K\|^2
        \left\|P_n f^\delta_\alpha-f^\delta_\alpha\right\|_{F}^2 + 2\|K\| \,\left\|P_n f^\delta_\alpha-f^\delta_\alpha\right\|_ {F} \left\|Kf^\delta_\alpha-g^\delta\right\|_{G}\\
        & + {\alpha} \left\| P_n f^\delta_\alpha-f^\delta_\alpha \right\|_{F}^2 + 2\alpha \left\|P_n f^\delta_\alpha-f^\delta_\alpha \right\|_{F} \left\|f^\delta_\alpha \right\|_{F}\,.
    \end{aligned} \label{eq:JalphaPn}
\end{equation*}
From Eq.~(\ref{eq:Pnfalpha}), we conclude $J_\alpha(P_n f^\delta_\alpha)-J_\alpha(f^\delta_\alpha)\to 0$, as $n\to \infty$.
Thus, $J_\alpha(f_{\alpha,n}^\delta)\to J_\alpha(f^\delta_\alpha)$ as $n\to \infty$, indicating that $f_{\alpha,n}^\delta$ is a minimizing sequence. Because $J_\alpha(f_{\alpha,n}^\delta)$ is convergent, it is bounded. The coercivity $\alpha \|f_{\alpha,n}^\delta\|_F^2 \leqslant J_\alpha(f_{\alpha,n}^\delta)$ then implies $\|f_{\alpha,n}^\delta\|_F$ is uniformly bounded in $n$. Consequently, there exists a subsequence $f_{\alpha,n_k}^\delta$ and a $v$ in $F$, such that $f_{\alpha,n_k}^\delta\rightharpoonup v$, as $k\to \infty$.

Since $K:F\to G$ is a bounded linear operator, weak convergence in $F$ implies weak convergence in $G$: $Kf_{\alpha,n_k}^\delta\rightharpoonup Kv$, as $k\to \infty$. By the weak lower semicontinuity of the norm, $J_\alpha(v)\leqslant \liminf_{k\to \infty} J_\alpha(f_{\alpha,n_k}^\delta)$. Since $J_\alpha(f_{\alpha,n}^\delta)\to J_\alpha(f_\alpha^\delta)$ as $n\to \infty$, the same holds for any subsequence. In particular, $\lim_{k\to \infty} J_\alpha(f_{\alpha,n_k}^\delta)=J_\alpha(f^\delta_\alpha)$. Hence, $J_\alpha(v)\leqslant \liminf_{k\to \infty} J_\alpha(f_{\alpha,n_k}^\delta)=J_\alpha(f^\delta_\alpha)$. 
This shows that $v$ is the minimizer of $J_\alpha$ over $F$. By Theorem \ref{thm:minimization-functional}, the minimizer is unique, so $v=f^\delta_\alpha$. Hence, 
\begin{equation}
    f_{\alpha,n_k}^\delta\rightharpoonup f^\delta_\alpha,\quad\text{weakly in } F,\quad k\to \infty.
\end{equation}

\emph{Strong convergence.}
Suppose, for contradiction, that the convergence of norms does not hold, i.e., $\limsup_{k\to \infty} \|f_{\alpha,n_k}^\delta\|_{F}> \|f^\delta_\alpha\|_{F}$.
%Then there exists a subsequence such that $\lim_{j\to \infty} \|f_{\alpha,n_j}^\delta\|_{F}> \|f^\delta_\alpha\|_{F}$.
By the weak lower semicontinuity,
\begin{equation}
    \begin{aligned}
        \|Kf^\delta_\alpha-g^\delta\|_{G}^2
        &\leqslant \liminf_{j\to \infty}\|Kf_{\alpha,n_j}^\delta-g^\delta\|_{G}^2 
        \leqslant \limsup_{j\to \infty}\|Kf_{\alpha,n_j}^\delta-g^\delta\|_{G}^2 \\
        &= \limsup_{j\to \infty} \left(J(f_{\alpha,n_j}^\delta)-{\alpha}\|f_{\alpha,n_j}^\delta\|_{F}^2\right) 
        < J(f^\delta_\alpha)-{\alpha} \|f^\delta_\alpha\|_{F}^2\\
        &=\|Kf_{\alpha}^\delta-g^\delta\|_{G}^2 \,,
    \end{aligned}
\end{equation}
which is a contradiction. Hence, $\limsup_{k\to \infty} \|f_{\alpha,n_k}^\delta\|_{F} \leqslant \|f^\delta_\alpha\|_{F}$. Together with the weak lower semicontinuity of the norm in $F$, we have
\begin{equation}
    \begin{aligned}
        \|f^\delta_\alpha\|_{F}
        \leqslant \liminf_{k\to \infty} \|f_{\alpha,n_k}^\delta\|_{F}
        \leqslant \limsup_{k\to \infty} \|f_{\alpha,n_k}^\delta\|_{F}
        \leqslant \|f^\delta_\alpha\|_{F}\,.
    \end{aligned}
\end{equation}
Thus, $\|f_{\alpha,n_k}^\delta\|_{F}\to \|f^\delta_\alpha\|_{F}$ as $k\to \infty$. 
%Since $f_{\alpha,n_k}^\delta\rightharpoonup f^\delta_\alpha $ as $k \to \infty$ weakly and and their norms converge,
We obtain strong convergence in $F$. This completes the proof.
\end{proof}

%% file: main.bbl
\begin{thebibliography}{99}

%\cite{Wilson:1974sk}
\bibitem{Wilson:1974sk}
K.~G.~Wilson,
``Confinement of Quarks,''
Phys. Rev. D \textbf{10} (1974), 2445-2459
%doi:10.1103/PhysRevD.10.2445
%7054 citations counted in INSPIRE as of 05 May 2026

%\cite{Shifman:1978bx}
\bibitem{Shifman:1978bx}
M.~A.~Shifman, A.~I.~Vainshtein and V.~I.~Zakharov,
``QCD and Resonance Physics. Theoretical Foundations,''
Nucl. Phys. B \textbf{147} (1979), 385-447
%doi:10.1016/0550-3213(79)90022-1
%5986 citations counted in INSPIRE as of 04 Nov 2025

%\cite{Shifman:1978by}
%\bibitem{Shifman:1978by}
M.~A.~Shifman, A.~I.~Vainshtein and V.~I.~Zakharov,
``QCD and Resonance Physics: Applications,''
Nucl. Phys. B \textbf{147} (1979), 448-518
%doi:10.1016/0550-3213(79)90023-3
%3290 citations counted in INSPIRE as of 04 Nov 2025

%\cite{Colangelo:2000dp}
\bibitem{Colangelo:2000dp}
P.~Colangelo and A.~Khodjamirian,
``QCD sum rules, a modern perspective,''
%doi:10.1142/9789812810458{\_}0033
[arXiv:hep-ph/0010175 [hep-ph]].
%830 citations counted in INSPIRE as of 04 Nov 2025

%\cite{Roberts:1994dr}
\bibitem{Roberts:1994dr}
C.~D.~Roberts and A.~G.~Williams,
``Dyson-Schwinger equations and their application to hadronic physics,''
Prog. Part. Nucl. Phys. \textbf{33} (1994), 477-575
%doi:10.1016/0146-6410(94)90049-3
[arXiv:hep-ph/9403224 [hep-ph]].
%1249 citations counted in INSPIRE as of 04 Nov 2025

%\cite{Chen:2016qju}
\bibitem{Chen:2016qju}
H.~X.~Chen, W.~Chen, X.~Liu and S.~L.~Zhu,
``The hidden-charm pentaquark and tetraquark states,''
Phys. Rept. \textbf{639} (2016), 1-121
%doi:10.1016/j.physrep.2016.05.004
[arXiv:1601.02092 [hep-ph]].
%1315 citations counted in INSPIRE as of 04 Nov 2025

%\cite{Li:2020xrz}
\bibitem{Li:2020xrz}
H.~N.~Li, H.~Umeeda, F.~Xu and F.~S.~Yu,
``$D$ meson mixing as an inverse problem,''
Phys. Lett. B \textbf{810} (2020), 135802
%doi:10.1016/j.physletb.2020.135802
[arXiv:2001.04079 [hep-ph]].
%44 citations counted in INSPIRE as of 04 Nov 2025


\bibitem{HnLi-spectrum}
H.~n.~Li and H.~Umeeda,
``QCD sum rules with spectral densities solved in inverse problems,''
Phys. Rev. D \textbf{102} (2020), 114014
%doi:10.1103/PhysRevD.102.114014
[arXiv:2006.16593 [hep-ph]].
%17 citations counted in INSPIRE as of 04 Nov 2025

H.~n.~Li,
``Dispersive analysis of glueball masses,''
Phys. Rev. D \textbf{104} (2021), 114017
%doi:10.1103/PhysRevD.104.114017
[arXiv:2109.04956 [hep-ph]];
%22 citations counted in INSPIRE as of 04 Nov 2025

%\cite{Li:2024fko}
%\bibitem{Li:2024fko}
H.~n.~Li,
``Dispersive Analysis of Excited Glueball States,''
Chin. Phys. Lett. \textbf{41} (2024), 101101
%doi:10.1088/0256-307X/41/10/101101
[arXiv:2408.06738 [hep-ph]].
%8 citations counted in INSPIRE as of 05 May 2026

S.~Zeng, H.~n.~Li and F.~Xu,
``Light baryon static properties in dispersive approach,''
[arXiv:2603.15989 [hep-ph]].
%0 citations counted in INSPIRE as of 05 May 2026

%\cite{Chew:1957zz}
\bibitem{Chew:1957zz}
G.~F.~Chew, M.~L.~Goldberger, F.~E.~Low and Y.~Nambu,
``Application of Dispersion Relations to Low-Energy Meson-Nucleon Scattering,''
Phys. Rev. \textbf{106} (1957), 1337-1344
%doi:10.1103/PhysRev.106.1337
%307 citations counted in INSPIRE as of 06 May 2026

%\cite{Mandelstam:1958xc}
\bibitem{Mandelstam:1958xc}
S.~Mandelstam,
``Determination of the pion - nucleon scattering amplitude from dispersion relations and unitarity. General theory,''
Phys. Rev. \textbf{112} (1958), 1344-1360
%doi:10.1103/PhysRev.112.1344
%609 citations counted in INSPIRE as of 06 May 2026

%\cite{Dobado:1992ha}
\bibitem{Dobado:1992ha}
A.~Dobado and J.~R.~Pelaez,
``A Global fit of pi pi and pi K elastic scattering in ChPT with dispersion relations,''
Phys. Rev. D \textbf{47} (1993), 4883-4888
%doi:10.1103/PhysRevD.47.4883
[arXiv:hep-ph/9301276 [hep-ph]].
%269 citations counted in INSPIRE as of 06 May 2026

%\cite{Lin:2021xrc}
\bibitem{Lin:2021xrc}
Y.~H.~Lin, H.~W.~Hammer and U.~G.~Mei{\ss}ner,
``New Insights into the Nucleon{\textquoteright}s Electromagnetic Structure,''
Phys. Rev. Lett. \textbf{128} (2022), 052002
%doi:10.1103/PhysRevLett.128.052002
[arXiv:2109.12961 [hep-ph]].
%118 citations counted in INSPIRE as of 06 May 2026

%\cite{Khodjamirian:2010vf}
\bibitem{Khodjamirian:2010vf}
A.~Khodjamirian, T.~Mannel, A.~A.~Pivovarov and Y.~M.~Wang,
``Charm-loop effect in $B \to K^{(*)} \ell^{+} \ell^{-}$ and $B\to K^*\gamma$,''
JHEP \textbf{09} (2010), 089
%doi:10.1007/JHEP09(2010)089
[arXiv:1006.4945 [hep-ph]].
%560 citations counted in INSPIRE as of 06 May 2026

%\cite{Aoyama:2020ynm}
\bibitem{Aoyama:2020ynm}
T.~Aoyama, N.~Asmussen, M.~Benayoun, J.~Bijnens, T.~Blum, M.~Bruno, I.~Caprini, C.~M.~Carloni Calame, M.~C{\`e} and G.~Colangelo, \textit{et al.}
``The anomalous magnetic moment of the muon in the Standard Model,''
Phys. Rept. \textbf{887} (2020), 1-166
%doi:10.1016/j.physrep.2020.07.006
[arXiv:2006.04822 [hep-ph]].
%1827 citations counted in INSPIRE as of 06 May 2026

%\cite{Balitsky:1989ry}
\bibitem{Balitsky:1989ry}
I.~I.~Balitsky, V.~M.~Braun and A.~V.~Kolesnichenko,
``Radiative Decay Sigma+ ---{\ensuremath{>}} p gamma in Quantum Chromodynamics,''
Nucl. Phys. B \textbf{312} (1989), 509-550
%doi:10.1016/0550-3213(89)90570-1
%647 citations counted in INSPIRE as of 29 Dec 2025

%\cite{Braun:1988qv}
\bibitem{Braun:1988qv}
V.~M.~Braun and I.~E.~Filyanov,
``QCD Sum Rules in Exclusive Kinematics and Pion Wave Function,''
Z. Phys. C \textbf{44} (1989), 157
%doi:10.1007/BF01548594
%513 citations counted in INSPIRE as of 29 Dec 2025


%\bibitem{brezis2011functional}
%Brezis, H. and Brézis, H., 2011. Functional analysis, Sobolev spaces and partial differential equations (Vol. 2, No. 3, p. 5). New York: Springer.

%\bibitem{Application in sci and eng}
%Lesnic, Daniel. ``Inverse problems with applications in science and engineering''. Chapman and Hall/CRC, 2021.

%\bibitem{inverse problem application}
%Kabanikhin, Sergey I. ``Inverse and ill-posed problems: theory and applications." Inverse and Ill-posed Problems. de Gruyter, 2011.


%\bibitem{Engl-1996}
%H.W. Engl and A.~Hanke, M.and~Neubauer, ``Regularization of inverse problems'', Mathematics and its Applications, vol. 375, Kluwer Academic Publishers Group, Dordrecht, 1996.

%\bibitem{approximate theory}
%Anselone PM, Davis J. Collectively compact operator approximation theory and applications to integral equations. (No Title). 1971 Dec.

%\bibitem{functional analysis}
%Yosida K. Functional analysis. Springer Science \& Business Media; 2012 Dec 6.

% \bibitem{linear integral equations}
% Kress R. Linear integral equations. New York: springer; 1989.

% \bibitem{Tik for Fredholm equation}
% Groetsch, Charles.``The theory of Tikhonov regularization for Fredholm equations." Boston Pitman Publication 104 (1984).

% \bibitem{Regularization-tools}
% Hansen, Per Christian.``Regularization tools: A Matlab package for analysis and solution of discrete ill-posed problems." Numerical algorithms 6.1 (1994): 1-35.

% \bibitem{Modern regularization methods}
% Benning, Martin, and Martin Burger. ``Modern regularization methods for inverse problems." Acta numerica 27 (2018): 1-111.

% \bibitem{Computational methods for IP}
% Vogel, Curtis R. ``Computational methods for inverse problems''. Society for Industrial and Applied Mathematics, 2002.

% \bibitem{Singular values and condition numbers}
% Allen Jr, Richard C., et al.``Singular values and condition numbers of Galerkin matrices arising from linear integral equations of the first kind." Journal of mathematical analysis and applications 109.2 (1985): 564-590.

% \bibitem{The posterier parameter choice}
% Engl, Heinz W., and Helmut Gfrerer. ``A posteriori parameter choice for general regularization methods for solving linear ill-posed problems." Applied numerical mathematics 4.5 (1988): 395-417.

% \bibitem{HPC-1993}
% DP O’Leary,  Hansen P C . ``The Use of the L-Curve in the Regularization of Discrete Ill-Posed Problems[J]''. SIAM Journal on Scientific Computing, 1993, 14(6):1487-1503.

% \bibitem{L curve for discrete ill-posed problem}
% Hansen, Per Christian. ``Analysis of discrete ill-posed problems by means of the L-curve." SIAM review 34.4 (1992): 561-580.

%\cite{Xiong:2025obq}
% \bibitem{Xiong:2025obq}
% A.~S.~Xiong, J.~Hua, T.~Wei, F.~S.~Yu, Q.~A.~Zhang and Y.~Zheng,
%``Ill-Posedness in Limited Discrete Fourier Inversion and Regularization for Quasi Distributions in LaMET,''
% [arXiv:2506.16689 [hep-lat]].
%3 citations counted in INSPIRE as of 04 Nov 2025

%\bibitem{finite element method}
% Dhatt G, Lefrançois E, Touzot G. Finite element method. John Wiley \& Sons; 2012 Dec 27.

 %\bibitem{Intro to elm}
% Nikishkov GP. Introduction to the finite element method. University of Aizu. 2004:1-70.

\bibitem{Kirsch-2011}
A.~Kirsch, ``An introduction to the mathematical theory of inverse
problems'', second ed., Applied Mathematical Sciences, vol. 120, Springer, New
York, 2011.

\bibitem{Oldest1}
A.~N.~Tikhonov, ``On the stability of inverse problems," Dokl. akad. nauk sssr. Vol. 39. No. 5. 1943.

\bibitem{Oldest2}
A.~N.~Tikhonov, ``Solution of incorrectly formulated problems and the regularization method," Sov Dok 4 (1963): 1035-1038.



\bibitem{Regularization_tools}
P.~C.~Hansen,  ``Regularization tools: A Matlab package for analysis and solution of discrete ill-posed problems," Numerical algorithms 6.1 (1994): 1-35.

\bibitem{Wei_Ting}
T.~Wei and Y.~H.~Luo, ``A generalized quasi-boundary value method for recovering a source in a fractional diffusion-wave equation," Inverse Problems 38.4 (2022): 045001.

\bibitem{Yan_Xiong_Bin}
X.~B.~Yan,  Z.~Q.~Xu and M.~Zheng, ``Bayesian Inversion with Neural Operator (BINO) for modeling subdiffusion: Forward and inverse problems," Journal of Computational and Applied Mathematics 454 (2025): 116191.

%\cite{Mutuk:2025lak,Zhao:2024drr}
\bibitem{Mutuk:2025lak}
H.~Mutuk,
``Reappraisal of the rho meson in nuclear matter by the inverse QCD sum rules method,''
Phys. Rev. D \textbf{111} (2025), 094029
%doi:10.1103/PhysRevD.111.094029
[arXiv:2503.10343 [hep-ph]].
%4 citations counted in INSPIRE as of 05 May 2026


%\cite{}
\bibitem{Zhao:2024drr}
Z.~X.~Zhao, Y.~P.~Xing and R.~H.~Li,
``A progress in the inverse matrix method in QCD sum rules,''
Eur. Phys. J. C \textbf{84} (2024) 1105
%doi:10.1140/epjc/s10052-024-13452-8
[arXiv:2407.09819 [hep-ph]].
%6 citations counted in INSPIRE as of 05 May 2026


%\cite{Li:2022qul}
\bibitem{HnLi-LCDA}
H.~n.~Li,
``Dispersive derivation of the pion distribution amplitude,''
Phys. Rev. D \textbf{106} (2022), 034015
%doi:10.1103/PhysRevD.106.034015
[arXiv:2205.06746 [hep-ph]].
%13 citations counted in INSPIRE as of 04 Nov 2025

%\cite{Li:2022jxc}
\bibitem{HnLi-mixing}
H.~n.~Li,
``Dispersive analysis of neutral meson mixing,''
Phys. Rev. D \textbf{107} (2023), 054023
%doi:10.1103/PhysRevD.107.054023
[arXiv:2208.14798 [hep-ph]].
%21 citations counted in INSPIRE as of 05 May 2026

\bibitem{HnLi-SMparameters}
H.~n.~Li,
``Dispersive constraints on fermion masses,''
Phys. Rev. D \textbf{107} (2023), 094007
%doi:10.1103/PhysRevD.107.094007
[arXiv:2302.01761 [hep-ph]].
%12 citations counted in INSPIRE as of 05 May 2026

H.~n.~Li,
``Dispersive determination of electroweak-scale masses,''
Phys. Rev. D \textbf{108} (2023), 054020
%doi:10.1103/PhysRevD.108.054020
[arXiv:2304.05921 [hep-ph]].
%11 citations counted in INSPIRE as of 05 May 2026

H.~n.~Li,
``Dispersive determination of neutrino mass ordering,''
Nucl. Phys. B \textbf{1018} (2025), 116978
%doi:10.1016/j.nuclphysb.2025.116978
[arXiv:2306.03463 [hep-ph]].
%10 citations counted in INSPIRE as of 05 May 2026

\bibitem{HnLi-NP}
H.~n.~Li,
``Dispersive determination of fourth generation quark masses,''
Phys. Rev. D \textbf{109} (2024), 115024
%doi:10.1103/PhysRevD.109.115024
[arXiv:2309.15602 [hep-ph]].
%9 citations counted in INSPIRE as of 05 May 2026


%\cite{Xiong:2025obq}
\bibitem{Xiong:2025obq}
A.~S.~Xiong, J.~Hua, Y.~F.~Ling, T.~Wei, F.~S.~Yu, Q.~A.~Zhang and Y.~Zheng,
``Ill-posedness in limited discrete Fourier inversion and regularization for quasi distributions in LaMET,''
Eur. Phys. J. C \textbf{85} (2025), 1409
%doi:10.1140/epjc/s10052-025-15130-9
[arXiv:2506.16689 [hep-lat]].
%11 citations counted in INSPIRE as of 06 May 2026

%\cite{Ling:2025olz}
\bibitem{Ling:2025olz}
Y.~F.~Ling, M.~H.~Chu, J.~Liang, J.~Hua, A.~S.~Xiong and Q.~A.~Zhang,
``Approaches to the inverse Fourier transformation with limited and discrete data,''
Eur. Phys. J. C \textbf{86} (2026), 379
%doi:10.1140/epjc/s10052-026-15528-z
[arXiv:2511.03593 [hep-lat]].
%4 citations counted in INSPIRE as of 06 May 2026

%\cite{Xiong:2025bmd}
\bibitem{Xiong:2025bmd}
A.~S.~Xiong, Q.~W.~Yuan, M.~Z.~Liu, F.~S.~Yu, Z.~W.~Liu and L.~S.~Geng,
``Solving the Inverse Source Problem in Femtoscopy with a Toy Model,''
[arXiv:2512.06904 [hep-ph]].
%3 citations counted in INSPIRE as of 06 May 2026


%\cite{Candido:2024hjt}
\bibitem{Candido:2024hjt}
A.~Candido, L.~Del Debbio, T.~Giani and G.~Petrillo,
``Bayesian inference with Gaussian processes for the determination of parton distribution functions,''
Eur. Phys. J. C \textbf{84} (2024) no.7, 716
%doi:10.1140/epjc/s10052-024-13100-1
[arXiv:2404.07573 [hep-ph]].
%24 citations counted in INSPIRE as of 19 May 2026

%\cite{Liang:2019frk}
\bibitem{Liang:2019frk}
J.~Liang \textit{et al.} [XQCD],
``Towards the nucleon hadronic tensor from lattice QCD,''
Phys. Rev. D \textbf{101} (2020) no.11, 114503
%doi:10.1103/PhysRevD.101.114503
[arXiv:1906.05312 [hep-ph]].
%78 citations counted in INSPIRE as of 19 May 2026

%\cite{Backus:1968svk}
\bibitem{Backus:1968svk}
G.~Backus and F.~Gilbert,
``The Resolving Power of Gross Earth Data,''
Geophys. J. Int. \textbf{16} (1968) no.2, 169-205
%doi:10.1111/j.1365-246X.1968.tb00216.x
%128 citations counted in INSPIRE as of 19 May 2026

%\cite{Asakawa:2000tr}
\bibitem{Asakawa:2000tr}
M.~Asakawa, T.~Hatsuda and Y.~Nakahara,
``Maximum entropy analysis of the spectral functions in lattice QCD,''
Prog. Part. Nucl. Phys. \textbf{46} (2001), 459-508
%doi:10.1016/S0146-6410(01)00150-8
[arXiv:hep-lat/0011040 [hep-lat]].
%614 citations counted in INSPIRE as of 19 May 2026

%\cite{Burnier:2013nla}
\bibitem{Burnier:2013nla}
Y.~Burnier and A.~Rothkopf,
``Bayesian Approach to Spectral Function Reconstruction for Euclidean Quantum Field Theories,''
Phys. Rev. Lett. \textbf{111} (2013), 182003
%doi:10.1103/PhysRevLett.111.182003
[arXiv:1307.6106 [hep-lat]].
%216 citations counted in INSPIRE as of 19 May 2026

\bibitem{YanfeiYang}
Y.~Wang, A.~G.~Yagola, C.~Yang, ``Computational Methods for Applied Inverse Problem'', Higher Education Press, Beijing, 2012.


%%%%%%%%%%%%%%%%%%%%%%%%%%%%%%%%%%%%%%%%%%%%%%%






\end{thebibliography}
